\documentclass[aps,prb,rsi,amsmath,amssymb,reprint]{revtex4-2}

\newenvironment{supplemental}

\usepackage[dvipdfmx]{graphicx}
\usepackage{color}
\usepackage{dcolumn}
\usepackage{bm}

\usepackage{midpage}
\usepackage{here}
\usepackage{float}
\usepackage{makecell}

\begin{document}


%

%
%

\title[Manuscript submitted to J. Chem. Phys.]
{
Quantum fluctuation-driven transport crossover between two liquid states in distinguishable helium-4
}

\author{Mika Tanabe}
\altaffiliation[]{Present affiliation: Sumitomo Electric Industries, Ltd., Osaka 541-0041, Japan.}
\author{Momoko Tsujimoto}
\altaffiliation[]{Present affiliation: KYOCERA Communication Systems Co., Ltd., Kyoto 612-8450, Japan.}
\author{Kenichi Kinugawa}%
\email{Author to whom correspondence should be addressed.  kinugawa@cc.nara-wu.ac.jp}
\affiliation{ 
Department of Chemistry, Graduate School of Humanities and Sciences, Nara Women's University, Nara 630-8506, Japan
}

\date{\today}

\begin{abstract}

We show the emergence of a quantum fluctuation-driven transport crossover between two liquid states in distinguishable helium-4 obeying Boltzmann statistics, in the absence of atomic exchange.
Using path integral centroid molecular dynamics simulations over 0.1–3.3 K and 1–60 bar, we investigate the transport properties of two distinct liquid states: the low quantum dispersion liquid (LQDL) and the high quantum dispersion liquid (HQDL).
While LQDL exhibits conventional liquid behavior consistent with the Stokes–Einstein (SE) relation, HQDL emerges at lower temperatures and displays anomalous gas-like transport characterized by superdiffusion and ultralow viscosity, accompanied by a breakdown of the SE relation.
This counterintuitive emergence of gas-like dynamics upon cooling reflects the dominant role of nuclear quantum fluctuations, in contrast to thermal fluctuations at higher temperatures.
Across the LQDL–HQDL boundary, we identify a transport crossover marked by a qualitative change in the velocity autocorrelation function (VAF), a transition in the Prandtl number, and the emergence of transport minima in shear and kinematic viscosities, thermal conductivity, and thermal diffusivity.
These minima reflect a crossover from liquid-like to gas-like transport upon cooling in the low-temperature subcritical region, in addition to the universal transport minima observed in the supercritical regime.
The transition from oscillatory to monotonic VAF defines a second Frenkel line, distinct from the conventional Frenkel line observed in the supercritical region.
LQDL is a heat-transport-dominated dissipative fluid, whereas HQDL is a momentum-dominated inertial fluid.
These results demonstrate that nuclear quantum fluctuations alone induce gas-like liquid behavior and provide a unified picture of transport phenomena
in distinguishable helium-4 without superfluidity.

\end{abstract}

\maketitle

\section{\label{sec:introduction}INTRODUCTION}

Liquid helium-4 ($^4$He) has long served as a prototypical quantum liquid that exhibits macroscopic quantum phenomena such as superfluidity, leading to transport properties markedly different from those of classical liquids \cite{london1938,london1938PR,wilks1970,wilks1967,feynman1972,Feynman1954}.
According to Landau's two-fluid theory, below the lambda-transition temperature ($T_{\lambda}=2.17$ K at saturated vapor pressure (SVP)), $^4$He exhibits the superfluid phase (He II), which consists of a normal fluid component with nonzero viscosity and a superfluid component with zero viscosity; the fraction of the latter increases continuously upon lowering the temperature from $T_{\lambda}$.  
In He II, the thermal conductivity $\lambda$ diverges.
The prevailing view holds that He II, with its extreme fluidity, emerges only in the presence of particle exchange among indistinguishable bosons \cite{feynman1972}.
Superfluidity is generally attributed to the interplay of weak interatomic van der Waals interactions, nuclear quantum effects (NQEs), and atomic exchange arising from Bose statistics.

What state, then, would $^4$He exhibit if atomic exchange effects due to Bose statistics were absent, i.e., if the atoms obeyed Boltzmann statistics and only NQEs were present?
In the high-temperature limit, Bose statistics naturally reduce to Boltzmann statistics as exchange effects vanish.
As noted by Feynman \cite{feynman1972,boninsegni2012}, in the zero-temperature limit, the ground-state wave function of bosons is identical to that of distinguishable particles, as it is positive definite and nodeless.
This suggests that distinguishable $^4$He at atmospheric pressure is likely to remain a fluid at $T=0$, as is the case for bosonic $^4$He.
However, this argument provides no direct predictive insight into the nature of the state and macroscopic transport properties at low but finite temperatures ($T>0$).
In the absence of exchange effects, distinguishable $^4$He is expected to exhibit behavior different from that of real $^4$He governed by Bose statistics.
However, beyond an early phase diagram \cite{boninsegni2012}, the detailed nature of the finite-temperature ($T>0$) states remains unexplored.

\begin{figure}
\includegraphics[width=8.5cm]{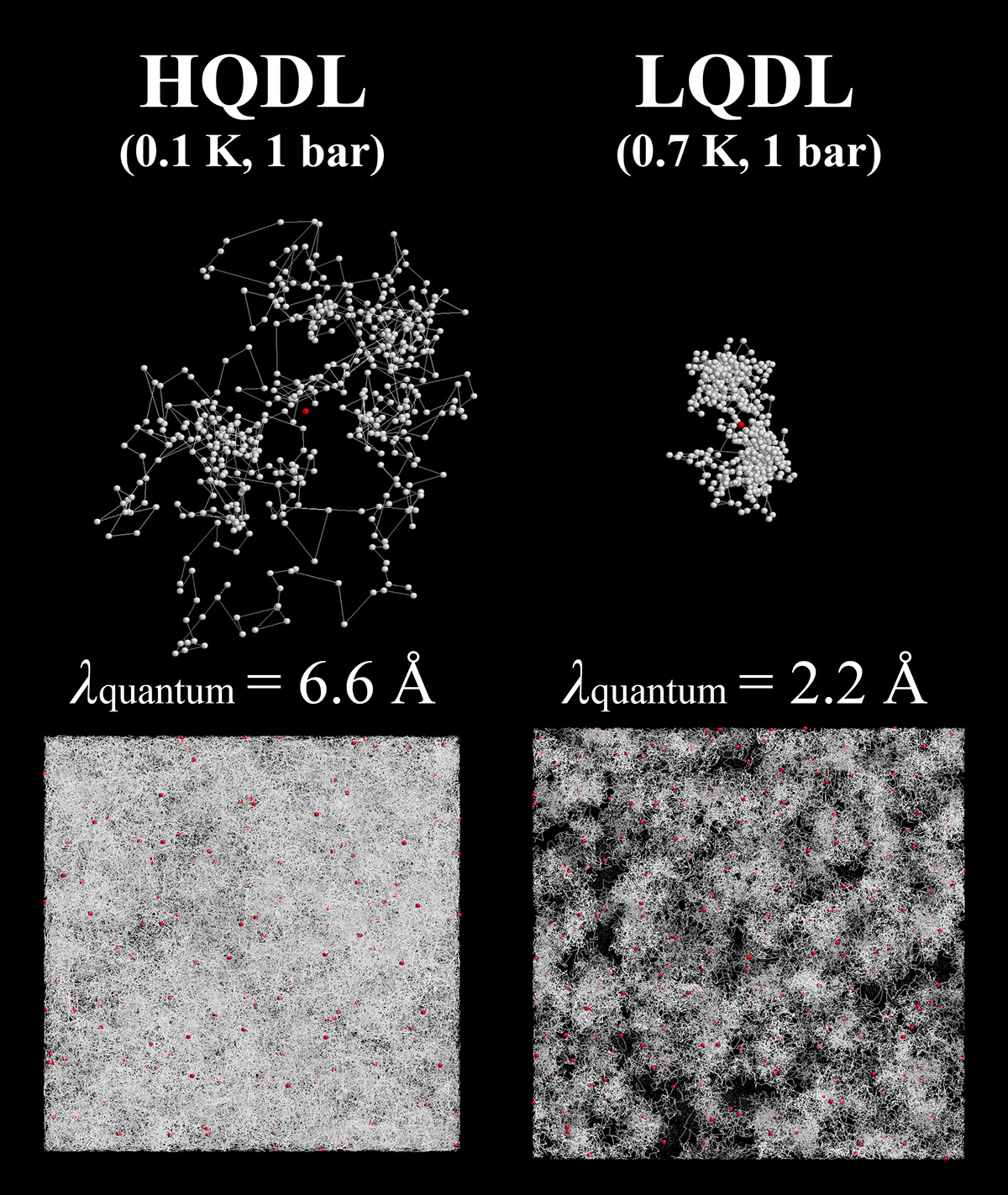}
\caption{\label{fig:graphics} 
Representative $xy$-projected snapshots of an atomic necklace and its centroid (top panels) and of all atomic necklaces and centroids in the simulation box (bottom panels) for HQDL (0.1 K, 1 bar; left) and LQDL (0.7 K, 1 bar; right).
Red and white spheres denote centroids and beads, respectively.
The left and right panels are shown on the same length scale, whereas the sizes of the centroids and beads are arbitrary.
The atomic necklaces in HQDL are substantially more spatially delocalized than those in LQDL and exhibit significant overlap among neighboring necklaces.
The indicated values of the quantum wavelength $\lambda_{\rm quantum}$ are averages for the corresponding thermodynamic conditions.
}
\end{figure}

To address this question, we recently performed extensive path integral centroid molecular dynamics (CMD) \cite{cao1993,cao5093,cao5106,cao6157,cao6168,Jang1999} simulations of distinguishable $^4$He over a wide range of temperatures and pressures, yielding a detailed state diagram on the pressure-temperature ($P$-$T$) plane \cite{tsujimoto2024}.
We revealed that two distinct liquid states, arising solely from NQEs, exist below the experimental lambda-transition temperature.
These states are referred to as the low quantum dispersion liquid (LQDL) and the high quantum dispersion liquid (HQDL).
Quantum dispersion of atoms is reflected in the spatial extent of atomic necklaces, which is quantified by the atomic quantum wavelength \cite{takemoto2018,kinugawa2021,tsujimoto2024} $\lambda_{\rm quantum}$ (approximately twice the radius of gyration $R_{\rm g}$ of the path integral ring polymer, or ``atomic necklace'' consisting of ``beads'').
For reference, Fig.~\ref{fig:graphics} illustrates the characteristic necklace configurations in the LQDL and HQDL states through representative snapshots of individual necklaces and of all necklaces in the corresponding simulation box.
Similar configurations were presented in our previous paper \cite{tsujimoto2024}.
The HQDL is characterized by a pronounced spatial delocalization of atoms; $\lambda_{\rm quantum}$ in the HQDL is substantially larger than that in the LQDL \cite{kinugawa2021,tsujimoto2024}.
The LQDL, which exists at higher temperatures than the HQDL, is nearly equivalent to real He I.
Thus, HQDL represents a distinct quantum liquid state with no direct analog in real $^4$He.
The conditions under which the LQDL and HQDL states emerge are summarized in the state diagram presented later in Fig.~\ref{fig:phase} and are discussed in detail in Sec.~\ref{sec:state}.

The transition between LQDL and HQDL is not a thermodynamic transition but a continuous crossover.
It is described by the expansion factor \cite{kinugawa2021,tsujimoto2024} $\alpha_{\lambda}=\lambda_{\mathrm{quantum}}/\lambda_{\mathrm{dB}}$, where the de Broglie thermal wavelength is given by $\lambda_{\mathrm{dB}}=\hbar\sqrt{2\pi\beta/m}$, with $\beta=1/(k_{\rm{B}}T)$, $m$ the atomic mass, and $k_{\rm{B}}$ the Boltzmann constant.
A similar two-state behavior also appears in the glassy regime \cite{kinugawa2021,tsujimoto2024}, yielding the low quantum dispersion amorphous solid (LQDA) and the high quantum dispersion amorphous solid (HQDA).\footnote{In Ref.~\onlinecite{kinugawa2021}, these phases were abbreviated as ``LDA'' and ``HDA,'' respectively. However, these abbreviations may be confused with the low- and high-density amorphous states of classical systems (e.g., water \cite{mishima3961998}). To avoid this ambiguity, we adopt the abbreviations LQDA and HQDA, following Ref.~\onlinecite{tsujimoto2024}.}
The emergence of two distinct states in both the liquid and glassy regimes can be naturally interpreted as a manifestation of {\it quantum polyamorphism}, arising from changes in the spatial extent of atomic quantum delocalization. 
This is distinct from classical polyamorphism, which involves density and structural changes evidenced by diffraction experiments \cite{mishima3961998,Poole1997,Katayama2004}.
In fact, LQDA and HQDA in compressed conditions \cite{kinugawa2021} correspond to the {\it trapped} and {\it tunneling} regimes in the quantum Lennard-Jones glass, respectively \cite{markland2011,markland2012}.
Furthermore, these two regimes can be viewed as analogous to the two-level systems or double-well potential model proposed by Phillips and Anderson for low-temperature glasses \cite{phillips1972,anderson1972}.  
This correspondence suggests that the two-state description is not specific to distinguishable $^4$He but instead likely reflects a general feature of quantum systems in the absence of atomic exchange.

In our recent study \cite{tsujimoto2024}, a preliminary analysis of the velocity autocorrelation function (VAF) suggested that the HQDL may possess dynamical characteristics distinct from those of the LQDL.
However, the transport properties of these two liquid states remain essentially unexplored.
Calculating the collective dynamical properties of quantum liquids is considerably more challenging than evaluating their static properties. 
To estimate such properties, quantum dynamical methods such as CMD \cite{yonetani2003,yonetani2004,imaoka2017} and ring polymer molecular dynamics (RPMD) \cite{craig2004,Habershon2013,sutherland2021} must be employed. 
CMD has been successfully applied to the real-time dynamics of quantum liquids, yielding quantitative agreement with the experimental dynamic structure factors $S(k,\omega)$ for systems such as liquid para-H$_2$ \cite{kinugawa1998,bermejo2000} and He I \cite{miura1999}.
Furthermore, transport properties such as shear viscosity $\eta_{\rm{s}}$ and thermal conductivity $\lambda$ have been estimated using CMD combined with the centroid approximation of the Green-Kubo formula \cite{imaoka2017,yonetani2003,yonetani2004}.
In particular, this approach reproduces the experimental temperature dependence quantitatively for liquid para-H$_2$ \cite{yonetani2003,yonetani2004} and He I \cite{imaoka2017}.
These findings demonstrate that CMD can reliably capture both microscopic dynamics and macroscopic transport properties in quantum liquids.
Accordingly, the transport properties of the present distinguishable quantum liquid can be reliably evaluated using CMD and the centroid-approximated Green-Kubo scheme.

On the basis of these considerations, we aim to investigate the dynamical and transport properties of the two distinct liquid states of distinguishable $^4$He. 
We employ CMD simulations in conjunction with the centroid approximation of the Green–Kubo formalism to evaluate the self-diffusion coefficient $D$, shear viscosity $\eta_{\rm{s}}$, kinematic viscosity $\nu$, thermal conductivity $\lambda$, and thermal diffusivity $\alpha$ over a wide $P$-$T$ range ($T=$0.1-3.3 K and $P=$1-60 bar).

In this paper, we show that HQDL, a unique liquid state of distinguishable $^4$He, exhibits gas-like quantum fluidity driven solely by NQEs, despite the absence of superfluidity arising from atomic exchange. 
Furthermore, analysis of the relaxation timescales reveals a decoupling between single-particle diffusion and collective shear relaxation, indicating a breakdown of the Stokes–Einstein (SE) relation.
We further demonstrate that the LQDL–HQDL transition can be interpreted as a liquid–liquid transport crossover across a second Frenkel line in the low-temperature subcritical regime. 
Here, the Frenkel line denotes a dynamical crossover between oscillatory (liquid-like) and monotonic (gas-like) decay of the VAF, and was originally proposed to distinguish these two dynamical regimes in supercritical fluids by Brazhkin \textit{et al.} \cite{brazhkin2012,brazhkin2013}
These results provide a unified picture of quantum-driven transport without superfluidity and highlight the fundamental role of NQEs in determining transport properties.

The paper is organized as follows.
Section~\ref{sec:method} describes the methodology of the CMD simulations and the evaluation of the transport properties.
The calculated results are presented in Sec.~\ref{sec:results}.
Section~\ref{sec:analysis} provides further analysis based on the obtained transport properties, and Sec.~\ref{sec:discussion} presents a comprehensive discussion.
A schematic illustration of the transport crossover is provided in Fig.~\ref{fig:schematic}, which summarizes the overall picture discussed in this work.
Finally, the conclusions are summarized in Sec.~\ref{sec:conclusions}.
A substantial amount of supplementary data is provided in the Supplementary Material.

\section{\label{sec:method}METHOD}

\subsection{\label{sec:CMDsimulation}CMD methodology}

We employed essentially the same CMD methodology as in our previous 
studies \cite{kinugawa1997,kinugawa1998,yonetani2003,yonetani2004,imaoka2017,takemoto2018,kinugawa2021,tsujimoto2024}. 
For clarity, we summarize the computational formalism used in this work.

In the discretized Feynman path integral representation, 
the canonical partition function of a quantum Boltzmann system composed of $N$
distinguishable atoms is
\begin{eqnarray}
\label{eq:partition}
Z_{NVT}&=&\frac{1}{N!}\left(\frac{mN_{\rm{b}}}{2\pi\beta\hbar^2}\right)^{\frac{3NN_{\rm{b}}}{2}}\\ \nonumber
&\times&\int\cdots\int
\prod^{N}_{i=1}\prod^{N_{\rm{b}}}_{n=1}
d{\textit{\textbf{r}}}_{i}^{(n)}\exp\left[-\beta H_{\rm{system}}\right],
\end{eqnarray}   
where the discretized atomic coordinates $\{{\textit{\textbf{r}}}_{i}^{(n)}\}$
(beads)  enter the isomorphic classical Hamiltonian \cite{chandler1981}
\begin{eqnarray}
\label{eq:hamiltonian}
H_{\rm{system}}
\left(
\{{\textit{\textbf{r}}}_{i}^{(l)}\}
\right)
&=&\sum_{i=1}^{N} \sum_{n=1}^{N_{\rm{b}}}
\frac{1}{2}k_{\rm{s}}\left({\textit{\textbf{r}}}_i^{(n)}-{\textit{\textbf{r}}}_i^{(n+1)}\right)^2 \nonumber\\
&+&\frac{1}{N_{\rm{b}}}
\Phi\left(\{\textit{\textbf{r}}_i^{(l)}\}\right)
.
\end{eqnarray}   
Here, $N_{\rm{b}}$ is the Trotter number, $\Phi$ is the system potential,
$m$ is the atomic mass, $\tau_n=n\beta\hbar/N_{\rm{b}}$ $(0<{\tau_n}\leq\beta\hbar)$ is the imaginary time, $k_{\rm{s}}=mN_{\rm{b}}/\beta^2\hbar^2$,
and the cyclic boundary condition 
 ${\textit{\textbf{r}}}_i^{(N_{\rm{b}}+1)}= {\textit{\textbf{r}}}_i^{(1)}$
 is imposed.
The potential $\Phi$ is given by the sum over all pairs of the two-body potential $\phi$: $\Phi=\sum_{i>j}\phi(r_{ij})$.
Thus, the quantum system is mapped onto a classical system of $N$ closed polymers
(necklaces), each consisting of $N_{\rm{b}}$ beads connected sequentially by harmonic springs with the Hooke constant $k_{\rm{s}}$.

The spatial extent of atomic necklaces is quantified by the quantum wavelength,
which is defined as \cite{tsujimoto2024,kinugawa2021,takemoto2018}
\begin{eqnarray}
\label{eq:wavelength}
\lambda_{\rm quantum}=\sqrt{\frac{1}{N}\sum_{i=1}^N
\langle |{\textit{\textbf{r}}}_{i}^{(n)}-{\textit{\textbf{r}}}_{i}^{(n+N_{\rm b}/2)}|^2 \rangle_n}
\end{eqnarray}   
where the average is taken over all bead indices $n$ and sampled configurations.
The quantum wavelength represents the effective diameter of an atomic necklace and is approximately twice the radius of gyration \cite{rubinstein2003} $R_{\rm g}$ \cite{kinugawa2021}.

Introducing the centroid coordinates $\{{\it{\bf{r}}}_{{\rm{c}}i}\}$,
the partition function can be rewritten as 
\begin{eqnarray}
\label{eq:partitionrewritten}
Z_{NVT}=\frac{1}{N!}
\int\cdots\int
\prod^{N}_{i=1}
d{\textit{\textbf{r}}}_{{\rm{c}}i}\rho_{{\rm{c}}}
(\{{\textit{\textbf{r}}}_{{\rm{c}}i}\}),
\end{eqnarray}   
where $\rho_{{\rm{c}}}$ is the centroid density distribution given by
\begin{widetext}
\begin{eqnarray}
\label{eq:centroiddensityNVT}
\rho_{{\rm{c}}}
\left(\{{\textit{\textbf{r}}}_{{\rm{c}}i}\}\right)&=&
\left(\frac{mN_{\rm{b}}}{2\pi\beta\hbar^2}\right)^{\frac{3NN_{\rm{b}}}{2}}\int\cdots\int\prod^{N}_{i=1}\prod^{N_{\rm{b}}}_{j=1}
d{\textit{\textbf{r}}}_i^{(j)} \delta({\textit{\textbf{r}}}_{{\rm{c}}i}-{\bar{\textit{\textbf{{r}}}}}_{i})\exp\left[-{\beta}H_{\rm{system}}(\{\textit{\textbf{r}}_{i}^{(l)}\})\right],
\end{eqnarray}  
\end{widetext}
with 
${\bar{\textit{\textbf{{r}}}}}_{i}=N_{\rm{b}}^{-1}\sum_{n=1}^{N_{\rm{b}}}{\textit{\textbf{r}}}_i^{(n)}$. 
The corresponding potential of mean force (centroid potential) is defined as
\begin{eqnarray}
\label{eq:centroidpotential}
V_{{\rm{c}}}\left(\{{\textit{\textbf{r}}}_{{\rm{c}}i}\}\right)&=&-\frac{1}{\beta}
\log\left\lbrack\left(\frac{2\pi\beta}{m}\right)^{\frac{3N}{2}}\rho_{{\rm{c}}}
\left(\{{\textit{\textbf{r}}}_{{\rm{c}}i}\}\right)\right\rbrack.
\end{eqnarray}

By introducing the centroid  momenta $(\{{\textit{\textbf{p}}}_{{\rm{c}}i}\})$,
the quasi-classical Hamiltonian becomes 
\begin{eqnarray}
\label{eq:centroidhamiltonian}
H_{\rm c}\left(\{{\textit{\textbf{p}}}_{{\rm{c}}i}\},\{{\textit{\textbf{r}}}_{{\rm{c}}i}\}\right)
=\sum_{i=1}^{N}\frac{{\textit{\textbf{p}}}_{{\rm{c}}i}^2}{2m}+V_{{\rm{c}}}\left(\{{\textit{\textbf{r}}}_{{\rm{c}}i}\}\right).
\end{eqnarray}  
The corresponding quasi-classical partition functions in the canonical (NVT),
microcanonical (NVE), and isothermal-isobaric (NPT) ensembles are summarized as follows:
\begin{subequations}
\label{eq:partitionfunctions}
\begin{widetext}
\begin{align}
\label{eq:partitionNVT}
Z_{NVT}&=
\frac{1}{N!}\int\cdots\int
\prod_{i=1}^{N}\frac{{\textit{d\textbf{p}}}_{{\rm{c}}i}}{(2\pi)^3}{d\textit{\textbf{r}}}_{{\rm{c}}i}
\exp\left\lbrack-{\beta}
H_{\rm{c}}\left(\{{\textit{\textbf{p}}}_{{\rm{c}}i}\},\{{\textit{\textbf{r}}}_{{\rm{c}}i}\}\right)\right\rbrack\\
\label{eq:partitionNVE}
Z_{NVE}&=\frac{1}{N!}\int\cdots\int
\prod_{i=1}^{N}\frac{{\textit{d\textbf{p}}}_{{\rm{c}}i}}{(2\pi)^3}{d\textit{\textbf{r}}}_{{\rm{c}}i}
\delta\left(H_{\rm{c}}\left(\{{\textit{\textbf{p}}}_{{\rm{c}}i}\},\{{\textit{\textbf{r}}}_{{\rm{c}}i}\}\right)-E\right),\\
\label{eq:partitionNPT}
Z_{NPT}&=
\frac{1}{V_0}
{\int}
dV
\exp(-{\beta}P_{\rm{ex}}V)Z_{NVT}.
\end{align}
\end{widetext}
\end{subequations}
where  $E$ is the conserved energy of the quasi-classical system
and $V_0$ is reference volume that renders the partition function dimensionless \cite{frenkel2002}.
Equation~(\ref{eq:partitionNVT}) is nothing but the quasi-classical representation of the canonical partition function for a quantum Boltzmann system, as presented in Feynman’s textbooks \cite{feynman1965,feynman1972}.
We note that no additional approximations have been introduced so far beyond the discretization in Eq.~(\ref{eq:partition}).

The centroid force acting on the $i$-th centroid is defined as 
the negative gradient of the centroid potential $V_{\rm{c}}$,
\begin{eqnarray}
\label{eq:force}
{\textit{\textbf{F}}}_{{\rm{c}}i}
&=&
-\frac{{\partial}{V}_{\rm{c}}\left(
\{{\textit{\textbf{r}}}_{\rm{c}i}\}
\right)}
{{\partial}{\textit{\textbf{r}}}_{{\rm{c}}i}}
=
-\left\langle\frac{1}{N_{\rm{b}}}\sum_{n=1}^{N_{\rm{b}}}
\frac{{\partial}{\Phi}\left(
\{{\textit{\textbf{r}}}_{i}^{(l)}\}
\right)}
{{\partial}{\textit{\textbf{r}}}_{i}^{(n)}}\right\rangle   \\
&\cong&
-\frac{1}{N_{\rm{b}}}\sum_{n=1}^{N_{\rm{b}}}
\frac{{\partial}{\Phi}\left(
\{{\textit{\textbf{r}}}_{i}^{(l)}\}
\right)}
{{\partial}{\textit{\textbf{r}}}_{i}^{(n)}}
=\sum_{j\neq{i}}^N{\textit{\textbf{f}}}_{ij}^{({\rm c})}, \nonumber
\end{eqnarray}   
where angle brackets denote the average over bead configurations for a given centroid
configuration $\{{\textit{\textbf{r}}}_{i}^{(l)}\}$, while  ${\textit{\textbf{f}}}_{ij}^{\rm (c)}$ is the force exerted on the $i$-th centroid by the $j$-th centroid,
\begin{eqnarray}
\label{eq:eachforce}
{\textit{\textbf{f}}}_{ij}^{({\rm c})}=-\frac{1}{N_{\rm b}}\sum_{n=1}^{N_{\rm b}}\frac{{\partial}\phi(|{\textit{\textbf{r}}}_i^{(n)}-{\textit{\textbf{r}}}_j^{(n)}|)}{\partial{\textit{\textbf{r}}}_i^{(n)}}. 
\end{eqnarray}
In the normal mode CMD (NMCMD) algorithm \cite{kinugawa1997}, the statistical average appearing in Eq. (\ref{eq:force}) is approximated by the force generated at each instantaneous bead configuration, yielding the approximate expression in Eq. (\ref{eq:force}).

In this study, CMD simulations were performed in three statistical ensembles: NPT, NVT, and NVE. 
The centroid equation of motion (EOM) for the NPT ensemble is given by  \cite{takemoto2018,kinugawa2021,tsujimoto2024}
\begin{eqnarray}
\label{eq:eom}
m{\ddot{\textit{\textbf{{r}}}}}_{{\rm{c}}i}
&=&
{\textit{\textbf{F}}}_{{\rm{c}}i}-m\dot{\xi}_1\dot{{\textit{\textbf{r}}}}_{{\rm{c}}i}-\frac{m}{N}\dot{\epsilon}\dot{\textit{\textbf{r}}}_{{\rm{c}}i},\;\;
i=1,2,\cdots,N,
\end{eqnarray}   
where
$\dot{\xi}_1$ is the velocity of the first layer of the Nos\'e-Hoover chain
(NHC) thermostat \cite{martyna1992} attached to all centroids, and
$\dot{\epsilon}$ is the velocity of the Andersen barostat controlling the system volume.
The resulting CMD trajectory samples the NPT probability distribution
$\exp[-{\beta}(H_{\rm{c}}+P_{\rm{ex}}V)]$ in Eq. (\ref{eq:partitionNPT}).
For the NVT ensemble, the barostat term (the third term on the right-hand side
of Eq. (\ref{eq:eom})) is omitted, and the CMD trajectory samples the canonical distribution $\exp[-{\beta}H_{\rm{c}}]$ in Eq. (\ref{eq:partitionNVT}).
For the NVE ensemble, both the thermostat and barostat terms in Eq. (\ref{eq:eom}) are removed, yielding the microcanonical distribution $\delta\left(H_{\rm c}-E\right)$ in Eq. (\ref{eq:partitionNVE}).
Thus, for a given ensemble, the statistical averages of static configurations obtained from CMD are identical to those from RPMD and path integral Monte Carlo simulations within statistical error.

For all statistical ensembles, the bead coordinates $\{{\textit{\textbf{r}}}_{i}^{(j)}\}$ of each atom are transformed to 
the NMCMD framework \cite{kinugawa1997}. 
Among the resulting degrees of freedom, all modes except the centroid mode ($N_{\rm{b}}$-th normal mode) are coupled to massive NHC thermostats \cite{tobias1993}.
Equation (\ref{eq:eom}) is integrated together with the EOMs for the 
normal modes and the attached massive NHC thermostats in the framework of a multiple time scale integrator called the reference system propagator algorithm (RESPA) \cite{martyna1996}.

\subsection{\label{sec:transport}Centroid-based Green–Kubo formalism for transport properties}

The method for estimating the transport properties is the same as that adopted in our previous studies \cite{yonetani2003,yonetani2004,imaoka2017}.
Following the Green-Kubo formula in linear response theory \cite{kubo1991}, a 
transport property $K$ is expressed as the real-time integral of the canonical correlation function $C_{\rm{can}}(t)$ of the relevant operator $\hat{A}$,
\begin{eqnarray}
\label{eq:green}
K=F\int_0^{\infty}C_{\rm{can}}(t)dt,
\end{eqnarray}   
where $F$ is a property-dependent prefactor, and 
\begin{eqnarray}
\label{eq:canonical}
C_{\rm{can}}(t)=\frac{1}{\beta\hbar}\int_0^{\beta\hbar}\langle\hat{A}(t)\hat{A}(-i\tau)\rangle{d}\tau.
\end{eqnarray}   
In the centroid approximation \cite{yonetani2003,yonetani2004,imaoka2017}, the canonical correlation function is replaced by the time correlation function of the corresponding centroid variable $A_{\rm{c}}$.
\begin{eqnarray}
\label{eq:approx}
C_{\rm{can}}(t)\cong\langle A_{\rm{c}}(t)A_{\rm{c}}(0) \rangle
\equiv
C^{(\rm{c})}(t).
\end{eqnarray}  
If $\hat{A}$  is  linear in the atomic positions or momenta, the equality $C_{\rm{can}}(t)=C^{(\rm{c})}(t)$ holds exactly \cite{Jang1999}.

\subsubsection{\label{sec:self-diffusion}Self-diffusion coefficient}

The  self-diffusion coefficient $D$ is obtained from
\begin{eqnarray}
\label{eq:diffusion}
D=\frac{1}{3}\int_0^{\infty}C_{\rm v}^{\rm (c)}(t)dt,
\end{eqnarray}   
where  $C_{\rm v}^{\rm(c)}(t)$ is the centroid velocity 
autocorrelation function (VAF),
\begin{eqnarray}
\label{eq:vaf}
C_{\rm v}^{\rm(c)}(t)=\frac{1}{N}\sum_{i=1}^{N}\langle{\textbf{\textit{v}}}_{{\rm{c}}i}(t)\cdot{\textbf{\textit{v}}}_{{\rm c}i}(0)\rangle,
\end{eqnarray}   
where ${\textbf{\textit{v}}}_{{\rm c}i}=
(v^{({\rm c})}_{ix},v^{({\rm c})}_{iy},v^{({\rm c})}_{iz})=\dot{{\textit{\textbf{r}}}}_{{\rm{c}}i}$.
The normalized VAF is defined as
\begin{eqnarray}
\label{eq:normalizedvaf}
\bar{C}_{\rm v}^{\rm(c)}(t)=C_{\rm v}^{\rm (c)}(t)/C_{\rm v}^{\rm (c)}(0).
\end{eqnarray}

\subsubsection{\label{sec:shearviscosity_eq}
Shear viscosity}

The $\alpha\gamma$ component of the centroid stress tensor,
$\overleftrightarrow{\sigma}^{(\rm c)}_{\alpha\gamma}$, is defined as
\begin{eqnarray}
\label{eq:stresstensor}
\sigma^{(\rm c)}_{\alpha\gamma}(t)
&=&
m\sum_{i=1}^{N}v_{i\alpha}^{(\rm c)}(t)v_{i\gamma}^{(\rm c)}(t) \\
&+&\frac{1}{2}\sum_{i=1}^N\sum_{j\ne i}^N
\left[r_{i\gamma}^{(\rm c)}(t)-r_{j\gamma}^{(\rm c)}(t)\right]
f_{ij,\alpha}^{({\rm c})},
\nonumber
\end{eqnarray}
where
$({r}_{ix}^{(\rm c)},{r}_{iy}^{(\rm c)},{r}_{iz}^{(\rm c)})
=\textit{\textbf{r}}_i^{(\rm c)}$ 
and $(f_{ij,x}^{(\rm c)},f_{ij,y}^{(\rm c)},f_{ij,z}^{(\rm c)})=
{\textit{\textbf{f}}}_{ij}^{({\rm c})}$.

The shear viscosity is evaluated from 
the stress autocorrelation function averaged over the six off-diagonal components,
\begin{equation}
\label{eq:stresscorrelation}
C_{\rm s}^{(\rm c)}(t)
=
\langle
\sigma^{(\rm c)}_{\alpha\gamma}(t)\,
\sigma^{(\rm c)}_{\alpha\gamma}(0)
\rangle_{\alpha\gamma},
\end{equation}
as
\begin{eqnarray}
\label{eq:shearviscosity}
\eta_{\rm s}
=
\frac{1}{V k_{\rm B} T}
\int_0^{\infty}
C_{\rm s}^{(\rm c)}(t)\, dt
\equiv
\frac{S_{\eta}}{V k_{\rm B} T}.
\end{eqnarray}
The normalized stress autocorrelation function is defined as
\begin{equation}
\label{eq:normstresscorrelation}
\bar{C}_{\rm s}^{(\rm c)}(t)
=
C_{\rm s}^{(\rm c)}(t)/C_{\rm s}^{(\rm c)}(0).
\end{equation}

The high-frequency shear modulus is obtained from the initial value at $t=0$ of the stress autocorrelation function \cite{zwanzig1965},
\begin{eqnarray}
\label{eq:shearmodulus}
G_{\infty}
=\frac{1}{V k_{\rm B} T}C_{\rm s}^{(\rm c)}(0)=
\frac{1}{V k_{\rm B} T}
\langle
\sigma^{(\rm c)}_{\alpha\gamma}(0)^2
\rangle.
\end{eqnarray}

\subsubsection{\label{sec:thermalconductivity}Thermal conductivity}

The $\alpha$ component of the centroid energy current,
$J^{(\rm c)}_{\alpha}$, is given by
\begin{widetext}
\begin{eqnarray}
\label{eq:energycurrent}
J^{(\rm c)}_\alpha(t)
&=&
\sum_{i=1}^{N}
v_{i\alpha}^{(\rm c)}(t)
\left[
\frac{1}{2}m {\textit{\textbf{v}}}_{{\rm c}i}^2(t)
+\frac{1}{2N_{\rm b}}
\sum_{j\ne i}^N \sum_{n=1}^{N_{\rm b}}
\phi(|{\textit{\textbf{r}}}_i^{(n)}-{\textit{\textbf{r}}}_j^{(n)}|)
\right]\nonumber \\
&+&
\frac{1}{2}
\sum_{i=1}^N \sum_{j\ne i}^N
{\textit{\textbf{v}}}_{{\rm c}i}(t)
\cdot
\left(
{\textit{\textbf{r}}}_{{\rm c}i}(t)
-{\textit{\textbf{r}}}_{{\rm c}j}(t)
\right)
f_{ij,\alpha}^{({\rm c})},
\quad \alpha=x,y,z.
\end{eqnarray}
\end{widetext}

The thermal conductivity is obtained from the energy current autocorrelation function averaged over
the three Cartesian components,
\begin{equation}
\label{eq:energycurrentcorr}
C_{\rm e}^{(\rm c)}(t)
=
\langle
J^{(\rm c)}_\alpha(t)\,
J^{(\rm c)}_\alpha(0)
\rangle,
\end{equation}
as
\begin{eqnarray}
\label{eq:thermalconductivity}
\lambda
=
\frac{1}{V k_{\rm B} T^2}
\int_0^{\infty}
C_{\rm e}^{(\rm c)}(t)\, dt
\equiv
\frac{S_{\lambda}}{V k_{\rm B} T^2}.
\end{eqnarray}
The normalized energy current correlation function is defined as
\begin{equation}
\label{eq:normenergycurrent}
\bar{C}_{\rm e}^{(\rm c)}(t)
=
{C}_{\rm e}^{(\rm c)}(t)/{C}_{\rm e}^{(\rm c)}(0)
\end{equation}

\subsection{\label{sec:simulationprocedure}Simulation procedure}

We conducted NMCMD simulations for a bulk $^4$He system consisting of 256 atoms in a cubic box with periodic boundary conditions \cite{tsujimoto2024}.
For the helium-helium interaction potential $\phi$, we employed the Aziz HFD-B3-FCI1 pair potential \cite{aziz1995} as in our previous
studies \cite{takemoto2018,kinugawa2021,tsujimoto2024}.
The Trotter number $N_{\rm{b}}$ was set to 500 \cite{tsujimoto2024}.
The centroid time increment $\Delta{t}_{\rm{MD}}$  was set to 0.1 fs.
The time step  for the normal mode propagation was taken to be the same as $\Delta{t}_{\rm{MD}}$, while the effective period of the non-centroid normal modes was set to 100 fs \cite{tsujimoto2024}.

In our preceding study \cite{tsujimoto2024}, NPT-CMD simulations were performed for two million steps at thermal equilibrium for each $P$-$T$ condition.
In the present work, each NPT-CMD run was first continued until the instantaneous volume matched the average volume obtained in the preceding run.
Next, the barostatic control in Eq.(\ref{eq:eom}) was switched off,
thereby changing the statistical ensemble to NVT with a fixed volume.
The NVT-CMD simulation was then continued for 200,000 steps.
 Subsequently, the thermostat acting on the $N$ centroids (the second term in Eq.(\ref{eq:eom}))  was removed to render the ensemble NVE, and the CMD simulation was propagated for more than 50,000 steps until the system reached apparent equilibrium. 
 After equilibration, the NVE-CMD simulation was further continued  
for two million steps, from which the time correlation functions were 
computed to evaluate the transport properties.
The transport properties were evaluated in the 
NVE ensemble because both barostat and thermostat controls adversely affect the evaluation of the time correlation functions
of collective properties \cite{yonetani2003,yonetani2004,imaoka2017}.
As an example, in a two-million-step NVE run at the density corresponding to 1 bar and 3.3 K, the total energy of NMCMD \cite{kinugawa1997}, which consists of the mechanical energy of the physical system and that of the massive Nos\'e–Hoover chain thermostats attached to the normal modes, was conserved within a relative error on the order of 10$^{-5}$.

We evaluated the transport properties for 115 density–temperature ($\rho$–$T$) conditions, which are shown as symbols in Fig.~\ref{fig:phase}(a);
equivalently, the conditions are plotted on the $P-T$ and $V-T$ planes in 
Figs.~S1 and S2 in the Supplementary Material,
respectively.
The state at each point was identified using the same criteria as in Ref.~\onlinecite{tsujimoto2024} (see Sec.~S1 in the Supplementary Material).

Specifically, the system was classified as an ergodic liquid when the self-diffusion coefficient $D$, obtained from the time integral of the VAF (Sec.~\ref{sec:self-diffusion}), exceeded $2.0\times10^{-10}~\mathrm{m^2\,s^{-1}}$, and as a non-ergodic state (hereafter referred to as an amorphous solid or glass according to this diffusion-based operational criterion) otherwise.
We emphasize that this classification is based on an operational criterion using the self-diffusion coefficient and is not intended as a rigorous definition of a glass transition. 
A more stringent characterization using a relaxation time obtained, for example, from the self-part of the intermediate scattering function would be valuable. 
Such an investigation is beyond the scope of the present work and will be the subject of future studies.

Liquid states were further classified as LQDL or HQDL based on the centroid-centroid radial distribution function, $g_{\rm cc}$, and the expansion factor of the quantum wavelength, $\alpha_{\lambda}$ (for details, see Secs.~S3 and S5 in the Supplementary Material and Ref.~\onlinecite{tsujimoto2024}).

\section{\label{sec:results} RESULTS}

In this section, we first present Table \ref{tab:comparison}, which summarizes the main results and highlights the key differences between the LQDL and HQDL states, serving as a guide to the results presented below. 
We begin with the state diagram and then provide a systematic characterization of the two states, progressing from single-particle dynamics to collective transport properties. 
Building on the structural framework established in our previous study \cite{tsujimoto2024}, where nuclear quantum delocalization was characterized and quantified (Figs.~2 and 7 in Ref.~\onlinecite{tsujimoto2024}; Sec.~S3 in the Supplementary Material), we focus here on dynamical and transport behavior.
For the static characteristics, see Ref.~\onlinecite{tsujimoto2024}. 
For reference, Table \ref{tab:scf} summarizes the characteristic features of liquid-like and gas-like regimes in the supercritical fluid reported in our previous work \cite{takemoto2018}; this table is included for comparison and does not represent results obtained in the present study.

\begin{table*}
\centering
\caption{Comparison of characteristic properties of the LQDL and HQDL states of distinguishable $^4$He.}
\label{tab:comparison}
\setlength{\tabcolsep}{6pt} 

\begin{tabular*}{\textwidth}{@{\extracolsep{\fill}} llll}
\hline\hline
\textbf{Property / Feature} & \textbf{LQDL} & \textbf{HQDL} & \textbf{Section / Reference} \\
\hline
Temperature range\textsuperscript{a}        & $T \gtrsim T_{\rm L-H}$         & $T \lesssim T_{\rm L-H}$        & Secs.~\ref{sec:state}, \ref{sec:minimum}, Ref.~\onlinecite{tsujimoto2024} \\
Nuclear quantum delocalization   & Moderate                     & Strong                   &  Sec. S3, Ref.~\onlinecite{tsujimoto2024} \\
Atomic tunneling   &  Rare        &  Rare     & Sec.~\ref{sec:minimum}, Ref.~\onlinecite{tsujimoto2024} \\
Power-law exponent $\chi$ of $\lambda_{\rm quantum}{\sim}T^{\chi}$ & $\chi\approx -0.3$  & $\chi\approx -0.5$    & Sec.~\ref{sec:minimum},  Ref.~\onlinecite{tsujimoto2024} \\
VAF decay profile        & Oscillatory decay              & Monotonic decay              &  Sec.~\ref{sec:VAF}, Ref.~\onlinecite{tsujimoto2024} \\
Diffusion behavior       & Normal ($\gamma \approx 1$) & Superdiffusive ($\gamma > 1$) & Sec.~\ref{sec:MSD} \\
Stokes-Einstein relation & Valid ($\xi \approx 1$)  & Breakdown ($\xi \approx 0.2-0.3$) & Sec.~\ref{sec:breakdown} \\
Shear viscosity ($\eta_s$) & Comparable to He I & Ultralow   & Sec.~\ref{sec:shearviscosity} \\
Thermal conductivity ($\lambda$) & Comparable to He I & Unlike He I or II (no divergence) & Sec.~\ref{sec:thermal} \\
Prandtl number ($Pr$)    & $Pr < 1$                 & $Pr > 1$                 & Sec.~\ref{sec:kineticandthermal} \\
Faster transport    &  Heat              & Momentum               & Sec.~\ref{sec:kineticandthermal} \\
Relaxation time of VAF   & $\sim 0.1$ ps             & $\sim 1$ ps              & Sec.~\ref{sec:relaxation} \\
Degree of dynamical decoupling  &  Fairly weak            & Strong             & Secs.~\ref{sec:relaxation},~\ref{sec:discussion} \\
Overall dynamical character   & Ordinary liquid-like       &  Gas-like              & Sec.~\ref{sec:discussion} \\
Overall characterization
& \makecell[l]{Dissipative fluid \\ (heat-dominated)}
& \makecell[l]{Inertial fluid \\ (momentum-dominated)}
& Sec.~\ref{sec:discussion} \\
\hline\hline
\end{tabular*}
\vspace{1mm}
\begin{minipage}{0.95\textwidth}
\footnotesize
\raggedright
\textsuperscript{a} $T_{\rm L-H}$ is the LQDL-HQDL transition temperature,
which is identified as a continuous crossover by the change in nuclear spatial delocalization and
is about 0.5 K, depending on the pressure.  The changes in the other
dynamical indices occur at slightly lower temperatures, as shown in Figs.~\ref{fig:phase} and \ref{fig:schematic}.
\par
\end{minipage}
\end{table*}

\begin{table*}
\centering
\caption{
Characteristic features of liquid-like and gas-like regimes in the supercritical fluid of distinguishable $^4$He (from Ref.~\onlinecite{takemoto2018}).
\textsuperscript{a}
}
\label{tab:scf}
\setlength{\tabcolsep}{10pt} 
\begin{tabular*}{\textwidth}{@{\extracolsep{\fill}} lll}
\hline\hline
\textbf{Property / Feature} & \textbf{Liquid-like regime} & \textbf{Gas-like regime} \\
\hline
Temperature range        & $T < T_{\rm F}$ & $T > T_{\rm F}$ \\
Nuclear quantum delocalization   & Moderate & Weak \\
Power-law exponent $\chi$ of $\lambda_{\rm quantum}{\sim}T^{\chi}$ & $\chi\approx-0.3$ & $\chi\approx-0.5$ \\
VAF decay profile        & Oscillatory decay & Monotonic decay \\
Diffusion behavior       & Normal ($\gamma \approx 1$) & Superdiffusive ($\gamma > 1$) \\
Relation to present study &  Analogous to LQDL  &  Analogous to HQDL \\
\hline\hline
\end{tabular*}
\vspace{1mm}
\begin{minipage}{0.95\textwidth}
\footnotesize
\raggedright
\textsuperscript{a} 
This table is reproduced from Ref.~\onlinecite{takemoto2018} for reference.
Here, $T_{\rm F}$ denotes the Frenkel temperature.
The temperature ranges are classified using the Frenkel line, whereas the values of the power-law exponent $\chi$ are taken from the same reference, where a change in $\chi$ across the Widom line was identified.
Since atomic exchange effects in real $^4$He are negligible at supercritical temperatures, the characteristics listed here correspond to those of the real system.
\end{minipage}
\end{table*}

\subsection{\label{sec:state}State diagram}

\begin{figure*}
\includegraphics[width=12cm]{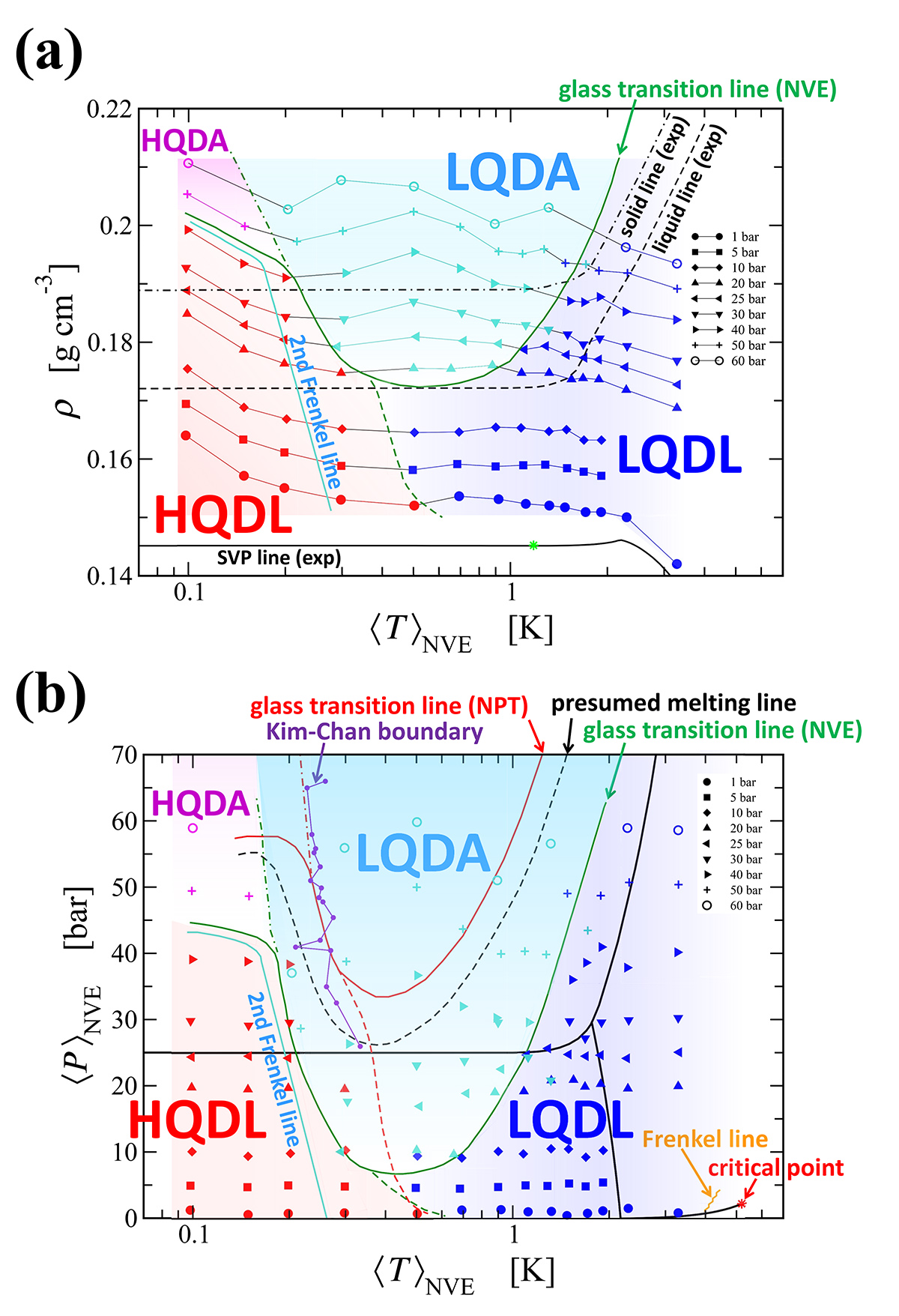}
\caption{\label{fig:phase} 
State diagrams of distinguishable $^4$He on (a) the density–temperature plane and (b) the pressure–temperature plane.
Symbols indicate the states identified in the present NVE-CMD simulations, plotted at the corresponding average temperature and density in (a) or pressure in (b) (see Sec.~\ref{sec:state}).
The color-coded regions are determined from the results of the present NVE simulations.
In the NVE simulations, the averaged pressures sometimes show deviations from those obtained in the preceding NPT simulations. 
In panel (a), the region between the experimental solid and liquid density lines \cite{swenson1950} corresponds to the solid–liquid coexistence region (hcp and He II) of real $^4$He \cite{beamish2020}. 
The green asterisk in (a) denotes the simulation conditions reported by Nakayama and Makri in Ref.~\onlinecite{nakayama2005}.
In panel (b), the pressure values indicated in the legend correspond to those of the preceding NPT simulations.
The glass transition and the melting lines are inferred from the NPT simulations \cite{tsujimoto2024}. 
The Kim–Chan boundary marks the locus at which nonclassical rotational inertia was observed, below which a ``supersolid'' phase was then suggested \cite{kim2004}.
The Frenkel line in the supercritical region is reproduced based on the predictions on the reduced $P/P_{\rm c}$–$T/T_{\rm c}$ plane in Ref.~\onlinecite{takemoto2018}, using the experimental critical point ($T_{\rm c}=5.20$ K and $P_{\rm c}=2.27$ bar). 
The second Frenkel line identified in the subcritical region is discussed in Sec.~\ref{sec:minimum}.
}
\end{figure*}

Figure~\ref{fig:phase}(b) shows the state diagram projected onto the $P$–$T$ plane.
The average pressure in the NVE simulations deviates significantly from that in the preceding NPT runs for some state points owing to changes in the thermodynamic state.
Upon switching from the NPT to the NVE ensemble, the amorphous solid (LQDA) region expands, and the glass transition line shifts to lower pressures relative to that in the NPT simulations \cite{tsujimoto2024}.

As a consequence, several LQDL state points originally situated in the solid–liquid coexistence region of real $^4$He move to the vicinity of the glass transition line, resulting in a state transition from LQDL to LQDA.
As a result, the glass transition line exhibits a more pronounced downward-convex shape toward the low-pressure region.

The above behavior may be attributed to the thermodynamic metastability expected in the present system, analogous to the two-phase coexistence observed in real $^4$He.
In contrast, the HQDL state remains stable and does not undergo any transition upon ensemble switching.
Notably, at $T \le 0.2$ K, the system retains the liquid (HQDL) state up to 40 bar, whereas it is transformed to LQDA at higher temperatures by isobaric heating.
Additional fundamental static properties, including the quantum wavelength $\lambda_{\rm quantum}$ and the expansion factor $\alpha_{\lambda}$, are presented in Figs.~S3 and S4 of the Supplementary Material.

We note that the present state diagram differs from the thermocrystallization scenario reported by Boninsegni et al. \cite{boninsegni2012}, who observed crystallization of distinguishable $^4$He at $T = 0.5$ K and $\rho=0.165$ $\mathrm{g\,cm}^{-3}$.
In contrast, no evidence of crystallization was found at the same state point in the present CMD simulations, as shown in Fig.~\ref{fig:phase}(a).
A possible interpretation of this discrepancy in terms of metastability was discussed in detail in our previous study \cite{tsujimoto2024}.
Briefly, distinguishable $^4$He may possess multiple metastable states whose appearance depends on the simulation method, statistical ensemble, and preparation protocol. 
At present, however, the mechanism governing this state selection remains unresolved, and a detailed examination of this issue is beyond the scope of the present work.

\subsection{\label{sec:VAF}Centroid velocity autocorrelation function}

\begin{figure}
\includegraphics[width=8.5cm]{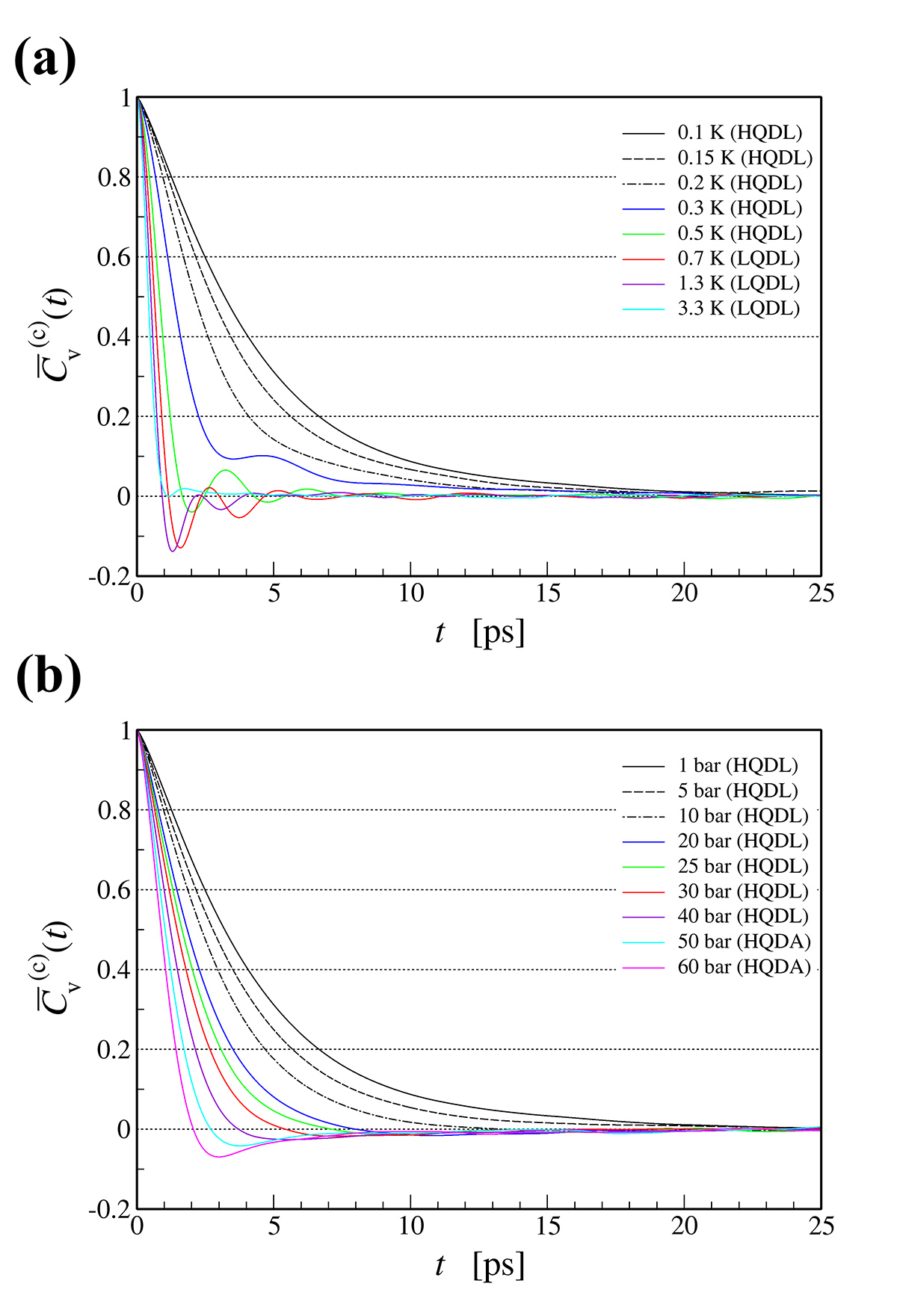}
\caption{\label{fig:vafnorm} 
Normalized centroid velocity autocorrelation functions 
$\bar{C}_v^{(\rm c)}(t)$ [Eq. (\ref{eq:normalizedvaf})] at
(a) densities corresponding to 1 bar and (b) at $T=0.1$ K,
obtained from NVE simulations.
}
\end{figure}

Figure~\ref{fig:vafnorm} shows the normalized centroid VAFs at densities corresponding to 1 bar and at $T=0.1$ K.
The complete set of results is provided in Figs.~S9-S11 in the Supplementary Material.
In Fig.~\ref{fig:vafnorm}, the VAFs of LQDL exhibit oscillatory relaxation characteristic of ordinary liquids,
whereas those of HQDL at $T\le0.2$ K decay slowly and monotonically.
Notably, even the solid HQDA state at 0.1 K and 50 bar shows only a weak remnant of the oscillatory relaxation typically observed in ordinary liquids, exhibiting a decay profile similar to that of HQDL.
Similar monotonic behavior is observed for HQDL at several $P$–$T$ points (5–20 bar, 0.1–0.2 K; 25–30 bar, 0.1–0.15 K; and 40 bar at 0.1 K), as shown in Figs.~S9-S11 in the Supplementary Material.

A monotonic VAF without oscillations is commonly regarded as a signature of a gas-like regime in supercritical fluids, in contrast to the oscillatory behavior of a liquid-like regime \cite{brazhkin2012,brazhkin2013}.
The crossover between these regimes is essentially associated with the Frenkel line, which separates gas-like and liquid-like dynamical regimes in a wide range of systems, including classical fluids \cite{brazhkin2012,brazhkin2013} and $^4$He \cite{takemoto2018}.
In this context, the dynamical crossover observed here resembles that across the Frenkel line.
However, in the present system, the gas-like HQDL state appears at lower temperatures than the LQDL state, which retains liquid-like oscillatory dynamics.

Compared with Fig.~11 of Ref.~\onlinecite{tsujimoto2024}, the present VAF results extend to substantially longer times and have improved statistical accuracy. 
Under the conditions shown in Fig.~\ref{fig:vafnorm}, no discernible differences attributable to the use of different ensembles are observed, and the overall dynamical behavior remains unchanged.

Finally, we compare our results with the centroid VAFs of distinguishable $^4$He reported by Nakayama and Makri \cite{nakayama2005}.
At $T=1.18$~K and density $\rho = 0.1452~\mathrm{g\,cm^{-3}}$ (see Fig.~\ref{fig:phase}(a)), corresponding to the experimental SVP, their results exhibit a monotonic decay \cite{nakayama2005}.
For comparison, we performed additional NVT CMD simulations at the same density ($0.145~\mathrm{g\,cm^{-3}}$) and at $T=1.1$ and $1.3$~K.
The resulting VAFs (Fig.~S12 in the Supplementary Material) show an oscillatory decay consistent with the LQDL state, rather than a monotonic decay.
We also carried out NPT CMD simulations at $P=0.01$~bar and at $T=1.1$ and $1.3$~K, which yielded a gas state ($V=8.02\times10^3$ and $1.03\times10^4$ cm$^3$mol$^{-1}$, respectively) and a decay-less VAF as shown in 
Fig.~S13 in the Supplementary Material.
However, this pressure, 0.01 bar, is significantly higher than the actual SVP ($\sim 10^{-4}$-$10^{-3}$~bar at 1.1-1.3 K \cite{donnelly1998}).
Therefore, under NPT conditions, the system is expected to preferentially sample the gas phase near the SVP conditions in our simulations.
While a given $P$–$T$ condition uniquely defines a thermodynamic phase, the SVP density on the $\rho$–$T$ plane lies at the boundary between the gas phase and the liquid–gas coexistence region \cite{landau1980}, making the sampled state notably sensitive to the statistical ensemble employed in the simulations.
This behavior is consistent with the state changes observed upon switching the statistical ensemble, as described in Sec.~\ref{sec:state}.
Accordingly, the monotonic decay reported by Nakayama \textit{et al.} 
may reflect sampling of a gas-like state or configurations near the liquid–gas coexistence region in their simulations.

\subsection{\label{sec:DOS}Phonon density of states}

We  evaluate the quantum vibrational power spectrum, namely the phonon density of states (DOS), by taking the Fourier transform of the centroid VAF \cite{cao5106,kubo1991},
\begin{eqnarray}
\label{eq:spectrum}
P(\omega)=\frac{\hbar\beta\omega}{2}\left[{\coth\left(\frac{\hbar\beta\omega}{2}\right)}+1\right]
\int_0^{\infty}\bar{C}_v^{\rm (c)}(t)\cos{{\omega}t}dt.
\nonumber \\
\end{eqnarray} 
The power spectra are presented in Figs.~S14 and S15 in the Supplementary Material.
The spectral profiles of HQDL and LQDL are clearly distinguishable.
The DOS of LQDL is characterized by a peak in the range of 12-15 cm$^{-1}$, whereas that of HQDL is shifted toward substantially lower frequencies, exhibiting a maximum at approximately 2-8 cm$^{-1}$.
Thus, the two liquid states possess distinctly different vibrational spectra.
Moreover, for both liquid states, the peak frequency systematically shifts toward lower frequencies with decreasing temperature, indicating a progressive redshift of the vibrational spectrum.
The low-frequency peak observed in HQDL at 0.1-0.15 K is consistent with the monotonic decay observed in the VAF.
Remarkably, this feature is also observed in HQDA, i.e., a glassy state at 50-60 bar and 0.1-0.15 K (Fig.~S15 in the Supplementary Material).
The absence of pronounced high-frequency vibrational components may suppress the backscattering of centroid velocities, leading to the monotonic decay of the VAF.
These results reflect the dominance of NQEs at ultralow temperatures, which effectively reduces the high-frequency components of the forces acting on the centroids.

\subsection{\label{sec:diffusion}Self-diffusion coefficient}

\begin{figure}
\includegraphics[width=8.5cm]{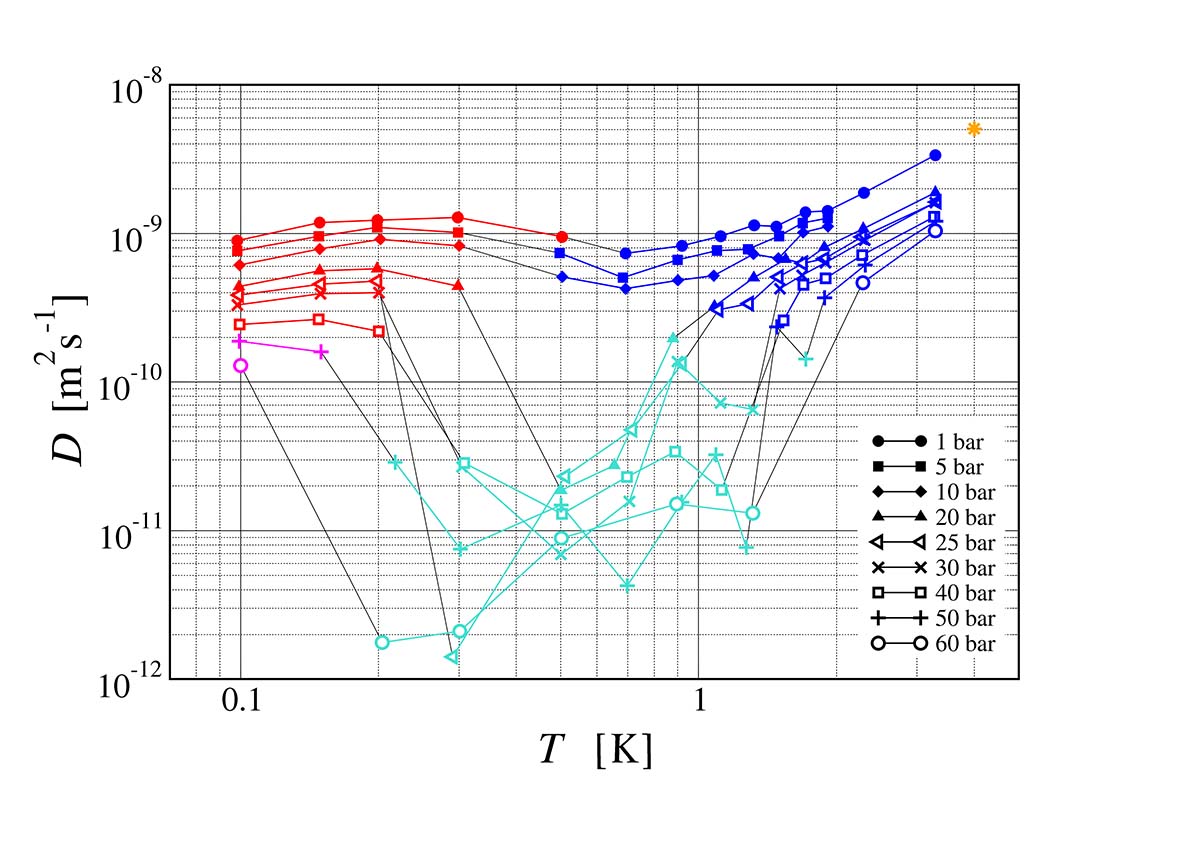}
\caption{\label{fig:dif} 
Temperature dependence of the self-diffusion coefficient $D$ 
 of distinguishable $^4$He: LQDL (blue), LQDA (cyan), HQDL (red), and HQDA (magenta).
The pressure values indicated in the legend denote those of the preceding NPT runs.
The LQDL above the experimental lambda-transition temperature is expected to be nearly equivalent to real He I.
The system with $D \leq 2.0\times10^{-10}~\mathrm{m^2\,s^{-1}}$ is identified as being in a glassy state.
The asterisk denotes the value of $D$ calculated by CMD at 4 K and at the saturated vapor pressure density in Ref.~\onlinecite{miura1999}.
The upper bound of the VAF integral is 25 ps.
}
\end{figure}

Figure~\ref{fig:dif} shows the self-diffusion coefficient estimated from Eq.~(\ref{eq:diffusion}).  
To our knowledge, no experimental measurements of the self-diffusion coefficient of liquid $^{4}$He have been reported because $^{4}$He has zero nuclear spin, making nuclear magnetic resonance (NMR) techniques inapplicable.  In Fig.~\ref{fig:dif}, the temperature dependence of $D$ along each isobar is non-monotonic.  
In general, $D$ for LQDL decreases as the temperature is lowered.  
Below the temperature at which $D$ reaches its minimum, the diffusion coefficient increases upon cooling, and the system enters the HQDL regime.
The initial value $C_{\rm v}^{\rm(c)}(0)$ in Eq. (\ref{eq:vaf})
decreases upon cooling according to the equipartition theorem 
$(\langle{\textbf{\textit{v}}}_{{\rm{c}}i}^2\rangle=3k_{\rm{B}}T/m)$.
Therefore, the increase in $D$ observed for HQDL is attributed to the gas-like slow relaxation of the VAFs.
A reentrant decrease in $D$ upon cooling is observed in the lowest temperature range.

Compared with Fig.~9 of Ref.~\onlinecite{tsujimoto2024}, the self-diffusion coefficients $D$ shown in Fig.~4 are generally lower, particularly in the LQDA region. 
This difference arises partly from the evaluation procedure: the values in Ref.~\onlinecite{tsujimoto2024} were {\it apparent self-diffusion coefficients} estimated from the short-time mean square displacement (MSD) over 0-4 ps, whereas the present values were evaluated by integrating the centroid VAF up to 25 ps using the centroid Green-Kubo relation. 
In addition, the downward displacement of some of the apparent isobars in the LQDA region reflects the ensemble-dependent vitrification of the corresponding LQDL states upon switching from the NPT to the NVE ensemble. 
Thus, the values reported here provide appropriate quantitative estimates of the long-time self-diffusion coefficient, whereas the previous values reported in Ref.~\onlinecite{tsujimoto2024} characterize short-time atomic mobility.

\subsection{\label{sec:MSD}Mean square displacement}

\begin{figure}
\includegraphics[width=8.5cm]{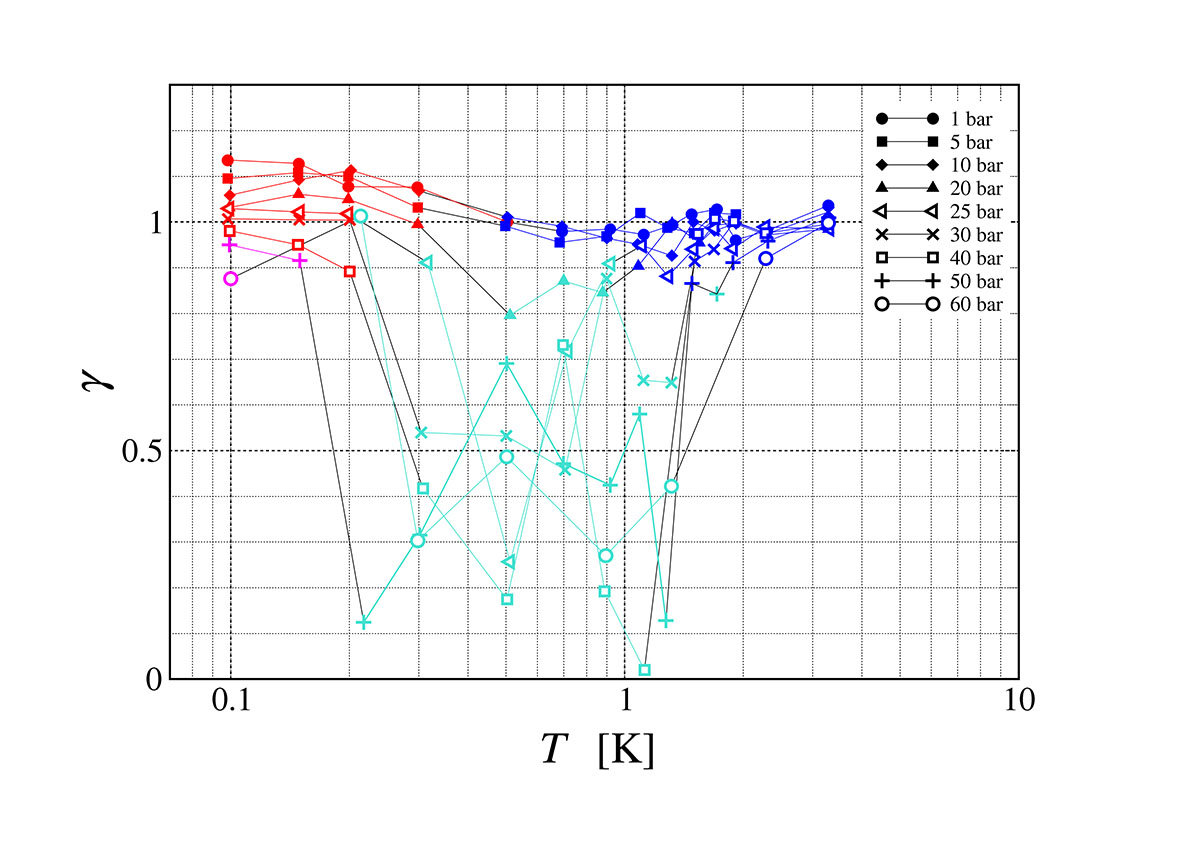}
\caption{\label{fig:gamma} 
Temperature dependence of the power-law exponent $\gamma$ of the mean square displacement (MSD) in Eq. (\ref{eq:msd}): LQDL (blue), LQDA (cyan), HQDL (red), and HQDA (magenta).
The pressure values indicated in the legend denote those of the preceding NPT runs.
}
\end{figure}

 The mean square displacement (MSD) of the centroids is defined as
\begin{eqnarray}
\label{eq:msd}
R^{{\rm (c)2}}(t)&=&\frac{1}{N}\sum_{i=1}^N{\langle}|{\textit{\textbf{r}}}_{{\rm c}i}(t)
-{\textit{\textbf{r}}}_{{\rm c}i}(0)|^2{\rangle}\nonumber \\
&{\sim}&t^{\gamma},\;\;\;t\gg0
\end{eqnarray} 
where the long-time behavior is characterized by the power-law exponent 
$\gamma$.  
The MSDs are shown in Figs.~S16-S18 in the Supplementary Material.
The exponent $\gamma$ is obtained by least-squares fitting in the long-time range ($t = 20$-60 ps).

Figure~\ref{fig:gamma} shows the temperature dependence of $\gamma$.
The LQDL at densities corresponding to $P \le 10$ bar exhibits $\gamma \simeq 1$, consistent with normal diffusion in ordinary liquids.
In contrast, HQDL exhibits $\gamma > 1$ over a wide pressure range (1–25 bar), indicating superdiffusive behavior \cite{takemoto2018}, which is characteristic of gas-like supercritical fluids, including $^4$He \cite{takemoto2018}.
This behavior is consistent with the monotonic decay of the VAF observed for HQDL (Sec.~\ref{sec:VAF}).

Both LQDA and HQDA exhibit $0 < \gamma < 1$, indicating subdiffusive behavior typical of supercooled liquids and glasses.
This behavior can be attributed to reduced free volume under high pressure, which suppresses atomic mobility compared with that at 1 bar at the same temperature.
In general, values of $\gamma < 1$ indicate the presence of {\it residual diffusion} in these glassy states \cite{bernu1987}.
Notably, $\gamma$ for HQDA is only slightly smaller than unity, whereas that for LQDA is significantly smaller.
This difference indicates that atomic dynamics in HQDA remain relatively mobile due to strong NQEs, 
while LQDA corresponds to a more strongly vitrified state, analogous to conventional glasses.

\subsection{\label{sec:NGP}Non-Gaussian parameter}

The non-Gaussian parameter (NGP) quantifies the deviation of the displacement distribution of individual atoms  from a Gaussian distribution and serves as an indicator of 
dynamical heterogeneity \cite{rahman1964,kob1997,goto2021,goto2023}.
The NGP of atomic centroids is defined as 
\begin{eqnarray}
\label{eq:NGP}
\alpha_2(t)=\frac{3R^{\rm (c)4}(t)}{5(R^{\rm (c)2}(t))^2}-1,
\end{eqnarray} 
where the fourth-order moment of the centroid displacement is defined as
\begin{eqnarray}
\label{eq:msd4}
R^{{\rm (c)4}}(t)&=&\frac{1}{N}\sum_{i=1}^N{\langle}|{\textit{\textbf{r}}}_{{\rm c}i}(t)
-{\textit{\textbf{r}}}_{{\rm c}i}(0)|^4{\rangle}.
\end{eqnarray}

The non-Gaussian parameters (NGP) are shown in Figs.~S19 and S20 in the Supplementary Material.
For HQDL at 0.1 K and 1 bar, $\alpha_2(t)$ remains close to zero over the entire time range, indicating that the displacement distribution is essentially Gaussian.
The LQDL at higher temperatures exhibits slight deviations from Gaussian behavior.
However, for both HQDL and LQDL, these deviations remain small (note the vertical scale).

Even in the pressurized HQDA state, $\alpha_2(t)$ at 0.1 K remains close to zero, and the NGP is nearly identical to that of HQDL at the same temperature.
This indicates that the vitrification from HQDL to HQDA does not lead to an enhancement of dynamical heterogeneity.
In contrast, Fig.~S20 shows that LQDA at 50 bar exhibits significantly larger $\alpha_2$ values,
indicating pronounced dynamical heterogeneity similar to that observed in conventional classical glasses.
These results demonstrate that HQDA represents a distinct glassy state with dynamical characteristics markedly different from those of both LQDA and classical glasses.

\subsection{\label{sec:shearviscosity}Shear viscosity}

\begin{figure}
\includegraphics[width=8.5cm]{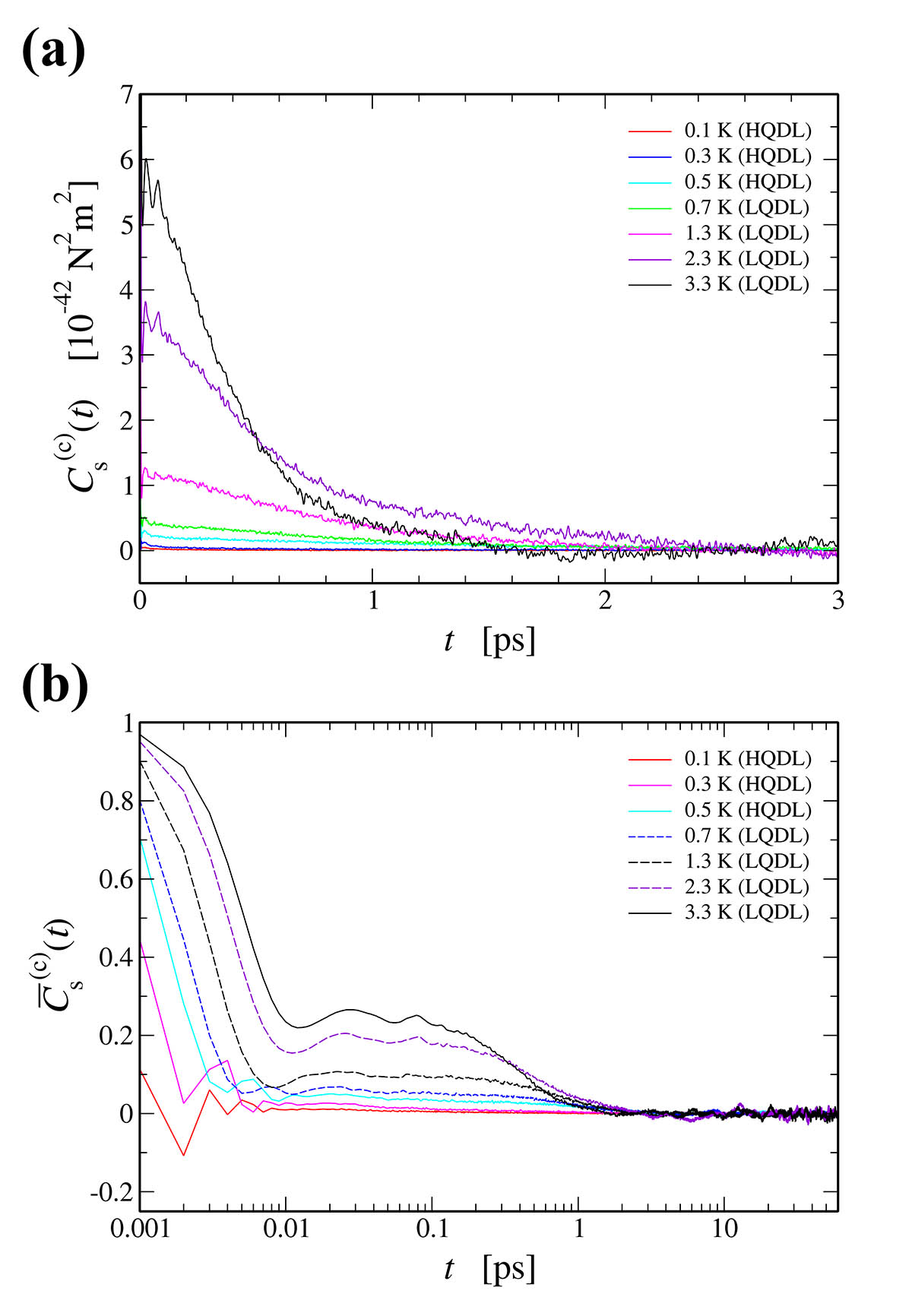}
\caption{\label{fig:stresscorr} 
Off-diagonal stress autocorrelation functions of distinguishable liquid $^4$He 
at densities corresponding to 1 bar for selected temperatures.
(a) Unnormalized function $C_{\rm s}^{\rm (c)}$ [Eq. (\ref{eq:stresscorrelation})]; (b) normalized function $\bar{C}_{\rm s}^{\rm (c)}$ [Eq. (\ref{eq:normstresscorrelation})].
}
\end{figure}

\begin{figure*}
\includegraphics[width=14cm]{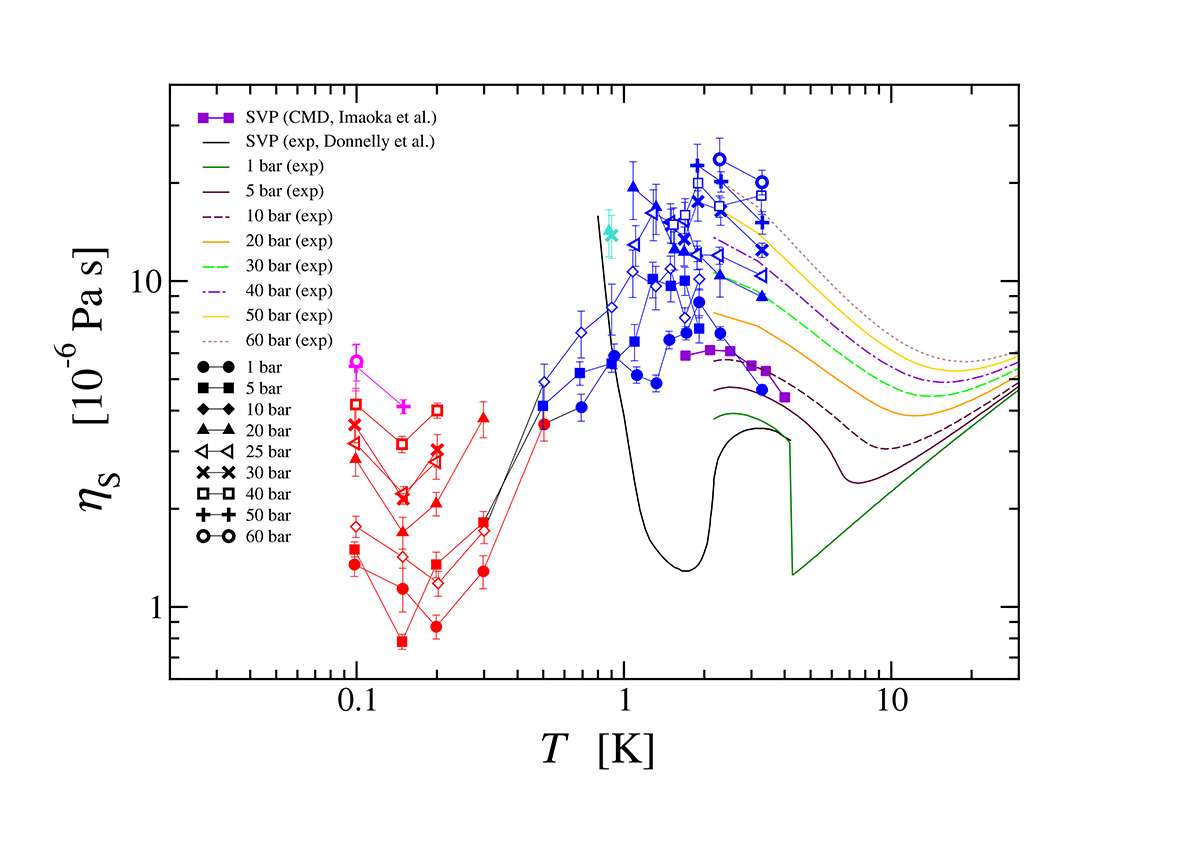}
\caption{\label{fig:vis} 
Temperature dependence of the shear viscosity $\eta_{\rm s}$ of distinguishable $^4$He in comparison with real $^4$He.
Simulation results are shown for LQDL (blue), LQDA (cyan), HQDL (red), and HQDA (magenta). 
The pressures (1--60 bar) indicated in the legend correspond to those used in the preceding NPT simulations.
Experimental data at the saturated vapor pressure (SVP) are taken from Donnelly and Barenghi \cite{donnelly1998}, representing $\eta_{\rm s}$ of He I ($T > T_\lambda = 2.17$ K) and the normal fluid component of He II ($T \le T_\lambda$).
Additional experimental data for He I at pressures between 1 and 60 bar are taken from Ref.~\onlinecite{nist}.
CMD results at SVP in the temperature range 1.7--4.0 K are taken from Ref.~\onlinecite{imaoka2017}.
Error bars indicate the larger of the deviations from the central value obtained by integrating the time correlation function up to 3 ps, relative to the values obtained using integration limits of 2 and 4 ps.
Most LQDA data are not plotted because the stress autocorrelation function did not converge to zero within the simulated time window, precluding a reliable evaluation of $\eta_{\rm s}$; only a few state points are shown.
}
\end{figure*}

Figure~\ref{fig:stresscorr} shows the off-diagonal stress autocorrelation function, $C_{\rm s}^{(\rm c)}(t)$, at densities corresponding to 1 bar; results at other pressures are provided in Figs.~S21-S23 in the Supplementary Material.
As shown in Fig.~S24 in the Supplementary Material, the initial value of the stress autocorrelation function, $C_{\rm s}^{(\rm c)}(0)$, decreases upon cooling.
At higher pressures, vitrification into LQDA results in an incomplete decay of $C_{\rm s}^{(\rm c)}(t)$.
While relaxation remains relatively fast in HQDA, it is too slow in LQDA to reach zero within the observation time window (see Figs.~S22 and S23 in the Supplementary Material), precluding the evaluation of $\eta_{\rm s}$ in this glassy regime.
In contrast, in Fig.~\ref{fig:stresscorr}(b), the time correlation function for HQDL decays to nearly zero within 0.01 ps, whereas that for LQDL decays much more slowly.
This behavior is further analyzed in Sec.~\ref{sec:relaxation}.

Figure~\ref{fig:vis} summarizes the temperature dependence of $\eta_{\rm s}$.
Since the major contribution to the integral of $C_{\rm s}^{\rm (c)}(t)$ in Eq.~(\ref{eq:shearviscosity}) arises within 3 ps, this time was adopted as the upper integration limit.
This uniform integration limit was applied to all state points to ensure
consistency and to avoid undue contributions from statistical noise
in the long-time tail.
While this may lead to a slight underestimation of $\eta_{\rm s}$,
the qualitative features discussed below remain robust.
The systematic uncertainty associated with this integration limit
is reflected in the error bars.
For LQDL, the estimated values are in good agreement with the experimental shear viscosity of He I above $T_\lambda$.
In contrast, HQDL exhibits systematically lower $\eta_{\rm s}$ than LQDL and shows a clear minimum at each pressure.
The lowest value obtained ($7.83\times10^{-7}$ Pa$\cdot$s at 0.15 K and 5 bar) is smaller than the experimental minimum of real He II at SVP ($1.29\times10^{-6}$ Pa$\cdot$s at 1.7 K).
To our knowledge, this value is one of the lowest shear viscosities reported for any atomic or molecular liquid.
Furthermore, up to 40 bar, the minimum $\eta_{\rm s}$ of HQDL remains comparable to or lower than that of real He I at SVP.

The Arrhenius plot (Fig.~S25 in the Supplementary Material) shows clear deviations from Arrhenius behavior \cite{eyring1936} for both LQDL and HQDL, implying a strongly temperature-dependent activation energy characteristic of quantum liquids such as liquid para-hydrogen \cite{yonetani2004}.

Notably, the shear viscosity $\eta_{\rm s}$ of HQDA at 0.1 K and 50–60 bar is comparable to that of LQDL at 1 bar.
Although HQDA is an amorphous solid characterized by a marked suppression of self-diffusion, the corresponding $\eta_{\rm s}$ indicates unexpectedly high fluidity, with values far below the conventional vitrification criterion ($\eta_{\rm s} \sim 10^{12}$ Pa$\cdot$s) \cite{ediger1996,angell1995,debenedetti2001}.
This apparent contradiction indicates a decoupling between the self-diffusion coefficient $D$ and $\eta_{\rm s}$, i.e., a breakdown of the Stokes–Einstein relation (see Sec.~\ref{sec:breakdown}).
Within the conventional viscosity-based criterion, all the states shown in Fig.~\ref{fig:vis} would be classified as liquids.
As shown in Sec.~\ref{sec:NGP}, the NGP of HQDA remains as low as that of HQDL despite vitrification.
Taken together, the low NGP and low viscosity indicate that HQDA is a glassy state that retains liquid-like transport properties except for strongly suppressed self-diffusion.

To elucidate the temperature dependence of $\eta_{\rm s}$, we separately examine the contributions of the time integral $S_{\eta}$ and the prefactor $1/(Vk_{\rm B}T)$ in Eq.~(\ref{eq:shearviscosity}).
The corresponding results are shown in Figs.~S26 and S27 in the Supplementary Material.
In Fig.~S26, the $S_{\eta}$ values back-calculated from the experimental $\eta_{\rm s}$ for the normal-fluid component are also plotted for comparison.
We note that in experiments, only the normal-fluid component is observable in the He II phase ($\eta_{\rm s}=0$ for the superfluid component), whereas He I consists entirely of this component.
The prefactor $1/(Vk_{\rm B}T)$ increases monotonically upon cooling, whereas the CMD-based $S_{\eta}$ for distinguishable $^4$He mostly decreases.
The resulting $\eta_{\rm s}$ reflects the competition between these opposing trends, giving rise to the minimum in the $\eta_{\rm s}$-$T$ curve.
The increase in $\eta_{\rm s}$ of HQDL in the lowest-temperature region is primarily driven by this prefactor.
In contrast, the experimental $S_{\eta}$ of the normal fluid component of real He I and II at SVP does not decrease monotonically with decreasing temperature.
Thus, although the $\eta_{\rm s}$-$T$ curve of distinguishable $^4$He at 1 bar superficially resembles that of the normal-fluid component in real $^4$He (He I and II) at SVP, the underlying contributions differ fundamentally, indicating distinct physical origins of shear viscosity in the two systems.

The high-frequency shear modulus $G_\infty$, defined in Eq.~(\ref{eq:shearmodulus}), is shown in Fig.~S28 in the Supplementary Material. 
$G_\infty$ increases monotonically upon cooling over the entire temperature range. 
This behavior indicates that the liquid becomes increasingly rigid at lower temperatures, suggesting that the potential (non-kinetic) contribution associated with the radial distribution and intermolecular interactions becomes dominant \cite{zwanzig1965,khrapak2020}. 
Notably, $G_\infty$ exhibits only a weak pressure dependence at a given temperature. 
For example, the value for HQDA (glass) at 0.1 K and 60 bar is comparable to that of HQDL at lower pressures. 
This implies that the instantaneous shear response of the glassy state is comparable to that of the liquid at the same temperature.

The shear viscosity is determined by the high-frequency shear modulus $G_{\infty}$ and the relaxation time $\tau_{\rm s}$ of the normalized off-diagonal stress autocorrelation function, defined as
\begin{eqnarray}
\label{eq:relaxationtime}
\tau_{\rm s} = 
\int_0^{\infty}\bar{C}_{\rm s}^{\rm (c)}(t)dt.
\end{eqnarray}
Then, the shear viscosity is expressed as \cite{dyre2006,bolmatov2013}
\begin{eqnarray}
\label{eq:etaproduct}
\eta_{\rm s} &=& G_{\infty} \tau_{\rm s}  \nonumber \\
&=&\frac{1}{Vk_{\rm B}T}\langle{\sigma}^{{\rm (c)}2}_{\alpha\gamma}(0)\rangle\tau_{\rm s}
=\frac{S_{\eta}}{Vk_{\rm B}T}.
\end{eqnarray}
Figure~S29 in the Supplementary Material shows the temperature dependence of $\tau_{\rm s}$. 
The relaxation time $\tau_{\rm s}$ decreases upon cooling, whereas $G_{\infty}$ increases, as shown in Fig.~S28 in the Supplementary Material. 
This counterintuitive decrease in $\tau_{\rm s}$ reflects the increasingly rapid decay of the stress correlation function at short times, indicating that stress relaxation becomes dominated by short time dynamics.
The increase in $G_{\infty}$ upon cooling contrasts with the decrease in the initial value of the stress correlation function, $C_{\rm s}^{\rm(c)}(0)$, shown in Fig.~S24 in the Supplementary Material.
This opposite tendency originates from the prefactor ${1}/{(Vk_{\rm B}T)}$ 
(Fig.~S27 in the Supplementary Material) included in $G_{\infty}$. 
The marked decrease of $\tau_{\rm s}$ upon cooling contributes significantly to the reduction of $\eta_{\rm s}$ in both HQDL and HQDA. 
As a result of the opposing temperature dependences of $G_{\infty}$ and $\tau_{\rm s}$, their product $\eta_{\rm s}$ exhibits a minimum, as clearly observed in Fig.~\ref{fig:vis}.
The emergence of the minimum will be discussed in Sec.~\ref{sec:minimum}.

For reference, we compare the magnitude of $G_\infty$ with that estimated for real helium.
As shown in Fig.~S28, $G_\infty$ for LQDL at 3.3 K and 1 bar is 41.7 MPa, which is approximately one order of magnitude larger than the theoretically estimated value for real helium, $G_\infty=7.0$ MPa at 20 MPa \cite{bolmatov2013}.
However, the experimentally accessible shear modulus may be significantly smaller than the initial value defined by Eq.~(\ref{eq:shearmodulus}) because of finite experimental time resolution.
As shown in Fig.~\ref{fig:stresscorr}(b), the stress autocorrelation function decays by nearly one order of magnitude within 1 ps.
Consequently, measurements probing the shear modulus on such timescales may yield values that are approximately one order of magnitude smaller than $G_\infty$ at $t=0$.

\subsection{\label{sec:thermal}Thermal conductivity}

\begin{figure}
\includegraphics[width=8.5cm]{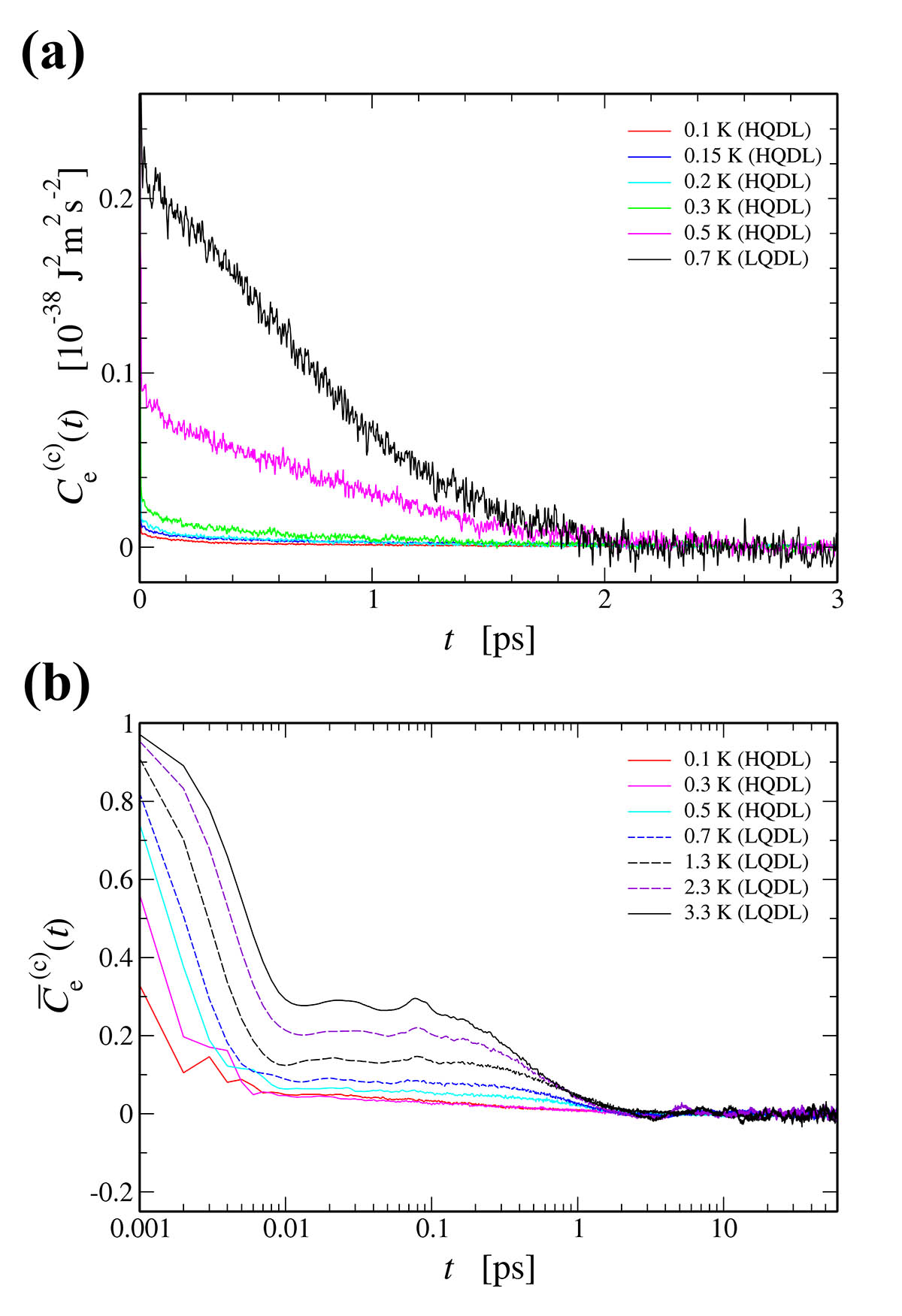}
\caption{\label{fig:1barenergycurrent} 
Energy current autocorrelation functions of distinguishable liquid $^4$He 
at densities corresponding to 1 bar for selected temperatures.
(a) Unnormalized function $C_{\rm e}^{\rm (c)}$ [Eq. (\ref{eq:energycurrentcorr})]; (b) normalized function $\bar{C}_{\rm e}^{\rm (c)}$ [Eq. (\ref{eq:normenergycurrent})].
}
\end{figure}

Figure~\ref{fig:1barenergycurrent} shows the energy current autocorrelation function at densities corresponding to 1 bar; results under other conditions are provided in Figs.~S30-S32 in the Supplementary Material.
As summarized in Fig.~S33 in the Supplementary Material, the initial value of the energy current time correlation function, $C_{\rm e}^{(\rm c)}(0)$, remarkably decreases upon cooling.

\begin{figure*}
\includegraphics[width=14cm]{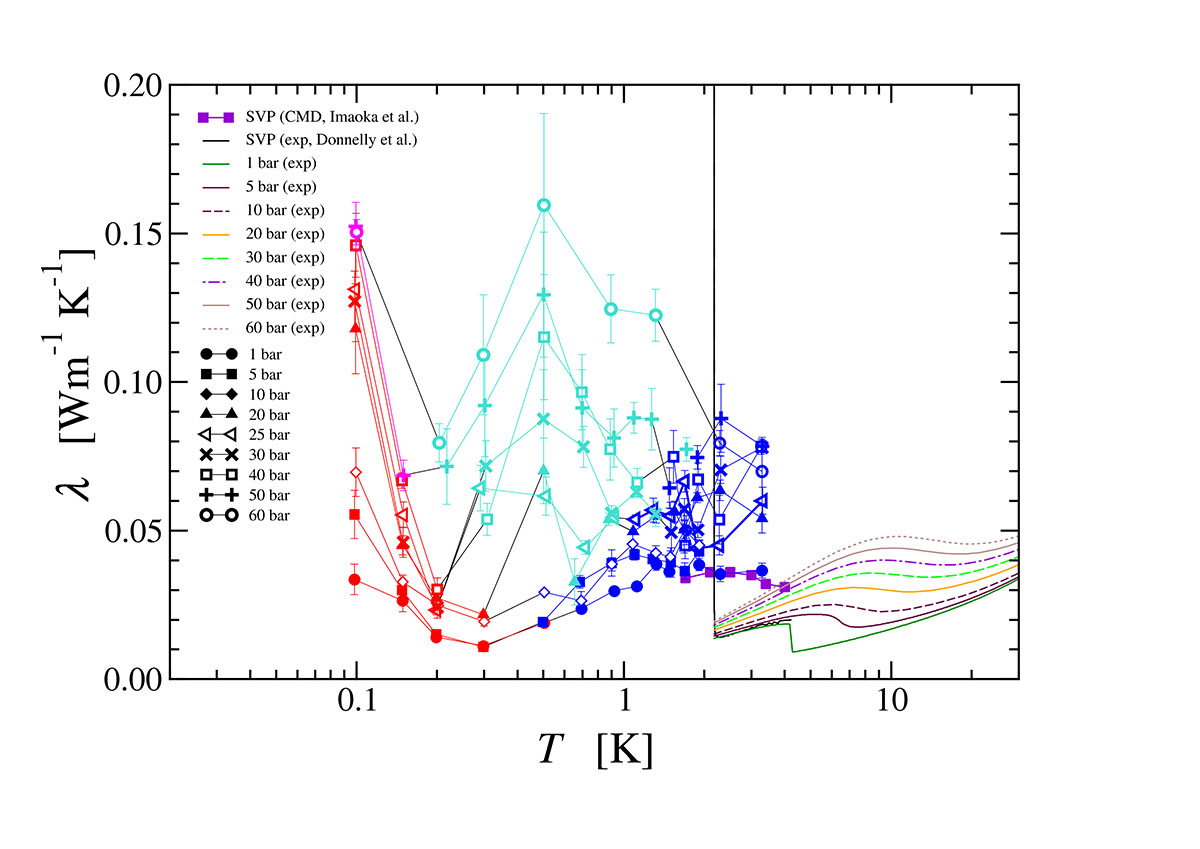}
\caption{\label{fig:thermal} 
Temperature dependence of the thermal conductivity $\lambda$ of distinguishable $^4$He in comparison with real $^4$He.
Simulation results are shown for LQDL (blue), LQDA (cyan), HQDL (red), and HQDA (magenta). 
The pressures (1--60 bar) indicated in the legend correspond to those used in the preceding NPT simulations.
Experimental data at the saturated vapor pressure (SVP) are taken from Donnelly and Barenghi \cite{donnelly1998}, exhibiting the divergence of $\lambda$ in He II below $T_{\lambda}=2.17$ K.
Additional experimental data for He I at pressures between 1 and 60 bar are taken from Ref.~\onlinecite{nist}.
CMD results at SVP in the temperature range 1.7--4.0 K are taken from Ref.~\onlinecite{imaoka2017}.
Error bars indicate the larger of the deviations from the central value obtained by integrating the time correlation function up to 3 ps, relative to the values obtained using integration limits of 2 and 4 ps.
}
\end{figure*}

Figure~\ref{fig:thermal} shows the estimated thermal conductivity $\lambda$. 
As in the case of $\eta_{\rm s}$, the time integral in Eq.~(\ref{eq:thermalconductivity}) was evaluated using an upper limit of 3 ps.
While some energy current autocorrelation functions, particularly in LQDA, remain weakly finite beyond this time, the dominant contribution to the integral is accumulated within this time window.
This common integration limit was adopted for all state points.
The associated uncertainty is reflected in the error bars and does not alter the qualitative conclusions discussed below.
The values of $\lambda$ at $T>T_{\lambda}$ are expected to correspond to those of real He I.  
However, the $\lambda$ values obtained from our CMD simulations at 1–60 bar and those obtained from CMD at SVP~\cite{imaoka2017} are somewhat overestimated compared with the experimental values of He I, although they remain within the same order of magnitude.
One possible reason for this discrepancy is the inclusion of the semiclassical centroid kinetic energy in Eq. (\ref{eq:energycurrent}).
A possible improvement may be achieved by adopting the new estimation scheme proposed by Sutherland \textit{et al.} \cite{sutherland2021}, which is based on the intermediate scattering function.

In Fig.~\ref{fig:thermal}, as the temperature decreases, $\lambda$ of distinguishable $^4$He liquid (LQDL and HQDL) first decreases, reaches a minimum, and then increases in the temperature range of HQDL.
This behavior is evidently unrelated to the divergence of $\lambda$ observed in superfluid He II, since the present CMD does not include atomic exchange.
The emergence of a minimum in the $\lambda$–$T$ curve of the present liquid will be discussed in Sec.~\ref{sec:minimum}.
On the other hand, $\lambda$ of LQDA (shown by cyan points in Fig.~\ref{fig:thermal}) is markedly higher than that of LQDL at the same temperatures.

This enhancement is attributed to the increased contribution
of higher-frequency phonon modes under compression, as shown in Figs.~S14 and S15 in the Supplementary Material (Sec.~\ref{sec:DOS}).
In contrast, $\lambda$ of HQDA is nearly identical to that of HQDL at the same temperature. This is consistent with the above-mentioned finding that the shear viscosity of HQDA is almost the same as that of HQDL in the present pressure range.

Figures~S34 and S35 in the Supplementary Material show
the time integral $S_{\lambda}$ and the prefactor $1/(Vk_{\rm{B}}T^2)$, respectively.
The $S_{\lambda}$ values for real $^4$He were back-calculated from the experimental $\lambda$ via Eq. (\ref{eq:thermalconductivity}).
The prefactor $1/(Vk_{\rm{B}}T^2)$ increases monotonically with decreasing temperature, 
whereas $S_{\lambda}$ for the present system decreases.
These opposite temperature dependencies give rise to the minimum of 
$\lambda$ in Fig.~\ref{fig:thermal}.
For real He I, $S_{\lambda}$ at $T>T_{\lambda}$ shows the same decreasing trend as LQDL upon cooling.
This indicates that the present CMD for LQDL above $T_{\lambda}$ captures the essential temperature dependence of $\lambda$ in real He I.
For HQDL, $S_{\lambda}$ exhibits only a weak temperature-dependence;
 therefore, the increase of $\lambda$ upon cooling is caused by the growth of the prefactor $1/(Vk_{\rm{B}}T^2)$.

Finally, we comment on the thermal conductivity of the glass states, LQDA and HQDA, formed under high pressure.
As shown in Fig.~\ref{fig:thermal}, the thermal conductivity $\lambda$ of LQDA is significantly larger than that of LQDL at the same temperature.
As demonstrated in Fig.~S36 in the Supplementary Material, this increase in $\lambda$ upon pressurization reflects a substantial increase in the relaxation time $\tau_{\rm e}$ of the energy current autocorrelation function.
This behavior suggests a shift in the dominant heat transport mechanism from collision-dominated (diffusive) transport to vibration-dominated transport.
In contrast, $\lambda$ of HQDA remains essentially unchanged from that of HQDL despite vitrification.
The absence of enhancement in $\lambda$ upon vitrification highlights the distinctive character of HQDA as a glass state.

\section{\label{sec:analysis}ANALYSIS OF CORRELATIONS BETWEEN TRANSPORT PROPERTIES}

\subsection{\label{sec:kineticandthermal}Kinematic viscosity, thermal diffusivity,
 and the Prandtl number}

The kinematic viscosity is defined as 
$\nu=\eta_{\rm{s}}/\rho$.
The thermal diffusivity is given by  
$\alpha=\lambda/({\rho}C_{P})$, where $C_P$ is the isobaric heat capacity evaluated from a fitted enthalpy–temperature curve (Figs.~S6 and S7 in the Supplementary Material).
Both $\nu$ and $\alpha$  have the same dimension $[\mathrm{m^2s^{-1}}]$, representing the diffusion 
rates of 
momentum and thermal energy, respectively.
In this subsection, we analyze $\nu$, $\alpha$, and the Prandtl number $Pr=\nu/\alpha$
to explore transport characteristics in more detail.

\begin{figure*}
\includegraphics[width=13cm]{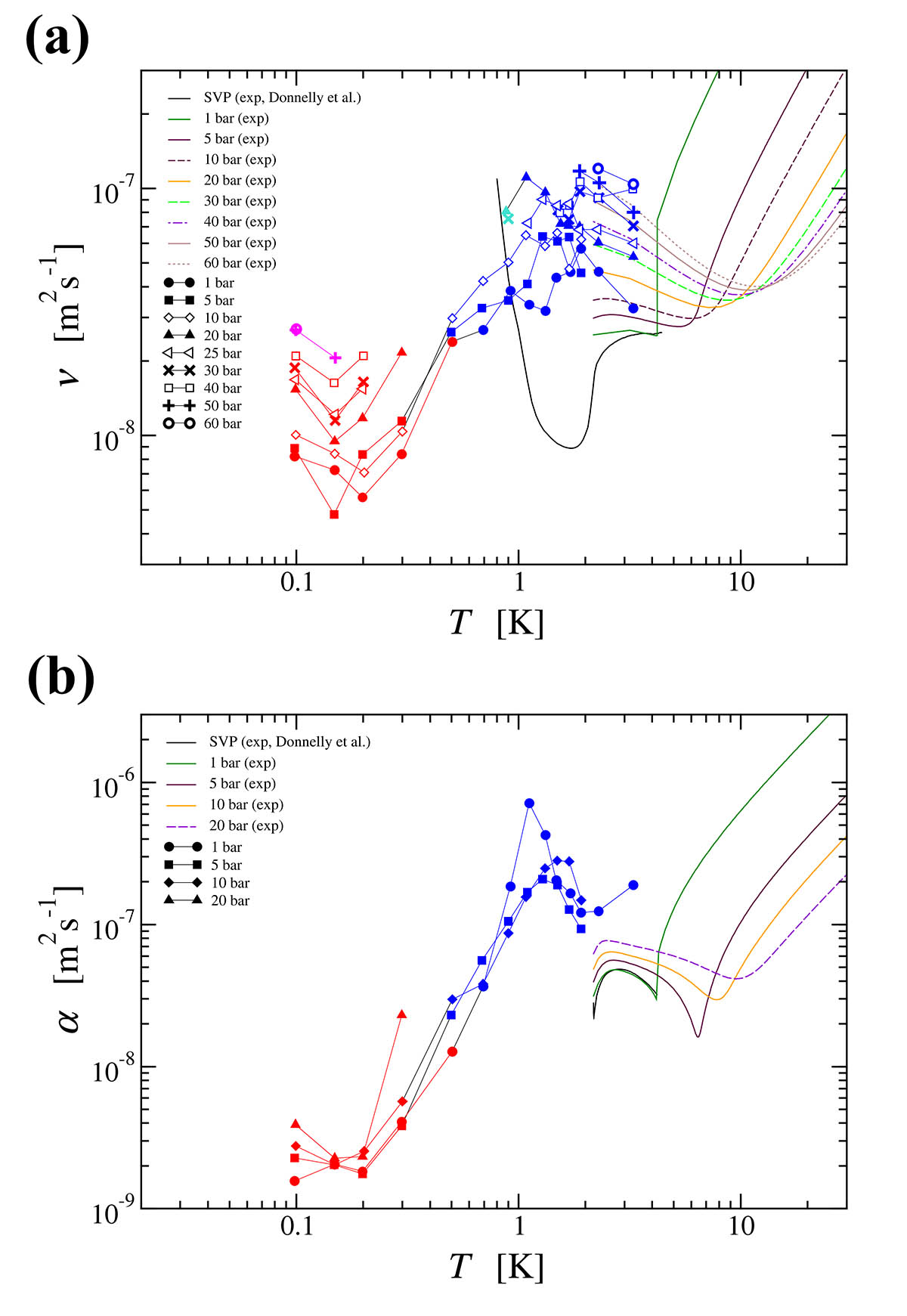}
\caption{\label{fig:ka} 
Temperature dependence of (a) the kinematic viscosity $\nu$ and (b) the thermal diffusivity $\alpha$ of distinguishable $^4$He in comparison with real $^4$He.
Simulation results are shown for LQDL (blue), LQDA (cyan), HQDL (red), and HQDA (magenta).
The pressures indicated in the legend correspond to those used in the preceding NPT simulations.
Experimental data at SVP are taken from
Ref.~\onlinecite{donnelly1998}; additional He I data are taken from
Ref.~\onlinecite{nist}.
The curve at 1 bar represents the properties of the liquid and gas phases below and above the melting point (4.21 K), respectively, whereas the curves at $P\geq 5$ bar correspond to fluid states without a melting transition, including the supercritical fluid state above the critical point ($T_{\rm c}=5.20$ K, $P_{\rm c}=2.27$ bar).
In panel (a), the experimental curve at SVP represents the kinematic viscosity $\nu$ of He I ($T>T_\lambda=2.17$ K) and the normal-fluid component of He II ($T\leq T_\lambda$).
}
\end{figure*}

Figure~\ref{fig:ka}(a) shows the temperature dependence of the calculated kinematic viscosity $\nu$.
Notably, the values of $\nu$ for HQDL at 1 bar are even lower than those of real $^4$He gas
at 1 bar and real supercritical fluid at 5 bar.
This feature indicates that HQDL is distinct
from a gas and a supercritical fluid, 
even though the superdiffusion observed for HQDL resembles the behavior of supercritical fluid (Secs.~\ref{sec:VAF} and \ref{sec:MSD}). 
The calculated $\nu$ exhibits
a minimum, which is considered a characteristic
feature of liquids in general \cite{trachenko2020,trachenko2023}.

\begin{figure*}
\includegraphics[width=13cm]{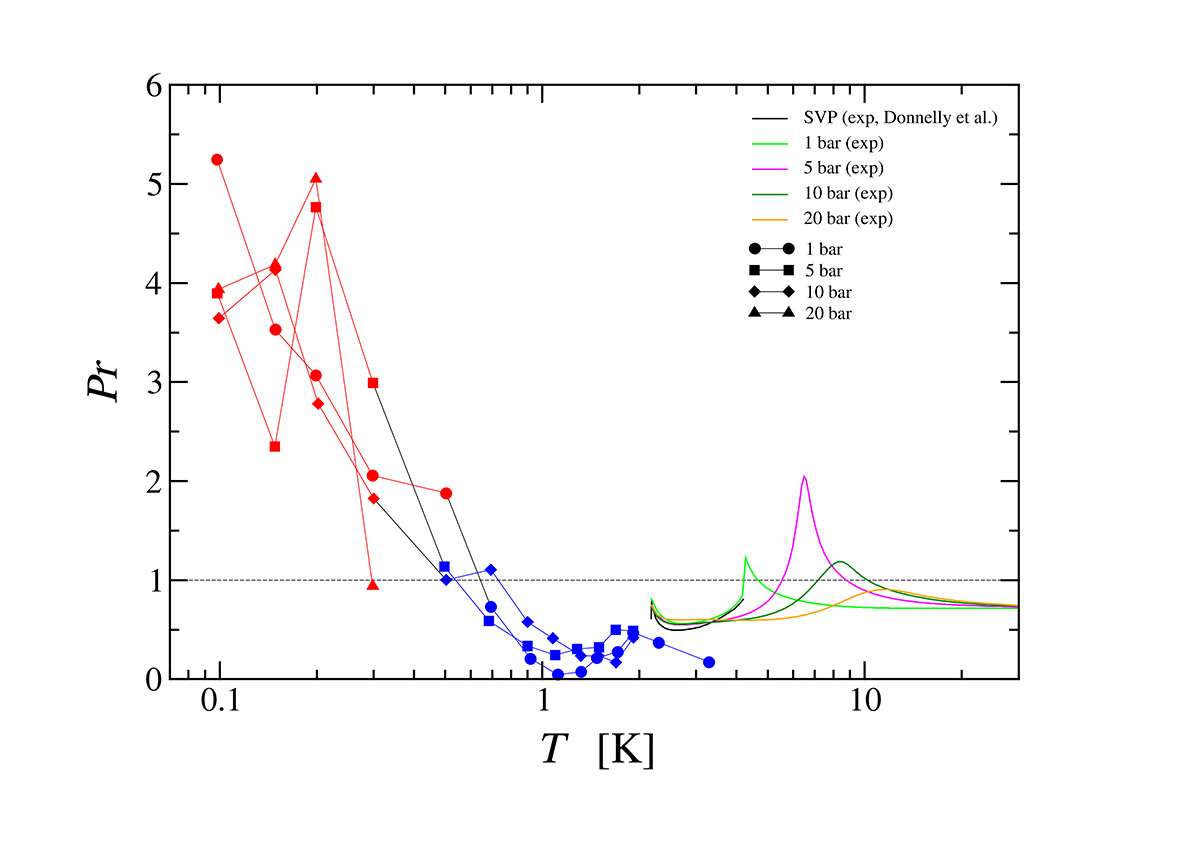}
\caption{\label{fig:prandtl} 
Temperature dependence of the Prandtl number $Pr=\nu/\alpha$ for LQDL (blue) and HQDL (red). 
Experimental curves for real $^4$He are constructed from data reported at the saturated vapor pressure (SVP) \cite{donnelly1998} and at pressures between 1 and 20 bar \cite{nist}.
}
\end{figure*}

Figure~\ref{fig:ka}(b) shows the thermal diffusivity $\alpha$ as a function of temperature. The value of $\alpha$ in HQDL is smaller than in the He I phase of real $^4$He. 
Figure~\ref{fig:prandtl} presents the Prandtl number, $Pr \equiv \nu/\alpha = \eta_{\rm s} C_P / \lambda$, calculated from Fig.~\ref{fig:ka}.
For supercritical $^4$He, we find $Pr < 1$ except near the Widom line, where $C_P$ exhibits a peak. 
For distinguishable $^4$He, $Pr$ crosses over from $Pr < 1$ to $Pr > 1$ in association with the LQDL–HQDL transition. 
This indicates that thermal diffusion dominates in LQDL, whereas momentum diffusion dominates in HQDL.
The result $Pr < 1$ in LQDL is consistent with that in supercritical $^4$He.
In contrast, momentum transport remains dominant in HQDL despite the increase in thermal diffusivity upon cooling, indicating a fundamental change in transport character.
The predominance of momentum transport is closely related to the remarkably short relaxation time of the stress autocorrelation function $\tau_{\rm s}$,
which contributes to rapid momentum transfer.
These results demonstrate that the LQDL–HQDL transition is well characterized as a transport crossover in terms of the Prandtl number.

\subsection{\label{sec:breakdown}
Breakdown of the Stokes-Einstein relation}

\begin{figure*}
\includegraphics[width=13cm]{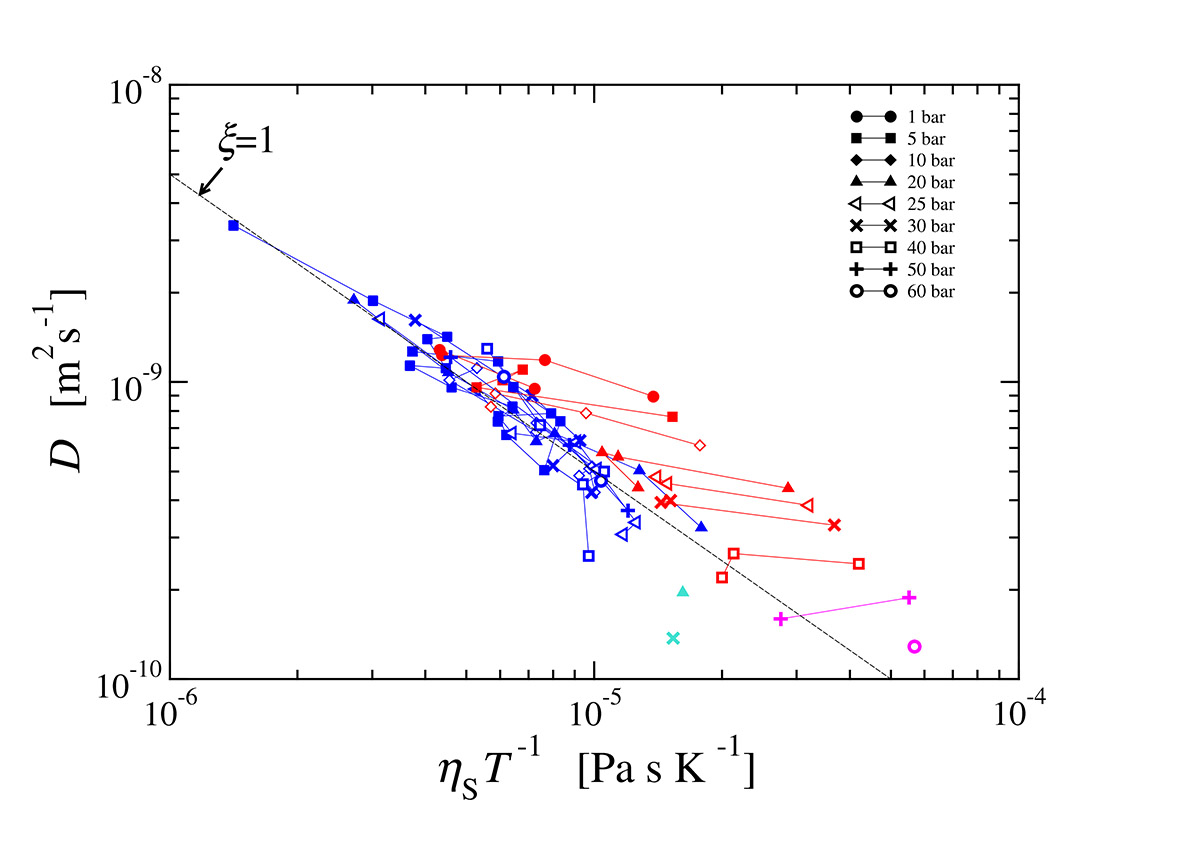}
\caption{\label{fig:fse} 
Fractional Stokes–Einstein relation for LQDL (blue), LQDA (cyan), HQDL (red), and HQDA (magenta). The dashed line indicates the slope corresponding to the Stokes–Einstein relation ($\xi=1$) as a guide to the eye. From the slope, the exponent $-\xi$ defined in Eq.~(\ref{eq:FSER}) is obtained. The pressures (1–60 bar) in the legend correspond to those used in the preceding NPT simulations.
}
\end{figure*}

The relation between $D$ and $\eta_{\rm s}$ is commonly described by the 
Stokes-Einstein (SE) relation with the slip boundary condition,
\begin{eqnarray}
\label{eq:SER}
D\eta_{\rm{s}}&=&\frac{k_{\rm{B}}T}{2\pi{d}},\;\;\;
\end{eqnarray} 
where $d$ is the hydrodynamic diameter of the atom. 
When the SE relation breaks down \cite{ohtori2017},
the deviation is often described by the fractional SE relation
\cite{becker2006,harris2009,harris2010jcp,harris2010jpc,kawasaki2017,pan2017,wei2018}
\begin{eqnarray}
\label{eq:FSER}
D\left(\eta_{\rm{s}}/T\right)^{\xi}&=&{\rm{const.}},
\end{eqnarray} 
where the exponent $\xi<1$ quantifies the deviation from the SE relation ($\xi=1$).

Figure~\ref{fig:fse} shows the fractional SE plots constructed from the calculated $D$ and $\eta_{\rm{s}}$. 
The slopes for HQDL and LQDL are clearly distinct,
and  the fitted $\xi$ values for each liquid 
state are summarized in Table~S1 in the Supplementary Material.
LQDL exhibits $\xi\approx1$, indicating that the SE relation is nearly satisfied regardless of pressure.
In contrast, HQDL exhibits  $\xi\approx0.2-0.3$, demonstrating a strong breakdown of the SE relation.
The upward shift of the HQDL (red) lines from the dashed reference line ($\xi=1$) in Fig.~\ref{fig:fse} indicates that atomic self-diffusion in HQDL is faster than expected from the SE relation with a constant hydrodynamic diameter $d$.

Fractional SE exponents ($\xi<1$) are typically observed for supercooled liquid water, in which dynamical heterogeneity plays a key role \cite{becker2006,kawasaki2017}.
However, no signature of supercooling is found in HQDL in this pressure range, as evidenced by $D$ and the VAF. 
Moreover, as shown in Sec.~\ref{sec:NGP}, there is also no indication of dynamical heterogeneity in HQDL despite the deviation from the SE relation.
These results demonstrate that the dynamical behavior of HQDL is highly unusual and is clearly distinct from that of LQDL. 
The sharp change in $\xi$ further indicates that the LQDL-HQDL transition is a dynamical and transport crossover, even though no  thermodynamic discontinuity is present \cite{tsujimoto2024}.

For the glass state, the 50 bar data for HQDA in Fig.~\ref{fig:fse} show a deviation from the SE relation similar to that observed in HQDL, suggesting that HQDA at 50 bar retains characteristics of its corresponding liquid state.
In contrast, $\xi$ for LQDA cannot be determined due to the limited number of available data points.
Furthermore, the shear viscosity $\eta_{\rm s}$ could not be evaluated in Sec.~\ref{sec:shearviscosity} because the time correlation function $C_{\rm s}^{({\rm c})}$ did not converge within the accessible simulation time window for most $\rho–T$ points of LQDA.

\begin{figure}
\centering
\includegraphics[width=8.5cm]{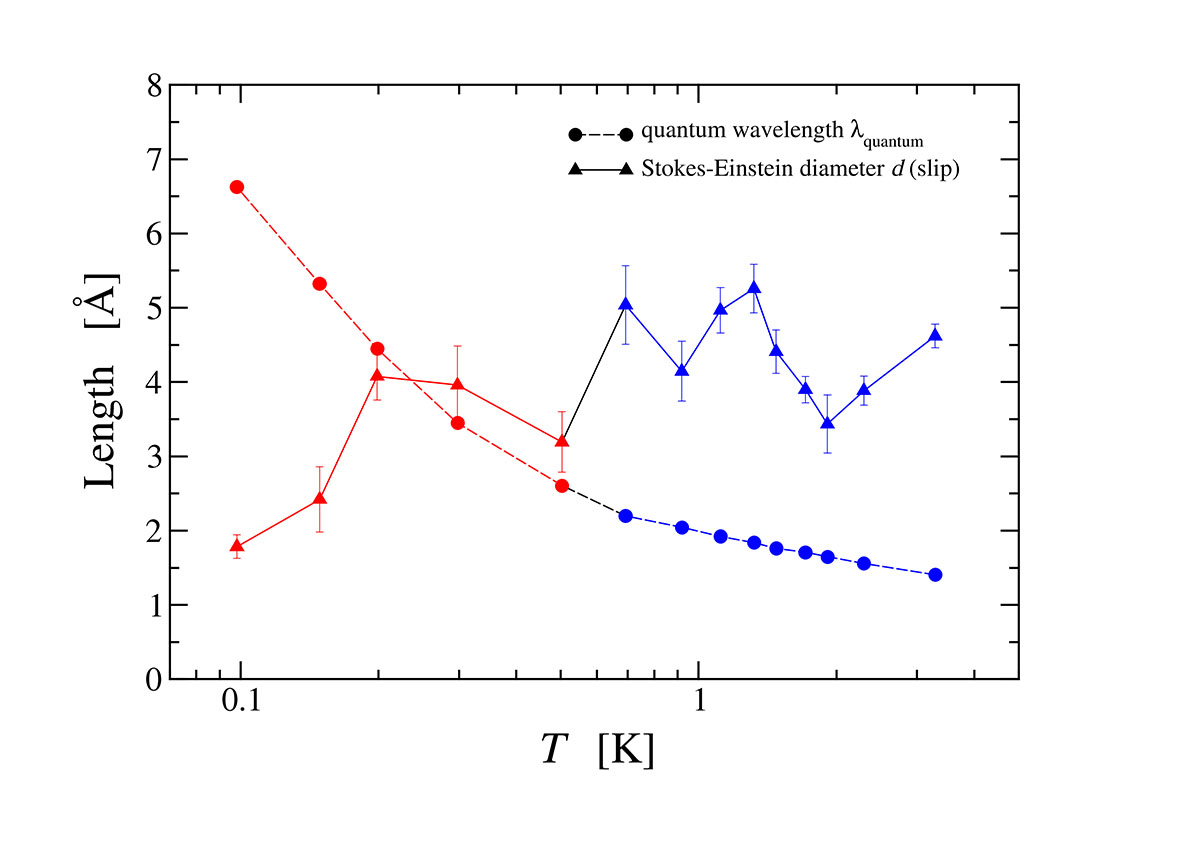}
\caption{\label{fig:dia} 
Apparent hydrodynamic diameter $d$ from the Stokes–Einstein relation compared with the quantum wavelength $\lambda_{\mathrm{quantum}}$ at 1 bar: LQDL (blue) and HQDL (red).  The diameter $d$ is estimated
from the self-diffusion coefficient $D$ and the
shear viscosity $\eta_{\rm{s}}$ via Eq.~(\ref{eq:SER}) using the slip boundary condition.
}
\end{figure}

Figure~\ref{fig:dia} shows the apparent hydrodynamic diameter $d$ inferred from the SE relation [Eq.~(\ref{eq:SER})], together with the quantum wavelength $\lambda_{\mathrm{quantum}}$, as a function of temperature at 1 bar.
For LQDL, the values of $d$ show no clear temperature dependence, consistent with the validity of the SE relation.
For HQDL, however, $d$ decreases upon cooling, whereas $\lambda_{\mathrm{quantum}}$ markedly increases.
The decrease of $d$ with decreasing temperature is consistent with the observed enhancement of $D$ relative to that expected from the SE relation with a constant $d$.
Since the temperature dependencies of $d$ and $\lambda_{\mathrm{quantum}}$ in HQDL 
are opposite, 
$\lambda_{\mathrm{quantum}}$, i.e., the diameter of atomic necklaces, cannot be identified with the hydrodynamic diameter and cannot by itself account for the observed $D$-$\eta$ relation.
This situation is quite different from that of classical polymers.
In bead–spring models for classical dilute polymer solutions, chains with a smaller radius of gyration 
$R_{\rm{g}}$, which serves as the counterpart of 
$\lambda_{\mathrm{quantum}}$ in our quantum system, exhibit lower viscosity \cite{yamakawa1971}.
This trend contrasts with our observations for LQDL and HQDL.

\subsection{\label{sec:relaxation}
Relaxation times of time correlation functions}

To further clarify the origin of the decoupling between $D$ and $\eta_{\rm s}$,
we examine the relaxation times of relevant correlation functions.

\begin{figure*}
\includegraphics[width=14cm]{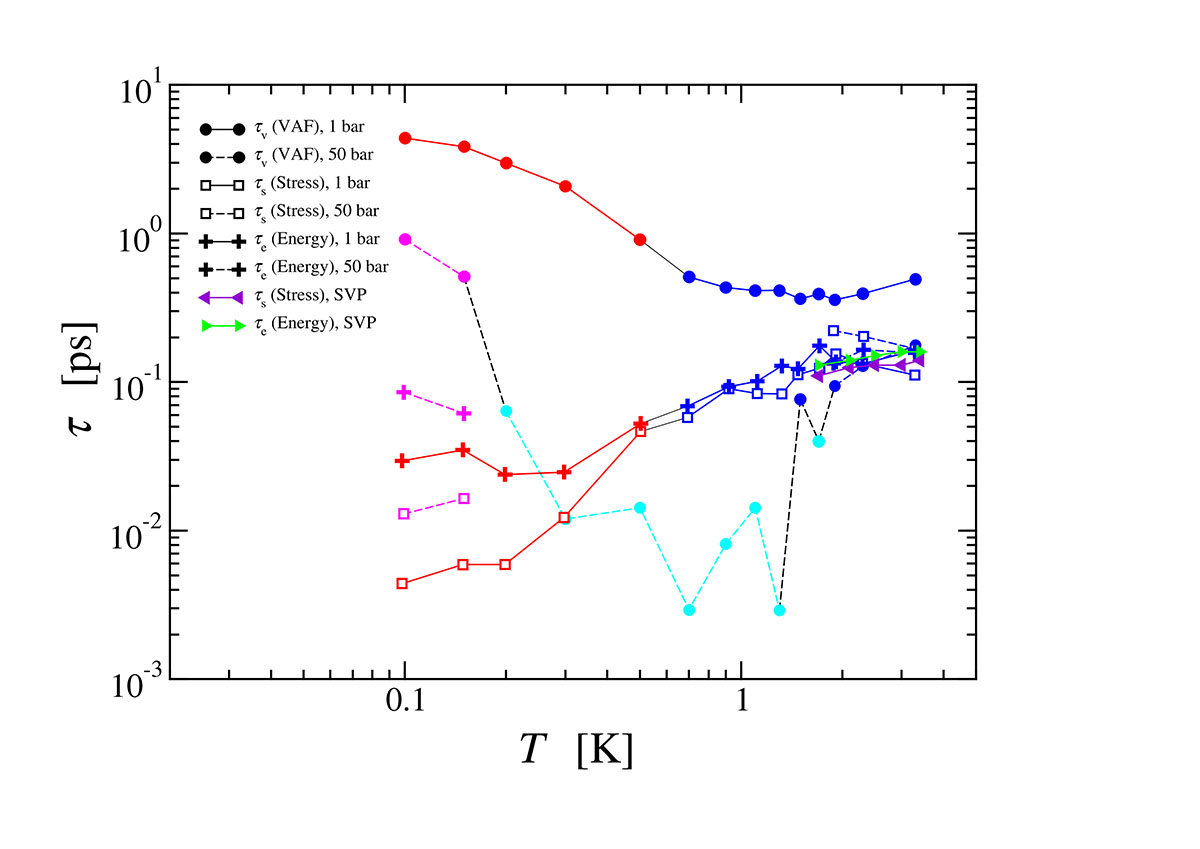}
\caption{\label{fig:relaxtime}
Comparison of relaxation times for three correlation functions of distinguishable $^4$He:
LQDL (blue), LQDA (cyan), HQDL (red), and HQDA (magenta).
The pressure values indicated in the legend denote those of the preceding NPT runs.
$\tau_{\rm v}$, $\tau_{\rm s}$, and $\tau_{\rm e}$ denote the relaxation times of the velocity autocorrelation function (VAF), the off-diagonal stress autocorrelation function, and the energy current autocorrelation function, respectively.
Solid and dashed lines represent data at 1 and 50 bar, respectively,
except for the data at saturated vapor pressure (SVP), which are taken from our previous CMD study~\cite{imaoka2017}.
}
\end{figure*}

Figure~\ref{fig:relaxtime} compares the relaxation times $\tau_{\rm v}$, $\tau_{\rm s}$, and $\tau_{\rm e}$, corresponding to the VAF, the stress autocorrelation function, and the energy current autocorrelation function, respectively.
The definition for $\tau_{\rm s}$ is given in Eq.~(\ref{eq:relaxationtime}), and analogous definitions were used for $\tau_{\rm v}$ and $\tau_{\rm e}$.
As the temperature decreases, the relaxation times become clearly separated;
at $0.1~\mathrm{K}$ and 1 bar, we find
$\tau_{\rm v} \simeq 4.4~\mathrm{ps} \gg \tau_{\rm e} (\sim 10^{-2}~\mathrm{ps}) > \tau_{\rm s} (\sim 10^{-3}~\mathrm{ps})$.
This pronounced separation of timescales provides clear evidence of the dynamical decoupling of $D$ and $\eta_{\rm s}$,
supporting the breakdown of the SE relation in HQDL, as discussed in Sec.~\ref{sec:breakdown}.
Further discussion of the relaxation times will be provided in Sec.~\ref{sec:discussion}.

\section{\label{sec:discussion}discussion}

\subsection{\label{sec:highfluidity}Distinct transport properties in HQDL}

The remarkably low, gas-like shear viscosity of the HQDL state indicates a distinct regime of quantum fluidity that emerges from nuclear quantum delocalization in the absence of superfluidity. 
The minimum shear viscosity reaches values below $1\times10^{-6}$~Pa$\cdot$s, lower than those of LQDL, He I, and the normal fluid component in He II.
HQDL appears to exhibit one of the lowest viscosities among atomic or molecular liquids, excluding the zero-viscosity superfluid component of He II.
These transport characteristics ultimately originate from NQEs.
At low temperatures, the highly delocalized necklace structures effectively smooth out the bare interatomic interactions represented by the Aziz potential. 
This quantum smoothing softens the effective forces acting on the atomic centroids, suppressing backscattering effects.
This leads to the monotonic decay of the VAF and consequently to the ultralow shear viscosity characteristic of HQDL.
For example, the quantum wavelength $\lambda_{\rm{quantum}}$ of atoms in HQDL at 0.1 K and 1 bar reaches $6.6~\rm{\AA}$ (see Fig.~S3 in the Supplementary Material) \cite{tsujimoto2024}.
Such a regime enters a range in which field-theoretic concept \cite{fredricksonbook} may become applicable owing to the spatial extension of quantum fluctuations over atomic length scales, although the present study is based on a fully atomistic description.

Theoretically, a universal lower bound on the shear viscosity has been proposed in terms of the ratio (the Kovtun-Son-Starinets (KSS) bound \cite{kovtun2005,schafer2009}
\begin{eqnarray}
\label{eq:viscositybound}
\eta_{\rm{s}}/(S/V)&\geq&\hbar/(4\pi{k}_{\rm{B}})=6.08\times10^{-13}\;{\rm{K\cdot{s}}} \\ \nonumber
&\simeq&0.0796\hbar/{k}_{\rm{B}},
\end{eqnarray}
where $S$ and $V$ are the entropy and volume, respectively.
A direct comparison with the KSS bound is not possible in the present study because the absolute entropy of the Boltzmann $^4$He system is not available.
Therefore, we cannot evaluate $\eta_{\rm s}/(S/V)$ quantitatively.
Nevertheless, it is instructive to discuss the possible order of magnitude of this ratio based on empirical values reported for known fluids.
The lowest shear viscosity observed in the HQDL state of this Boltzmann $^4$He system is $\eta_{\rm{s}} = 7.83\times10^{-7}\ {\rm{Pa\cdot s}}$ at 0.15 K and 5 bar.
For the fluids of $\rm{H_2O}$ and $^4$He, the values of $\eta_{\rm{s}}/(S/V)$ are reported to be 0.7-8.2 times $\hbar/k_{\rm{B}}$ \cite{schafer2009}.
Assuming, as a rough order-of-magnitude estimate, that $\eta_{\rm{s}}/(S/V)$ of the present system lies within the range $0.1$-$10\,\hbar/k_{\rm{B}}$, we estimate the entropy of HQDL at 0.15 K and 5 bar to be
$S=25.1$-$0.251~\rm{J\,K^{-1}mol^{-1}}$.
This range is consistent with the empirical entropy magnitude,
$S\approx R=8.31~\rm{J\,K^{-1}mol^{-1}}$,
commonly observed for many fluids \cite{schafer2009}.
However, it is far larger than the experimental absolute entropy of real He II at 0.15 K,
$9.19\times10^{-5}~\rm{J\,K^{-1}mol^{-1}}$ \cite{donnelly1998}.
This comparison suggests that HQDL remains a highly disordered state without Bose condensation, in marked contrast to the real superfluid phase, which is characterized by an almost vanishing entropy.

It was noted that the minimum kinematic viscosity and minimum thermal diffusivity of fluids are given by the same expression:
\begin{eqnarray}
\label{eq:minimal}
\nu_{\rm{m}}=\alpha_{\rm{m}}=\frac{1}{4\pi}\frac{\hbar}{\sqrt{m_{\rm{e}}m}},
\end{eqnarray}
where $m_{\rm{e}}$ is the electron mass \cite{trachenko2020,trachenko2023}.
This minimum arises from the crossover between liquid-like and gas-like regimes \cite{trachenko2020,trachenko2023}.
Following Eq.~(\ref{eq:minimal}), $\nu_{\rm{m}}$ for $^4$He is estimated to be $1.08\times10^{-7}\ {\rm{m^2\;s^{-1}}}$.
The minima of $\nu$ and $\alpha$ for HQDL in Fig.~\ref{fig:ka} are both on the order of $10^{-9}\ {\rm{m^2\;s^{-1}}}$, which is significantly smaller than $\nu_{\rm{m}}$ and $\alpha_{\rm{m}}$.
The origin of this discrepancy is currently unclear, but it may reflect the absence of atomic exchange in the present model.
By introducing Bose statistics, the HQDL state of this model system would not remain unchanged.

As shown in Sec.~\ref{sec:breakdown}, HQDL is one of the rare liquids obeying the fractional Stokes–Einstein (SE) relation with an exceptionally small exponent, $\xi\approx0.2$-$0.3$, despite exhibiting no dynamical heterogeneity. 
Although the breakdown of the SE relation is typically attributed to dynamical heterogeneity in supercooled liquids \cite{pastore2021}, it can also emerge from structural inhomogeneity, such as the three-dimensionally percolated network structures observed in metallic liquids far above the glass transition \cite{pan2017,pan2017pccp}.
In HQDL, such structural inhomogeneity is clearly manifested by the interpenetration of the spatially extended necklaces, as evidenced by the non-zero distribution at $r \approx 0$ in the centroid–centroid radial distribution function $g_{\rm cc}(r)$ (see Fig.~S8 in the Supplementary Material and Fig.~6 in Ref.~\onlinecite{tsujimoto2024}). 

This percolation-driven structural feature offers a possible explanation for the temporal decoupling between the transport properties.
In a conventional classical liquid, single-particle mass diffusion and collective shear relaxation are correlated with the same physical process, meaning that the VAF relaxation time $\tau_{\rm v}$ and the stress relaxation time $\tau_{\rm s}$ are almost on the same order of magnitude. 
In contrast, HQDL exhibits a significant time-scale separation: $\tau_{\rm v} \ (\approx 4.4~\textrm{ps}) \gg \tau_{\rm s} \ (\sim 10^{-3}~\textrm{ps})$ at 0.1~K (Sec.~\ref{sec:relaxation}). 
This mechanism may reflect the difference between individual atomic motion driven by nuclear quantum fluctuations and the collective modes of the percolated necklace network.
Under an applied macroscopic shear stress, the percolated network can dissipate the shear stress almost instantaneously ($\tau_{\rm s}$) without structural hindrance, since the relaxation does not require the actual spatial displacement of individual atoms. 
In contrast, the spatial translation of each individual atom proceeds more slowly ($\tau_{\rm v}$), as the necklaces must smoothly move on the centroid potential surface $V_{\rm c}$ that has been effectively smoothed out by the strong NQEs. 
This distinctive mechanism may be the physical reason for the breakdown of the SE relation in the absence of dynamical heterogeneity.

It is not straightforward to apply classical kinetic theory to the gas-like regime
of HQDL. 
At a basic level, the shear viscosity is expressed as \cite{poling2001}
\begin{equation}
\eta_{\rm s} \sim \rho m v L,
\end{equation}
where $v$ and $L$ denote the mean velocity and the mean free path, respectively. 
Indeed, the increase in $\eta_{\rm s}$ upon cooling resembles that observed in the normal liquid phase of real $^3$He prior to the superfluid transition at approximately 2.5 mK. 
Even if modified kinetic descriptions are introduced \cite{dobbs2000,Lifshitz1981,chapman1970}, it remains unclear whether such a kinetic description applies to the gas-like regime of HQDL.
Nevertheless, the transport crossover identified here is strongly supported by the present results.
The qualitative change in the relaxation behavior of the VAF, together with the emergence of transport minima, provides phenomenological evidence for a dynamical and transport crossover.
Consequently, HQDL can be regarded as a phenomenologically gas-like state (Table~\ref{tab:comparison}), irrespective of whether its behavior can be interpreted within gas kinetic theory.

Such gas-like characteristics, i.e., extremely low shear viscosity and ballistic-like atomic motion, reflect the inertial nature of the dynamics.
Accordingly, HQDL may be qualitatively compared to the superfluid component in real He II, where particles move without scattering, although the underlying mechanisms are fundamentally different.
In contrast to He II, however, HQDL shows no anomalous enhancement of thermal conductivity because the prohibition of particle exchange hinders the development of many-body correlations among atoms that would support coherent energy transport.
As $\tau_{\rm s}$ is nearly instantaneous, momentum transport is extremely fast and predominant, whereas coherent thermal transport is relatively suppressed, consistent with $Pr > 1$.
The former feature arises from the suppression of interparticle friction due to nuclear quantum delocalization of the atoms, whereas the latter reflects the prohibition of exchange correlations among expanded atomic wavepackets, i.e., necklaces.
HQDL can thus be regarded as a momentum-dominated inertial fluid with suppressed coherent thermal transport.
This is in contrast to LQDL, which behaves as an ordinary dissipative fluid that is heat-transport dominated ($Pr < 1$), similar to real He I (Table~\ref{tab:comparison}).

\subsection{\label{sec:minimum} Transport minima and a second Frenkel line}

\begin{table*}
\centering
\caption{Comparison of the two Frenkel lines in distinguishable $^4$He.}
\label{tab:frenkel_comparison}
\setlength{\tabcolsep}{8pt}
\begin{tabular*}{\textwidth}{@{\extracolsep{\fill}} lll}
\hline\hline
\textbf{Feature} & \textbf{Frenkel line (supercritical)} & \textbf{Second Frenkel line (subcritical)} \\
\hline
Origin  & Brazhkin \textit{et al.} \cite{brazhkin2012,brazhkin2013} & This work \\
Regime & Supercritical fluid &  Low-temperature subcritical liquid  \\
Primary driver & Thermal fluctuations & Nuclear quantum fluctuations (NQEs) \\
Dynamical and transport crossover & Liquid-like $\to$ Gas-like (upon heating) & Liquid-like $\to$ Gas-like (upon cooling) \\
Transport minima & Widely observed in supercritical fluids & Newly emerging in the subcritical region \\
\hline\hline
\end{tabular*}
\end{table*}

\begin{figure*}
\includegraphics[width=12cm]{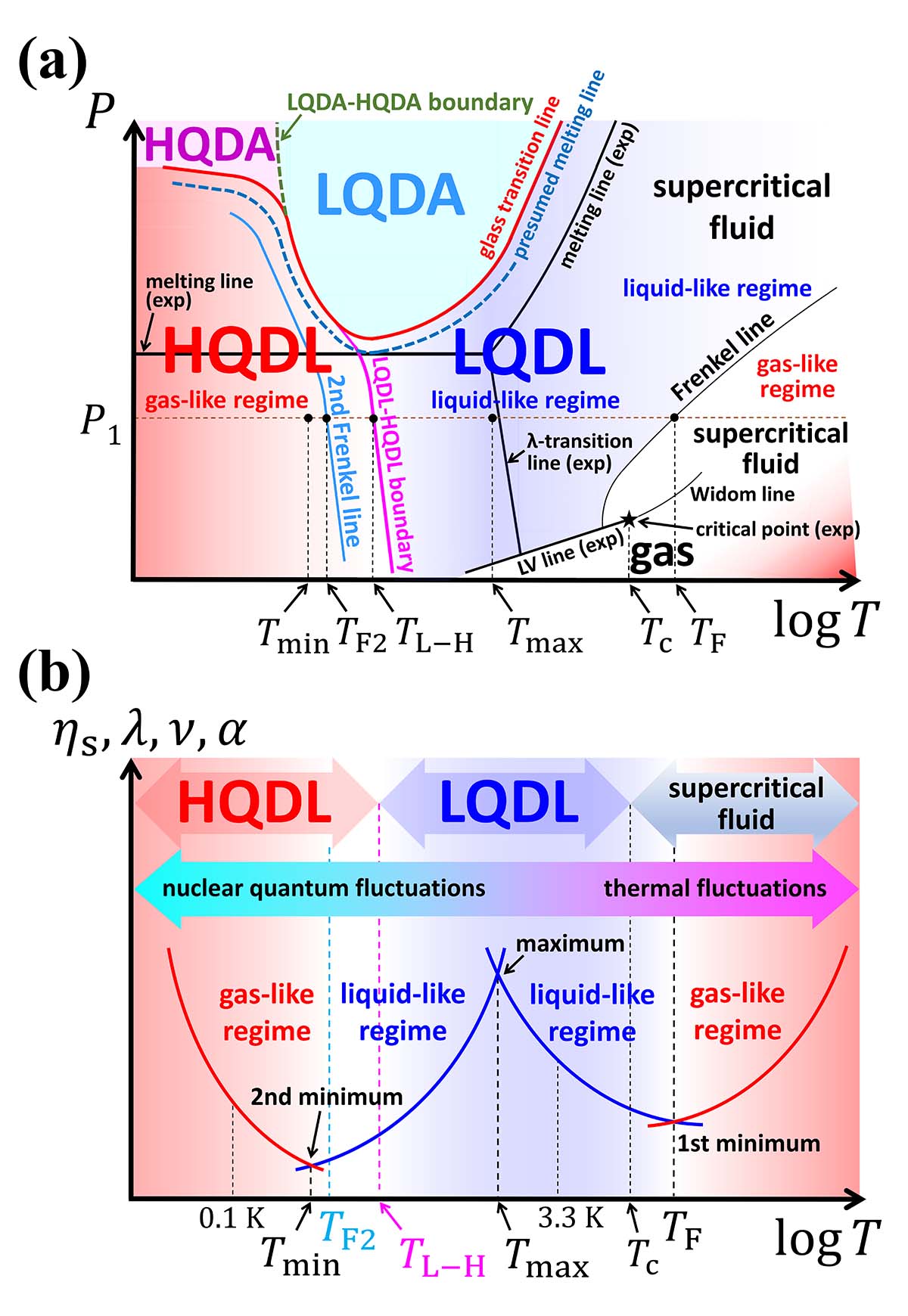}
\caption{\label{fig:schematic} 
Schematic pressure-temperature diagram and corresponding transport crossovers of distinguishable $^4$He.
(a) Schematic $P$--$T$ state diagram. Black solid lines indicate the phase or state boundaries experimentally observed for real $^4$He, while the Frenkel line in the supercritical region was identified in our previous study \cite{takemoto2018}.
In the subcritical temperature range, distinguishable $^4$He exhibits the LQDL, HQDL, LQDA, and HQDA regimes, which are distinct from the experimentally observed phases of real $^4$He.
(b) Schematic isobaric temperature dependence of four transport properties along an isobar at pressure $P_1$ ($P_1>P_{\rm c}$), indicated in panel (a).
The characteristic temperatures $T_{\rm c}$, $T_{\rm F}$, $T_{\rm F2}$, $T_{\rm min}$, and $T_{\rm max}$ denote the critical temperature, the Frenkel temperature associated with the Frenkel line introduced by Brazhkin \textit{et al.}, the second Frenkel temperature in the subcritical low-temperature region, and the temperatures at which a transport property exhibits a minimum and a maximum, respectively.
Two transport crossovers are identified: one associated with the Frenkel line in the supercritical region and the other with the second Frenkel line in the subcritical low-temperature region.
Each collective transport coefficient exhibits two distinct minima.
With decreasing temperature, nuclear quantum fluctuations become dominant, whereas thermal fluctuations dominate at higher temperatures.
}
\end{figure*}

In this subsection, we discuss the anomalous transport minima observed in the subcritical region.
To facilitate the following discussion, we first summarize the key differences between the newly identified second Frenkel line and the conventional one found in the supercritical regime in Table \ref{tab:frenkel_comparison}. 
Furthermore, a schematic illustration of these crossovers is provided in Fig.~\ref{fig:schematic}, which serves as a unified picture to guide our following discussions.
For clarity, we note that the Widom line is rigorously defined as the locus of maxima of the correlation length on the $P$–$T$ plane \cite{xu2005,stanley2008,mcmillan2010}, whereas the Frenkel line is defined as the dynamical boundary at which two characteristic timescales become equal \cite{brazhkin2012}, namely, the time required for a particle to move a distance comparable to its size and the period of transverse vibrations.
In the discussion below, following previous studies, we operationally identify the Widom line with the locus of $C_P$ maxima \cite{xu2005,stanley2008,mcmillan2010,takemoto2018} and the Frenkel line with the crossover from oscillatory to monotonic relaxation of the VAF \cite{brazhkin2013,yang2015,takemoto2018}.

Trachenko \textit{et al.} noted that the fluid state of all substances exhibits a minimum in each of $\eta_{\rm s}$, $\lambda$, $\nu$, and $\alpha$ as a function of temperature along isobars.
This is an intrinsic property of fluids in general, including in the supercritical region, and supercritical helium is no exception \cite{trachenko2020,trachenko2021,trachenkobook}.
They attributed the emergence of these minima to the continuous crossover between the liquid-like and gas-like regimes of particle dynamics.
This dynamical transition is commonly observed for supercritical fluids that do not undergo the liquid-gas phase transition.
The minimum exists in proximity to the Frenkel temperature $T_{\rm{F}}$, which marks the dynamical boundary between liquid-like and gas-like regimes in supercritical  fluids \cite{brazhkin2012,brazhkin2013}.
These two dynamical states across the Frenkel line are conveniently distinguished 
by whether the VAF exhibits monotonic relaxation (gas‑like) or oscillatory relaxation (liquid‑like) \cite{brazhkin2012,brazhkin2013}. 
We previously showed that the supercritical fluid of distinguishable $^4$He exhibits the same dynamical transition across the Frenkel line as classical fluids \cite{takemoto2018}.
In the present study, we extend this framework to the subcritical region, revealing a novel dynamical regime driven by quantum fluctuations rather than thermal ones.

Each collective transport coefficient exhibits two distinct minima along isobars in this system, as seen in the plots of  $\eta_{\rm{s}}$-$T$ (Fig.~\ref{fig:vis}), $\lambda$-$T$ (Fig.~\ref{fig:thermal}), $\nu$-$T$ (Fig.~\ref{fig:ka}(a)), and $\alpha$-$T$ (Fig.~\ref{fig:ka}(b)).
In the plotted pressure range, the minimum temperatures of $\nu$ and $\alpha$ are $T_{\rm{\nu,min}}=0.15-0.2~\rm{K}$ and $T_{\rm{\alpha,min}}=0.2~\rm{K}$, respectively, both slightly lower than the LQDL-HQDL transition temperature  ($T_{\rm{L-H}}=0.3$-0.7 K, Fig.~\ref{fig:phase}).
As shown in Sec.~\ref{sec:VAF}, the VAF profile changes from an oscillatory decay to a monotonic decay, broadly accompanying the LQDL-to-HQDL transition.
This supports the idea that the emergence of minima in this liquid is associated with the crossover from liquid-like to gas-like regimes, although this crossover occurs in the subcritical low-temperature region.
Accordingly, the temperature at which the VAF profile changes can be regarded as that of {\it a second Frenkel line}, $T_{\rm{F2}}$, introduced here as another dynamical transition.
This temperature lies close to, but slightly below,  $T_{\rm{L-H}}$.
The locus of $T_{\rm F2}$ defines the second Frenkel line on the $P$-$T$ plane, as shown in Fig.~\ref{fig:phase}.
The minimum temperatures $T_{\rm{\nu,min}}$ and $T_{\rm{\alpha,min}}$ are even below $T_{\rm F2}$ (as schematically illustrated in Fig.~\ref{fig:schematic}).
Indeed, also for supercritical fluids, the temperatures at which transport coefficients reach their minima are slightly shifted from $T_{\rm{F}}$ \cite{trachenko2020,trachenko2021,trachenkobook}.

Thus, two distinct transport minima can be identified along isobars
in distinguishable $^4$He: one is known to lie in the supercritical region of real $^4$He, where exchange effects are negligible and the system is therefore effectively equivalent to distinguishable $^4$He, and the other is newly found in the subcritical low-temperature region.
Notably, the crossover from liquid-like to gas-like regimes in this subcritical liquid is caused by decreasing temperature, not by heating.
Upon cooling along an isobar in distinguishable $^4$He, NQEs become increasingly prominent, leading to the delocalization of atomic positions,
i.e., an increase in the quantum wavelength
 $\lambda_{\mathrm{quantum}}$,
thereby effectively weakening intermolecular interactions.
Consequently,  the relaxation in the VAF becomes more gas-like or ballistic at lower temperatures, while liquid-like oscillatory behavior becomes more pronounced at higher temperatures.
At high temperatures in the supercritical region, gas-like dynamics is driven by enhanced thermal fluctuations, whereas at low temperatures in the subcritical region it emerges from enhanced quantum fluctuations (see Fig.~\ref{fig:schematic}(b)).
This explains the two distinct minima in the transport coefficients in the supercritical and subcritical regions (Figs.~\ref{fig:vis}, ~\ref{fig:thermal}, and ~\ref{fig:ka}).
In other substances, the second Frenkel line is absent because they solidify upon further cooling from the liquid state, owing to much stronger intermolecular interactions and significantly weaker NQEs than in this system.
Therefore, distinguishable $^4$He can be regarded as a prototypical example illustrating how transport crossovers can emerge as the dominant fluctuation mechanism changes from thermal to nuclear quantum fluctuations in a system that does not freeze down to absolute zero.

Upon isobaric cooling from the supercritical region toward 0.1 K,
the dynamical regime evolves through the sequence:  
${\rm G}{\rightarrow}T{_{\rm F}}{\rightarrow}{\rm L}{\rightarrow}T{_{\rm max}}{\rightarrow}{\rm L}{\rightarrow}T{_{\rm L-H}}{\rightarrow}T{_{\rm F2}}{\rightarrow}T{_{\rm min}}{\rightarrow}{\rm G}$, where G and L denote gas-like and liquid-like regimes, respectively, and $T_{\rm max}$ is the  temperature of the local maximum near 2 K.
Thus, the system reenters a gas-like regime at low temperature after passing through a liquid-like regime at intermediate temperatures.
This reentrant transport crossover is summarized schematically in Fig.~\ref{fig:schematic}, while the main differences between the two Frenkel lines are listed in Table~\ref{tab:frenkel_comparison}.
No indication of a thermodynamic transition,
such as a $C_P$ maximum, is observed along the isobar at
$P_1\,(>P_{\rm c})$ over the entire illustrated temperature range in Fig.~\ref{fig:schematic},
even though clear dynamical crossovers appear.
Consistently, the enthalpy–temperature curves shown in Fig.~S5 of the Supplementary Material remain smooth throughout this temperature range at $P\leq10$ bar, where vitrification does not occur (see also Ref.~\onlinecite{tsujimoto2024}).
In the supercritical region of $^{4}$He, a dynamical crossover is not observed across the Widom line, here identified operationally as the locus of $C_P$ maxima.
Instead, it is detected as a change in the VAF at the Frenkel line, as in classical supercritical fluids \cite{takemoto2018}.
Similarly, within the temperature range considered here,
the second Frenkel line is not accompanied by a $C_P$ maximum
but instead represents a purely dynamical crossover.
On the other hand, the LQDL-HQDL crossover, which emerges
at a slightly higher temperature than the second Frenkel line, coincides with a change in the temperature exponent $\chi$ of the quantum wavelength 
$\lambda_{\mathrm{quantum}}{\sim}T^{\chi}$ (Fig.~S3 in the Supplementary Material).
We find $\chi \approx -0.3$ for the LQDL regime, and
$-0.6 \le \chi \le -0.5$ for the HQDL regime \cite{tsujimoto2024}.
In fact, the former values are characteristic of the liquid-like regime of supercritical $^{4}$He below the Widom line, whereas the latter correspond to the gas-like regime beyond the line (Table~\ref{tab:scf}) \cite{takemoto2018}.
Therefore, the LQDL–HQDL crossover resembles the regime change
between liquid-like and gas-like supercritical fluids in terms of
the temperature dependence of the nuclear quantum delocalization.
Interestingly, the separation between $T_{\rm F2}$ and
$T_{\rm L-H}$ resembles that between the Frenkel temperature ($T_{\rm F}$),
defined by the VAF crossover, and the Widom temperature,
associated with the crossover in $\chi$, in the supercritical region \cite{takemoto2018}.

As noted in Sec.~\ref{sec:introduction}, Markland \textit{et al.} reported that a quantum Lennard–Jones glass exhibits two distinct glass states (the \textit{trapped} and \textit{tunneling} regimes) \cite{markland2011,markland2012}. 
We subsequently showed that these regimes correspond to LQDA and HQDA in distinguishable $^4$He, respectively \cite{kinugawa2021}.
In Ref.~\onlinecite{tsujimoto2024}, we demonstrated that LQDL and HQDL are the liquid counterparts of LQDA and HQDA, respectively; the latter were initially observed under compressed conditions \cite{kinugawa2021}.
Atomic tunneling is rare in LQDA, as well as in LQDL and HQDL \cite{tsujimoto2024}, whereas it has been observed only in highly compressed HQDA \cite{kinugawa2021}.
This suggests that atomic transport in this pressure range occurs predominantly without tunneling.
Nevertheless, LQDA and HQDA can be regarded as analogous to Markland’s two regimes in terms of nuclear quantum delocalization.
Furthermore, they identified signatures of inverse melting in the tunneling regime of their quantum model glass \cite{markland2012}, consistent with the inverse freezing from HQDL to LQDA at 40–50 bar \cite{tsujimoto2024}.
Consequently, the dynamical crossover identified in the present work, as illustrated in Fig.~\ref{fig:schematic}, may not be unique to distinguishable $^4$He.
More broadly, similar two-state behavior and associated transport crossovers may emerge in quantum liquids without atomic exchange, provided that freezing upon cooling is suppressed by strong NQEs and weak intermolecular interactions.

Recently, Trachenko claimed  that 
the lambda-transition temperature of 
real  $^4$He  ($T_{\lambda}$) is almost equal to the Frenkel temperature, i.e., $T_{\lambda}{\cong}T_{\rm{F}}$ \cite{trachenko2025,trachenko2023p},
based on the VAFs obtained from the numerical simulations of bosonic $^4$He by Nakayama {\it et al} \cite{nakayama2005}.
We note, however, that Boltzmann $^4$He  does not exhibit the 
lambda transition and that the second Frenkel temperature $T_{\rm{F2}}$ is 
far below the $T_{\lambda}$ of real $^4$He.
It is an important subject for future work to investigate this
issue by performing dynamical simulations for bosonic $^4$He.

\subsection{\label{sec:meltingline} Mirror symmetry in melting and transport crossovers}

We next discuss the geometric characteristics of the state diagram, highlighting a striking mirror symmetry shared between the macroscopic melting curve and the microscopic transport crossovers on the $P$–$T$ plane.

The successive state transitions, schematically illustrated in Fig.~\ref{fig:schematic}, appear to exhibit mirror symmetry. 
This observation is consistent with the mirror symmetry of the downward-convex shape of the melting line (and the glass transition line) in the state diagram (Fig.~\ref{fig:phase}). 
The LQDL–HQDL boundary line originates near the minimum of the presumed melting curve on the $P$–$T$ plane.
On the left side of this curve, inverse melting occurs upon isobaric cooling of HQDA \cite{tsujimoto2024}. 
Consistent with this behavior, the temperature range below the minimum of the melting curve is associated with the {\it inverse transport transition} of the liquid, 
namely, a transition from a liquid-like to a gas-like regime induced by cooling, as discussed in Sec.~\ref{sec:minimum}.

Trachenko argued that the melting line of any substance should be parabolic, potentially including a segment corresponding to inverse melting \cite{trachenko2024}.
This is consistent with our downward-convex melting curve.
When we apply  his theoretical equation Eq. (19) in Ref.~\onlinecite{trachenko2024} to real helium, it predicts that the  minimum would occur at a negative temperature.
For known materials, the portion of the melting curve with a negative slope lies entirely in the negative‑temperature domain and is therefore not observable.
Thus, this system appears to be one of the very few known exceptions in which the melting curve attains a minimum at a positive temperature.

\subsection{\label{sec:glass}Distinction between two glass states}

The contrasting transport behaviors of the two vitrified states, LQDA and HQDA, 
demonstrate that strong NQEs fundamentally alter the collective dynamics of the amorphous solid, preventing conventional vitrification.
As described in Secs.~\ref{sec:MSD}--\ref{sec:thermal}, HQDA exhibits several unusual features.
Even at 50-60 bar, the MSD exponent $\gamma$ is only slightly smaller than unity, indicating that the character of atomic motion remains close to that in the liquid states (HQDL and LQDL).
Furthermore, the single-particle self-diffusion is suppressed below our operational threshold of vitrification (Sec.~\ref{sec:simulationprocedure}), whereas the collective shear viscosity $\eta_{\rm s}$ is less than $10^{-5}$~Pa$\cdot$s, which is orders of magnitude smaller than the conventional vitrification criterion, $\eta_{\rm s} \sim 10^{12}$~Pa$\cdot$s \cite{ediger1996,angell1995,debenedetti2001}.
While individual atomic mass transport is effectively frozen on this timescale, the system macroscopically remains capable of rapidly dissipating shear stress.
This extraordinary decoupling indicates that HQDA should not be viewed as a conventional structural glass but rather resembles a non-ergodic liquid-like state, where the strongly suppressed diffusion of individual atoms and collective shear relaxation operate on strongly decoupled timescales due to zero-point motion.
As described in Secs.~\ref{sec:breakdown} and~\ref{sec:relaxation}, this decoupling is evidenced by the conformity to the fractional SE relation and the marked difference between the relaxation times $\tau_v$ and $\tau_{\rm s}$.
These features characterize HQDA as a glassy yet fluidic state driven by prominent NQEs, which rapidly release shear stress even in the absence of active atomic self-diffusion.

Despite this dynamical decoupling, however, the non-Gaussian parameter (NGP) does not indicate dynamical heterogeneity, in contrast to typical supercooled liquids and glasses.
In addition, HQDA has a thermal conductivity $\lambda$ comparable to that of HQDL, suggesting that its transport properties remain liquid-like.
Overall, HQDA retains many characteristics of the liquid state (HQDL) despite being classified as a glass by diffusion-based criteria.

In contrast, LQDA does not exhibit such anomalous transport behavior.
In this state, the stress autocorrelation function does not decay to zero within 3~ps (see Figs.~S22 and S23 in the Supplementary Material), which effectively prevents the estimation of $\eta_{\rm s}$ and implies a high viscosity.
Moreover, LQDA shows a significantly higher thermal conductivity $\lambda$ than LQDL (Sec.~\ref{sec:thermal}).
Unlike HQDA, therefore, LQDA does not resemble its corresponding liquid state, LQDL.
Consequently, the two glass states can be clearly distinguished in terms of their transport characteristics.

\subsection{\label{sec:realistic}Comparison with real quantum systems}

As the temperature decreases, atomic exchange effects arising from Bose statistics become increasingly pronounced in bulk $^4$He, rendering the distinguishable-atom model progressively unrealistic. 
However, when $^4$He is confined within extremely narrow nanopores of mesoporous materials, long-range Bose-exchange correlations and global phase coherence
are strongly suppressed.
This exchange-suppressed regime, experimentally investigated for $^4$He confined in nanopores of mesoporous materials, gives rise to markedly different phase diagrams, including a third liquid state known as the localized Bose–Einstein condensate (LBEC), or a BEC-like low-entropy state, in addition to the normal fluid and the superfluid states \cite{shirahama2008,shirahama2008ltp,yamamoto2004,yamamoto2008,yamamoto2008jpsj,taniguchi2010,taniguchi2011,taniguchi2013}.
Interestingly, such confined systems of $^4$He exhibit two distinct states below the normal-fluid temperature: the LBEC (or a BEC-like low-entropy state) and the superfluid state. 
Upon cooling, the normal liquid first transitions to the LBEC phase at $T_{\rm B}$, and subsequently to the superfluid phase at $T_{\rm S}$ (denoted $T_{\rm c}$ in the literature) \cite{shirahama2008,shirahama2008ltp,yamamoto2004,yamamoto2008,yamamoto2008jpsj}. 
Importantly, a broad maximum in $C_P$ appears at the higher temperature $T_{\rm B}$, whereas no corresponding maximum is observed at $T_{\rm S}$.
This feature of the transition from the LBEC to the superfluid state at $T_{\rm S}$ may bear resemblance to our model system, in which the LQDL–HQDL transition occurs as a continuous transport crossover without a $C_P$ maximum. 
At first glance, the two temperatures $T_{\rm B}$ and $T_{\rm S}$ may be regarded as analogous to the Widom and Frenkel temperatures (the former defined by the loci of $C_P$ maxima) in the supercritical region, respectively, in that the former reflects a thermodynamic boundary, whereas the latter is associated with a dynamical crossover.

We emphasize that this comparison is not intended to imply a direct correspondence between the confined $^4$He LBEC state and the present distinguishable $^4$He system. 
Rather, it is meant only to contrast the behavior of systems in which Bose exchange effects are partially suppressed with that of a system in which Bose exchange is entirely absent, without assuming a common microscopic origin.
While the macroscopic transport in real quantum systems is ultimately governed by phase
coherence, our observations raise the possibility that strong NQEs alone can give rise to transport characteristics qualitatively
reminiscent of some of those observed in 
confined $^4$He.
In this sense, the present results provide an exchange-free reference point for disentangling the respective roles of NQEs and Bose exchange in quantum transport.
Whether the observed similarities have a deeper physical connection remains an open question and is an important subject for future work.

It is worth expanding our perspective beyond helium to other fundamental quantum systems in which NQEs play a dominant role. 
Such quantum-fluctuation-driven phase behavior can be related to the reentrant melting observed in ultrahigh-pressure crystalline hydrogen \cite{scandolo2003,bonev2004,morales2010}. 
In high-pressure hydrogen, strong NQEs of light protons enhance zero-point motion, destabilizing the crystalline lattice and consequently favoring the liquid state. 
The present distinguishable $^4$He fluid exhibits a reentrant transport crossover ($\text{LQDL} \rightarrow \text{LQDA} \rightarrow \text{HQDL}$) upon isobaric cooling at $20\leq{P}\leq40$ bar. 
This provides a conceptual bridge to NQE-driven phase anomalies in high-pressure hydrogen. 
An important future direction is to investigate how $^4$He behaves in terms of transport properties when Bose statistics are fully included, in comparison with the Boltzmann system analyzed in the present study.

\subsection{\label{sec:note}Methodological considerations and robustness of the transport crossover}

Finally, we assess the validity and limitations of the CMD approximation within the Green–Kubo framework, confirming that our qualitative conclusions regarding the transport crossovers remain physically robust.

A rigorous theoretical justification for the centroid approximation in the Green–Kubo formula, Eq.~(\ref{eq:approx}), is generally not available. 
This approximation becomes exact when the physical variable $A$ is a linear function of atomic positions and/or momenta \cite{Jang1999}, whereas its validity for nonlinear variables largely relies on empirical validation. 
We have previously shown that the present scheme for evaluating $\eta_{\rm s}$ and $\lambda$ yields good agreement with experimental values for He I at SVP densities and above $T_{\lambda}$ \cite{imaoka2017}, as well as for liquid para-hydrogen \cite{yonetani2003,yonetani2004}. 
This supports the applicability of the same approach to the present study.

Although a rigorous theoretical justification for the nonlinear time correlation functions by the CMD remains lacking, it is nevertheless possible to rationalize why the calculated $\eta_{\rm s}$ reproduces the experimental trends so well. 
As shown in Eq.~(\ref{eq:etaproduct}), $\eta_{\rm s}$ can be expressed as the product of the high-frequency elastic modulus $G_{\infty}$, a static quantity, and the relaxation time $\tau_{\rm s}$ of the stress autocorrelation function.
Static properties are accurately evaluated within the framework of the static centroid approximation, regardless of the validity of the time dependence. 
The dynamic quantity $\tau_{\rm s}$ is generally expected to be relatively insensitive to the approximations introduced, as it reflects an integrated timescale rather than the detailed temporal features of the time correlation function.
Therefore, $\eta_{\rm s}$, as the product of these two quantities, can be evaluated with reasonable accuracy in practice.

On the other hand, the agreement for $\lambda$ was less satisfactory than that for $\eta_{\rm{s}}$.
An alternative and more accurate method for estimating $\lambda$ was provided by Sutherland \textit{et al.} \cite{sutherland2021}, which bypasses the nonlinear centroid current operators.
In the present work, we base our discussion on the transport coefficients computed consistently using our Green–Kubo scheme, and a quantitative comparison with available experimental data remains an important task for future work.

The observed transport properties are governed by the degree of spatial extension of the necklaces.
The extension of atomic necklaces is a static equilibrium quantity that is evaluated as a statistical average within the framework of the discretized path integral representation of the partition function [Eq.~(\ref{eq:partition})]. 
Therefore, this static property is largely independent of the choice of dynamical approximation (CMD or RPMD) and specific numerical parameters, provided that the simulations are carried out within the same path integral formalism. 
Accordingly, while the quantitative values of collective transport coefficients may exhibit minor variations depending on numerical details, the conclusion that the transport regime undergoes a qualitative transformation driven by NQEs is expected to be robust and independent of methodological details.
In this sense, the present quantum crossover originates from a static modification of the effective interaction due to equilibrium quantum delocalization.
This is in contrast to thermal fluctuations in classical supercritical fluids, 
where the intermolecular potential itself remains unchanged, 
and the crossover arises purely from the dynamical characteristics of particle motion and the associated timescale separation \cite{brazhkin2012,brazhkin2013}.

Finally, we briefly comment on the possible role of Bose statistics, which are not included in the present model. 
For Bose systems, the characteristic timescale for particle exchange, $\tau_{\rm Bose} \approx h/(k_{\rm B}T)$ \cite{zaccone2023}, is on the order of 100 ps at 0.1 K. 
This is significantly longer than the relaxation times $\tau_{\rm v}\simeq4.4~{\rm ps} \gg \tau_{\rm e}\sim10^{-2}~{\rm ps} > \tau_{\rm s}\sim10^{-3}~{\rm ps}$ of the relevant time correlation functions evaluated in this work. 
This clear separation of timescales suggests that exchange effects would likely be negligible within the present simulation time window and the Green–Kubo framework. 
However, this issue warrants further investigation through advanced simulations incorporating Bose statistics.

\section{\label{sec:conclusions}CONCLUSIONS}

For distinguishable $^4$He obeying Boltzmann statistics,
we have investigated the dynamical
and transport properties of two distinct liquid states, i.e., LQDL and HQDL, over a wide $P$-$T$ range ($0.1\leq{T}\leq{3.3}$ K and $1\leq{P}\leq{60}$ bar) by employing
 CMD simulations together with the centroid approximation to the Green-Kubo formula.
 This study clarifies how these two liquid states, which emerge in $^4$He in the absence of atomic exchange, exhibit remarkably unusual and distinctive transport properties.
The main conclusions are summarized as follows.

\begin{enumerate}
\renewcommand{\labelenumi}{(\arabic{enumi})}

\item 
The LQDL emerges at $T\geq 0.5$ K and becomes essentially 
equivalent to real He I for $T>T_{\lambda}$, where the atomic exchange effects are negligible.
The estimated shear viscosity for $T>T_{\lambda}$ agrees well
with experimental values for real He I ($\sim10^{-6}-10^{-5}$ Pa$\cdot$s).
Across  the entire temperature range, including below $T_{\lambda}$,
the LQDL exhibits dynamical and transport properties characteristic of ordinary liquids:
(i) normal diffusion with an MSD exponent  $\gamma\approx{1}$, indicating standard Brownian motion; 
(ii) an oscillatory decay of the VAF; 
(iii) conformity to  the SE relation;
and
(iv) a Prandtl number $Pr<1$.
Thus, LQDL behaves as an ordinary heat-transport-dominated dissipative fluid, similar to real He I.

\item 
The HQDL is an anomalous non-superfluid state characterized by 
(i) superdiffusion, manifested by an MSD exponent $\gamma > 1$; 
(ii) a monotonic decay of the VAF;
(iii) one of the lowest shear viscosities among known atomic or molecular liquids ($\eta_{\rm s} \sim 10^{-7}~\mathrm{Pa\cdot s}$);
(iv) consistency with the fractional SE relation ($0.23\leq\xi\leq0.30$), in clear deviation from the conventional SE relation; 
and 
(v) a Prandtl number $Pr>1$.
Although the thermal conductivity increases upon cooling, it does not diverge as in real He II.
Features (i) and (ii) are shared with gas-like supercritical fluids of both classical and quantum substances, including $^4$He. 
The ultralow but nonzero viscosity of HQDL does not correspond to the zero viscosity of the superfluid component in real He II. 
Instead, it indicates quantum fluidity without superfluidity. 
The inertial nature of HQDL may be compared to the superfluid component in real He II, where particles move without scattering, despite the fundamental differences in the underlying mechanisms. 
However, the divergence of thermal conductivity due to interparticle exchange correlations is absent in HQDL.
Thus, HQDL is a momentum-dominated inertial fluid with suppressed coherent thermal transport and exhibits a gas-like character.

\item 
As reported previously \cite{tsujimoto2024}, the LQDL–HQDL transition is a continuous crossover, manifested as a change in the expansion factor $\alpha_{\lambda}$ of $\lambda_{\mathrm{quantum}}$, and is not a thermodynamic phase transition. 
In this study, we show that this transition is also a transport crossover, characterized by a change from liquid-like to gas-like behavior, a crossover from the SE relation to the fractional SE relation, and a change in the Prandtl number.
Accordingly, the two liquids can be distinguished as distinct states through their transport properties.
This crossover reflects changes in both the spatial extension of necklaces and the resulting transport behavior.

\item 
In each temperature dependence of shear viscosity $\eta_{\rm s}$, thermal conductivity $\lambda$, kinematic viscosity $\nu$, and thermal diffusivity $\alpha$, a second minimum emerges in the subcritical region, in addition to the minimum in the supercritical region. 
These minima lie slightly below the LQDL–HQDL transition temperature.
They are attributed to a crossover from a liquid-like to a gas-like regime upon cooling.
On the $P$–$T$ state diagram, the gas-like/liquid-like dynamical crossover identified by the change in the VAF profile defines a second Frenkel line, whereas the corresponding transport minima appear at slightly lower temperatures.
This boundary is not accompanied by any indication of a maximum in $C_P$, in contrast to the Widom line; this is also the case in the supercritical region.

The transport crossover across this second Frenkel line in the subcritical region is opposite to that in the supercritical region, where the transition from liquid-like to gas-like behavior occurs upon heating. 
Instead, it arises from NQEs, which become increasingly pronounced at lower temperatures, whereas the gas-like dynamics in the supercritical region is driven by enhanced thermal fluctuations. 
The present system therefore provides a prototypical example in which a second minimum appears in each of $\eta_{\rm s}$, $\lambda$, $\nu$, and $\alpha$. 
In other substances, which possess stronger intermolecular interactions and much weaker NQEs, such a second minimum cannot emerge because they solidify upon further cooling. 
Thus, distinguishable $^4$He provides a unique opportunity to observe transport crossovers driven by a change in the dominant mechanism from thermal to nuclear quantum fluctuations without encountering freezing.

\item 
Consistently, even in the glassy state of HQDA identified by self-diffusion-based criteria, collective transport properties remain comparable to those of HQDL, and the viscosity remains low despite strongly suppressed self-diffusion. 
This marked decoupling between single-particle diffusion and collective stress relaxation gives rise to a glassy yet fluidic state driven by strong NQEs.

\end{enumerate}

We regard the quantum fluidity of HQDL as gas‑like transport behavior 
arising from the NQEs, irrespective of the absence of Bose condensation or 
superfluidity. 
This concept provides a unified framework for characterizing low‑viscosity quantum liquids beyond the conventional two-phase picture between normal fluid and superfluid. 
Accordingly, this study provides a basis for understanding quantum fluidity in non-superfluid systems and offers new perspectives on reexamining the known transport anomalies of real $^4$He.

As described in Sec.~\ref{sec:introduction}, under atmospheric pressure,  distinguishable $^4$He obeying Boltzmann statistics is expected to be a fluid at $T=0$.
We have shown that the Boltzmann liquid in the temperature range from 3.3 K down to 0.1 K does not remain in a single dynamical or transport state.
Rather, it bifurcates into two distinct dynamical states and, upon cooling, undergoes a crossover into a gas-like liquid state (HQDL), whose gas-like character is enhanced by nuclear quantum fluctuations.
This study reveals how a mere replacement of Bose statistics by Boltzmann statistics can dramatically transform the nature of $^4$He in the low-temperature regime, even before reaching absolute zero.
This finding also suggests that, in general, a Boltzmann liquid may not necessarily solidify upon cooling when NQEs are sufficiently strong. 
In such cases, cooling drives the system into a gas-like transport regime.
The present work thus provides an opportunity to reconsider, in a systematic way, what {\it freezing} of matter truly means.

\section*{\label{sec:SM}SUPPLEMENTARY MATERIAL}

 The Supplementary Material provides additional data and analyses supporting the results presented in the main text. 
 It includes the criteria used for state identification, simulation conditions, quantum wavelength and expansion factor, enthalpy and isobaric heat capacity, radial distribution functions, centroid velocity autocorrelation functions, phonon density of states, mean square displacements, and non-Gaussian parameters. 
 Additional results for stress and energy current autocorrelation functions, together with detailed analyses of shear viscosity and thermal conductivity, are also provided.
 A table summarizing the fitted exponents of the fractional Stokes-Einstein relation is included.

\section*{\label{sec:data}DATA AVAILABILITY}

 The data that support the findings of this study are available from the corresponding author upon reasonable request.

\appendix
\renewcommand{\appendixname}{Section}

\begin{acknowledgments}

We thank Professor Naoyuki Sakumichi for valuable discussions.
We also thank Moeha Amma for insightful discussions in our laboratory.
K.K. also thanks Professor Alessio Zaccone for drawing our attention to their work on dynamical indistinguishability in quantum fluids.
This work was supported by JSPS KAKENHI Grant Number 24K08354.

\end{acknowledgments}

\bibliography{Kinugawa.bib}

\clearpage
\onecolumngrid

\setcounter{equation}{0}
\setcounter{section}{0}
\setcounter{figure}{0}
\setcounter{table}{0}
\setcounter{page}{1}
\makeatletter
\begin{supplemental}

\setcounter{figure}{0}
\renewcommand{\thefigure}{S\arabic{figure}}
\setcounter{table}{0}
\renewcommand{\thetable}{S\arabic{table}}
\setcounter{equation}{0}
\renewcommand{\theequation}{S\arabic{equation}}
\renewcommand{\thesection}{S\arabic{section}}

\begin{center}
{\Large\bfseries Supplementary Material}

\vspace{2cm}

{\Large\bfseries Quantum fluctuation-driven transport crossover between two liquid states in distinguishable helium-4}

\vspace{1cm}

Mika Tanabe$^{}$, Momoko Tsujimoto$^{}$, and Kenichi Kinugawa$^{}$

\vspace{0.3cm}
{\it{Department of Chemistry, Graduate School of Humanities and Sciences, Nara Women's University, Nara 630-8506, Japan}} \\

\vspace{1cm}

\section*{Contents}

\begin{tabular}{lr}
Section S1. Criteria for state identification & \pageref{sec:SMcriteria}  \\
Section S2. Simulation conditions & \pageref{sec:SMcondition} \\
Section S3. Quantum wavelength and expansion factor & \pageref{sec:SMwavelength} \\
Section S4. Enthalpy and isobaric heat capacity & \pageref{sec:SMHandC} \\
Section S5. Radial distribution functions & \pageref{sec:SMRDF} \\
Section S6. Centroid velocity autocorrelation functions & \pageref{sec:SMVAF} \\
Section S7. Phonon density of states & \pageref{sec:SMPDS} \\
Section S8. Mean square displacements & \pageref{sec:SMMSD} \\
Section S9. Non-Gaussian parameters & \pageref{sec:SMNGP} \\
Section S10. Off-diagonal stress autocorrelation functions & \pageref{sec:SMSAF} \\
Section S11. Analysis of shear viscosity & \pageref{sec:SMASV} \\
Section S12. Energy current autocorrelation functions & \pageref{sec:SMECA} \\
Section S13. Analysis of thermal conductivity & \pageref{sec:SMATC} \\
Section S14. Table & \pageref{sec:SMTable} \\
\end{tabular}

\end{center}

\vspace{0.5cm}

\vspace{0.5cm}

\makeatletter
\def\section{\@startsection{section}{1}{\z@}%
  {-3.5ex \@plus -1ex \@minus -.2ex}%
  {2.3ex \@plus.2ex}%
  {\normalfont\large\bfseries}}
\makeatother

\clearpage

\onecolumngrid

\section{Criteria for state identification}
\label{sec:SMcriteria}

The criteria for state identification are as follows~\cite{tsujimoto2024}:

(1) When the self-diffusion coefficient calculated from Eq.~(\ref{eq:diffusion}) 
satisfies $D > 2.0 \times 10^{-10}$ m$^{2}$s$^{-1}$ 
(see Sec.~\ref{sec:diffusion}), the system is in a liquid state. 
Otherwise, the state is regarded as a glass because no crystalline states emerge in the present study.

(2) When a nonzero distribution of the centroid–centroid radial distribution function $g_{\rm cc}$ 
is observed at $r$ shorter than the closest distance in the bead–bead radial distribution function $g_{\rm bb}$, 
this indicates the high quantum-dispersion (HQD) state. Otherwise, the state is classified as the low quantum-dispersion (LQD) state.

(3) An increase and decrease in the expansion factor $\alpha_{\lambda}$ 
(Fig.~\ref{fig:SM_expansion}) under isobaric cooling correspond to the HQD and LQD states, respectively.

\section{\label{sec:SMcondition}Simulation conditions}

\twocolumngrid

\begin{figure}[H]
\includegraphics[width=9cm]{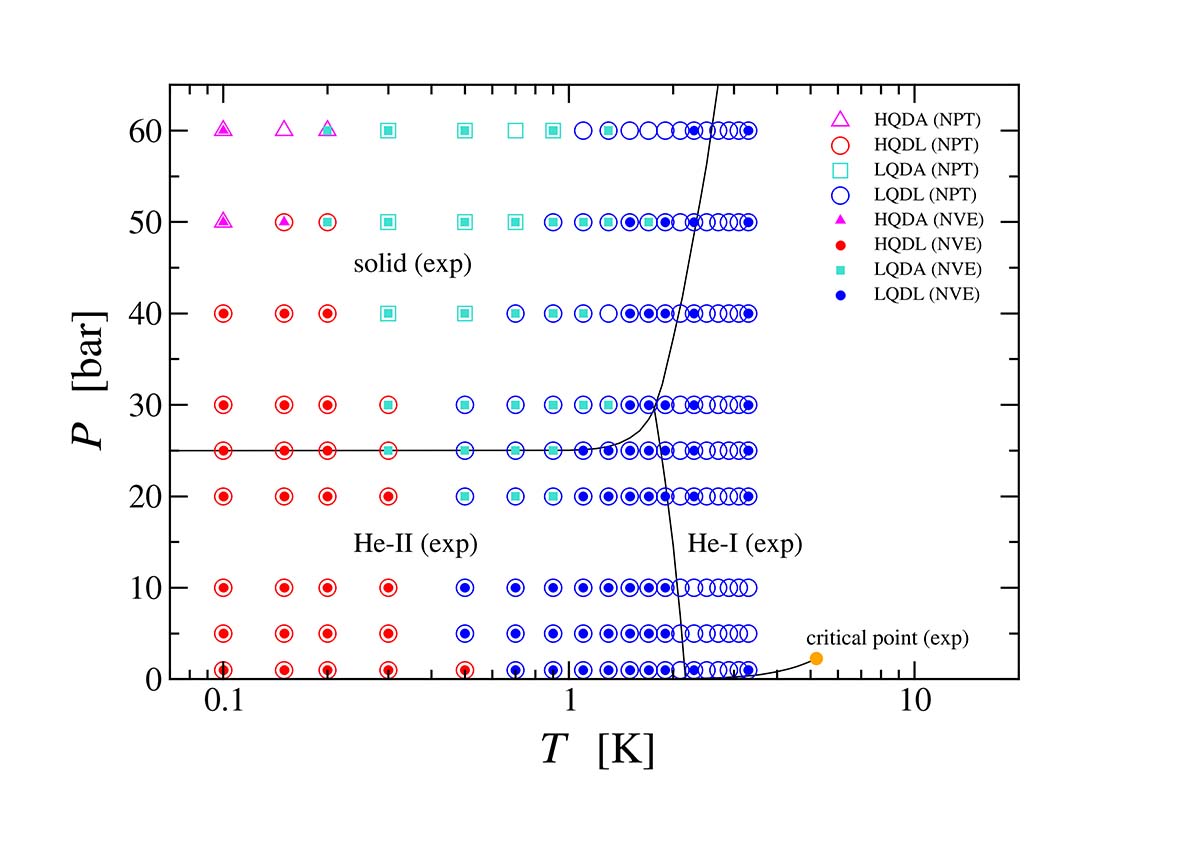}
\caption{\label{fig:conditon}
Pressure-temperature conditions used for evaluating the transport properties.
The plotted $P$--$T$ points correspond to those in the preceding NPT runs.
Closed symbols denote the conditions for the present NVE simulations, while open symbols indicate the conditions adopted in the NPT simulations in Ref.~\onlinecite{tsujimoto2024}, which precede the present NVE runs.
Solid lines represent the phase boundaries of real $^4$He.
Colors denote the identified states in the CMD simulations for each ensemble.
}
\vspace{2mm} \noindent\textbf{Alt Text:} Figure S1 shows the pressure-temperature conditions used for the transport-property calculations. The identified liquid and amorphous states occupy distinct regions, and differences between the preceding NPT simulations and the present NVE simulations are illustrated.
\end{figure}

\begin{figure}[H]
\includegraphics[width=9cm]{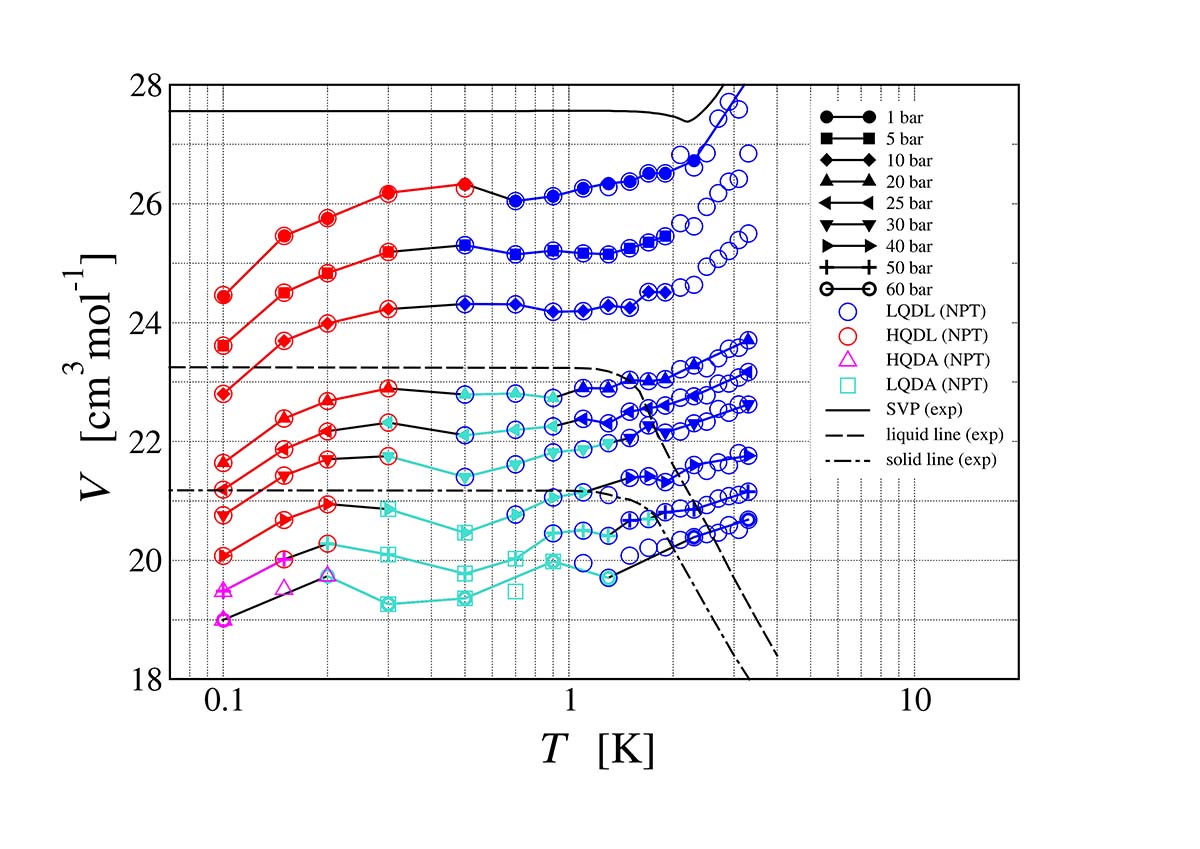}
\caption{\label{fig:conditonVT}
Volume–temperature conditions used for evaluating the transport properties.
Closed symbols denote the conditions for the present NVE simulations, while open symbols indicate the conditions adopted in the NPT simulations in Ref.~\onlinecite{tsujimoto2024}, which precede the present NVE runs.
Solid lines represent the phase boundaries of real $^4$He.
Experimental data at the saturated vapor pressure (SVP) are taken from Donnelly and Barenghi ~\cite{donnelly1998}.
Colors denote the identified states in the CMD simulations for each ensemble.
The region between the experimental solid and liquid volume lines ~\cite{swenson1950} corresponds to the solid–liquid coexistence region (hcp and He II) of real $^4$He.
The pressure values indicated in the legend correspond to those of the preceding NPT runs.
}
\vspace{2mm} \noindent\textbf{Alt Text:} Figure S2 shows the volume-temperature conditions used for the transport-property calculations. The state classification obtained from CMD simulations is displayed together with the experimental phase boundaries of real helium-4.
\end{figure}

\clearpage

\onecolumngrid
\section{Quantum wavelength and expansion factor}
\label{sec:SMwavelength}

\begin{figure}[H]
\centering
\includegraphics[width=11cm]{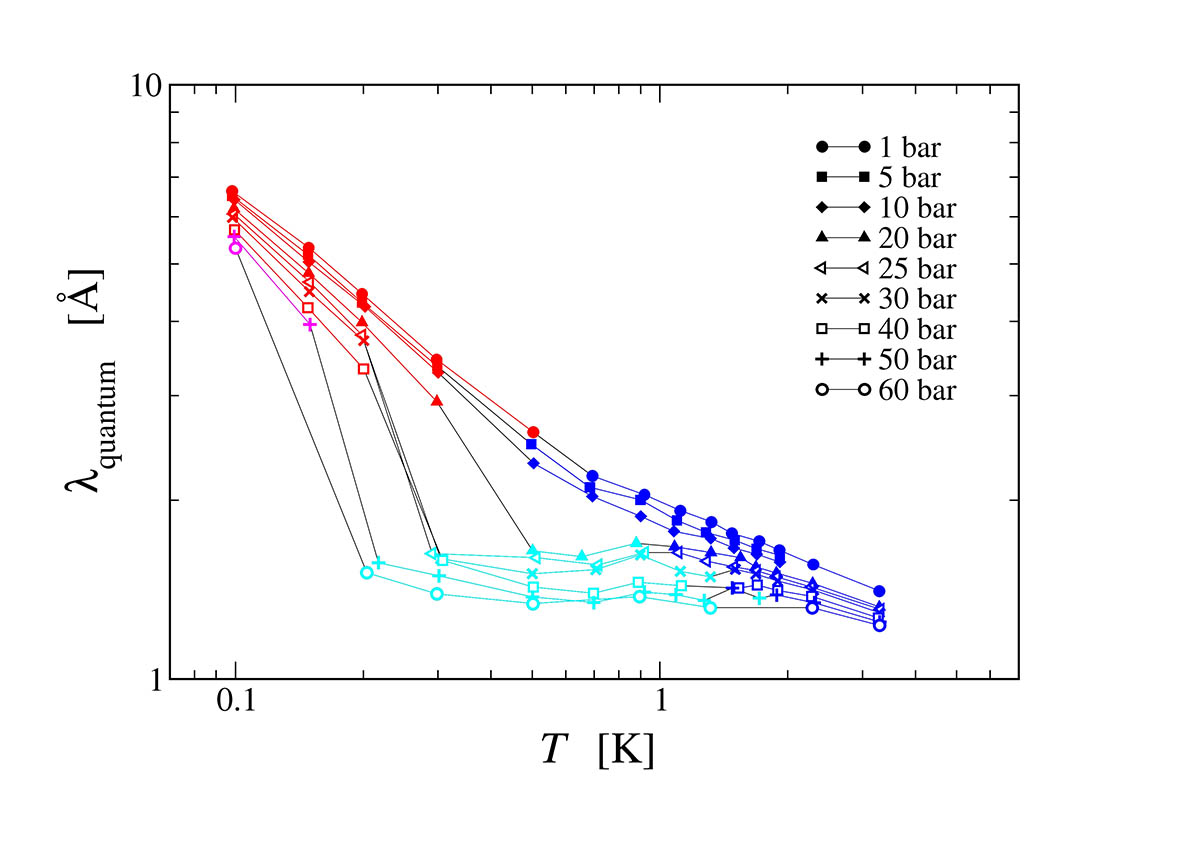}
\caption{\label{fig:SM_lambda}} 
Temperature dependence of quantum wavelength $\lambda_{\rm quantum}$ in the NVE runs: LQDL (blue), LQDA (cyan), HQDL (red), and HQDA (magenta).
The slopes of the plots for LQDL and HQDL are clearly distinct. 
Fitting the power-law relation $\lambda_{\mathrm{quantum}}\sim T^{\chi}$ yields $\chi \approx -0.3$ for LQDL and $-0.6 \le \chi \le -0.5$ for HQDL~\cite{tsujimoto2024}.

\vspace{2mm} \noindent\textbf{Alt Text:} Figure S3 plots the quantum wavelength as a function of temperature. The high quantum dispersion states exhibit longer quantum wavelengths than the low quantum dispersion states, reflecting enhanced quantum delocalization.
\end{figure}

\begin{figure}[H]
\centering
\includegraphics[width=11cm]{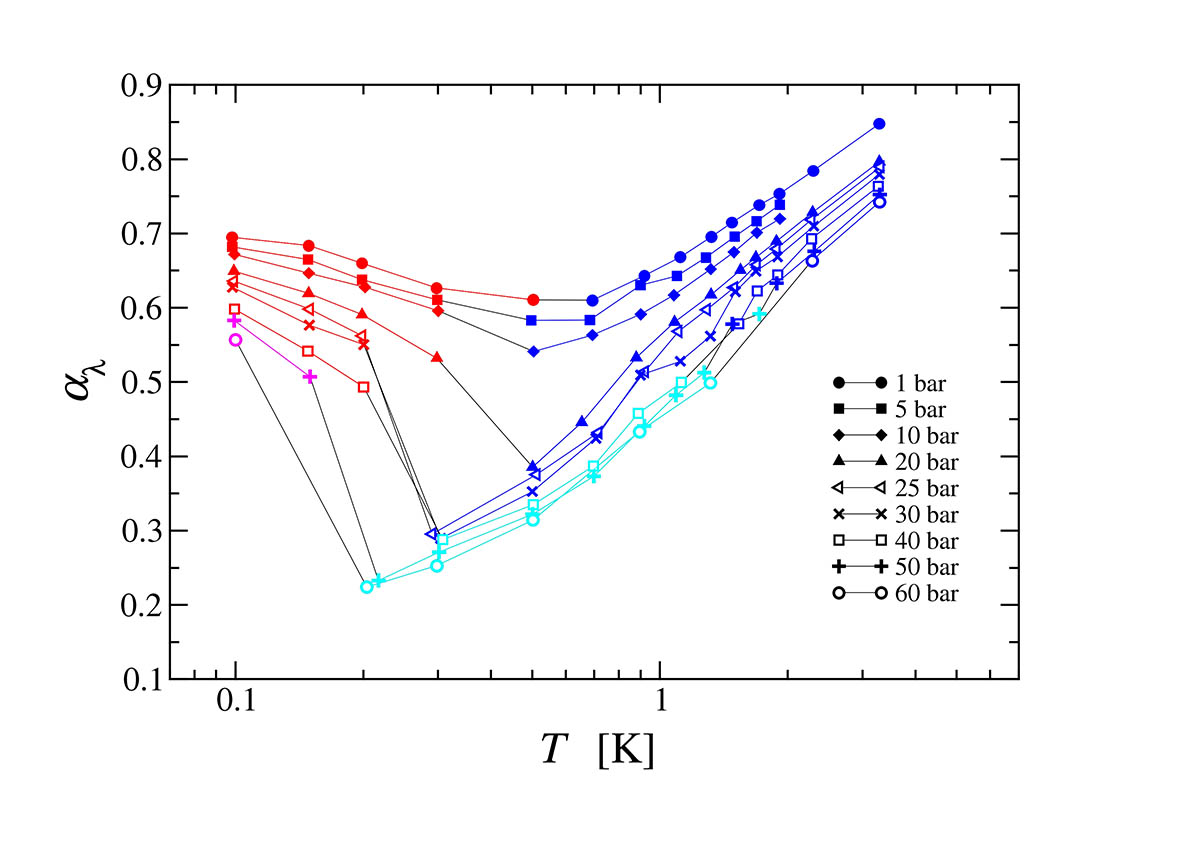}
\caption{\label{fig:SM_expansion}} 
Temperature dependence of expansion factor $\alpha_{\lambda}$ in the NVE runs: LQDL (blue), LQDA (cyan), HQDL (red), and HQDA (magenta).

\vspace{2mm} \noindent\textbf{Alt Text:} Figure S4 plots the expansion factor of the quantum wavelength against temperature. Upon cooling, the factor decreases in the low quantum dispersion states and increases in the high quantum dispersion states, providing a criterion for state identification.
\end{figure}

\clearpage

\onecolumngrid
\section{Enthalpy and isobaric heat capacity}
\label{sec:SMHandC}

\begin{figure}[H]
\centering
\includegraphics[width=11cm]{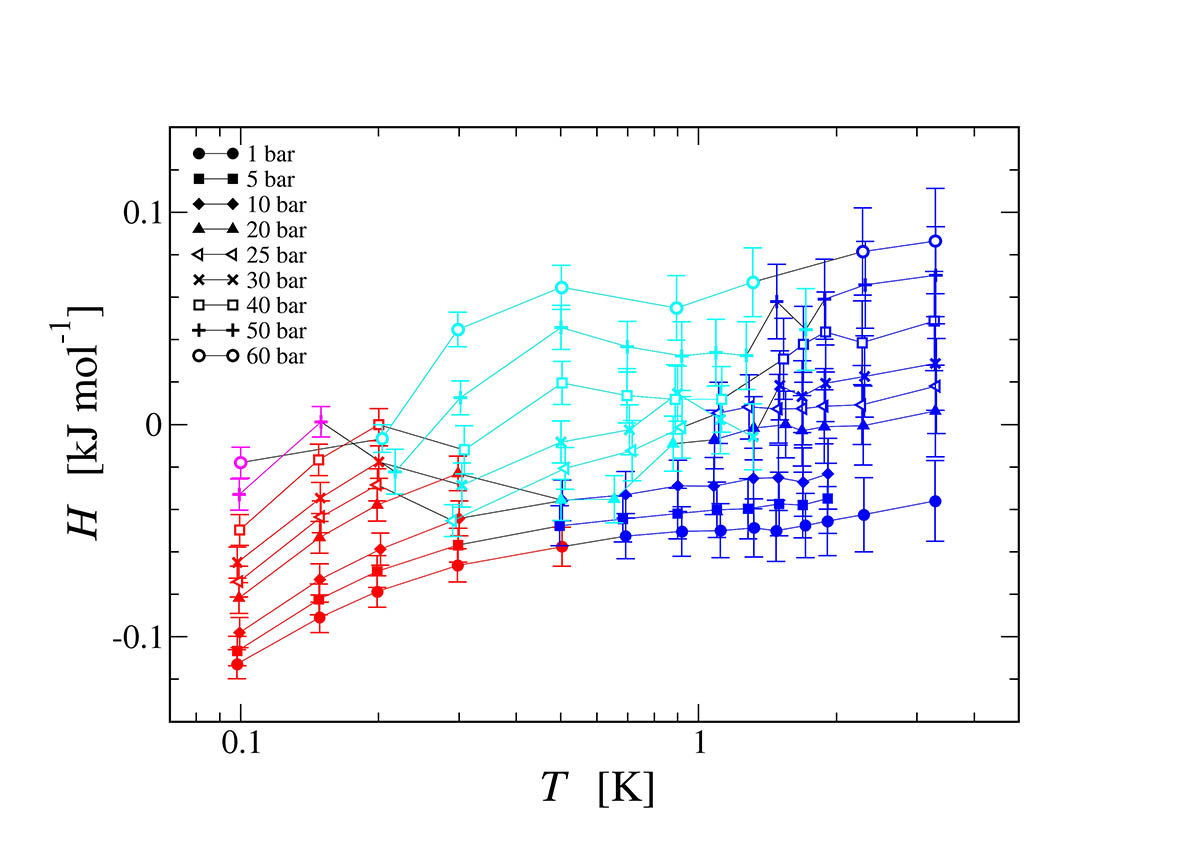}
\caption{\label{fig:SM_enthalpy}} 
Temperature dependence of enthalpy $H$ in the NVE runs: LQDL (blue), LQDA (cyan), HQDL (red), and HQDA (magenta).

\vspace{2mm} \noindent\textbf{Alt Text:} Figure S5 shows the temperature dependence of enthalpy for the different liquid and amorphous states. The enthalpy varies smoothly with temperature without indications of a discontinuous phase transition.
\end{figure}

\begin{figure}[H]
\centering
\includegraphics[width=11cm]{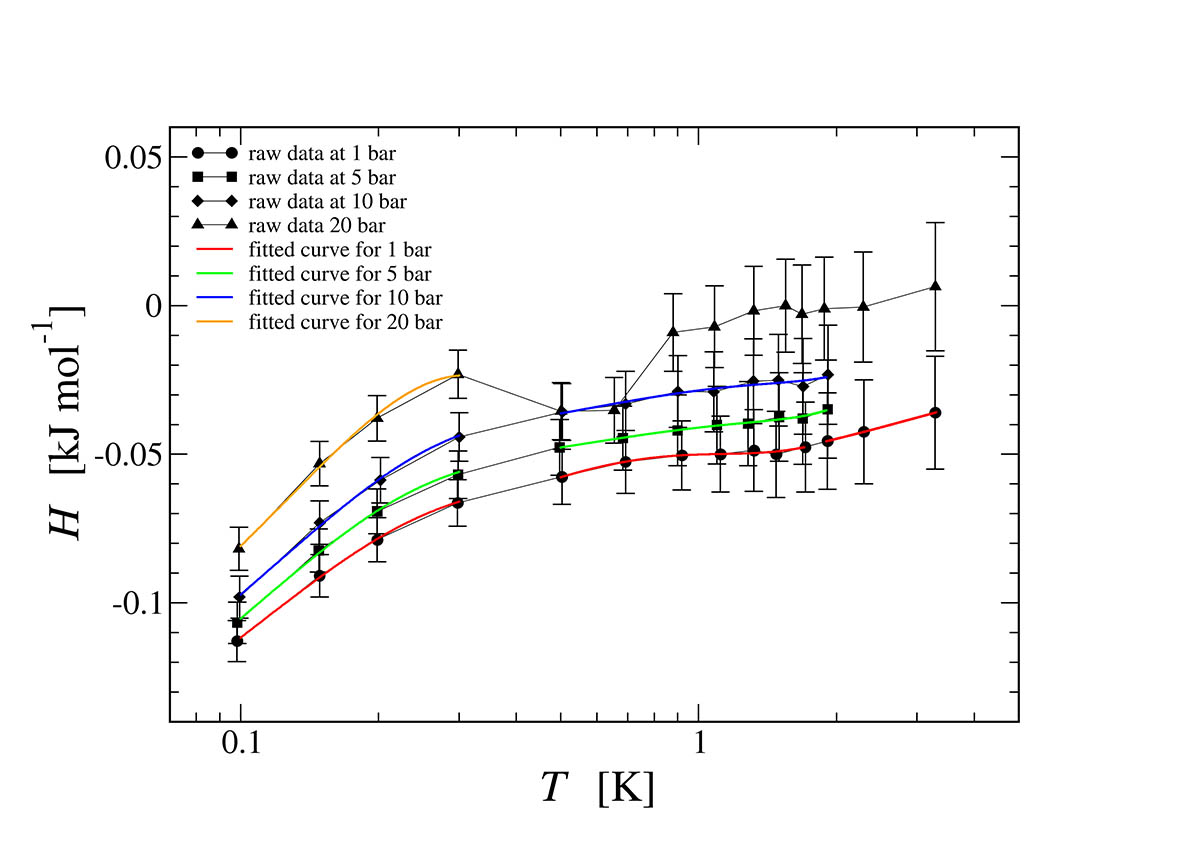}
\caption{\label{fig:SM_fittedenthalpy}} 
Temperature dependence of enthalpy $H$ with fitted curves in the NVE runs.

\vspace{2mm} \noindent\textbf{Alt Text:} Figure S6 compares the calculated enthalpy data with fitted curves. The fitted functions reproduce the smooth temperature dependence used for subsequent thermodynamic analysis.
\end{figure}

\begin{figure}[H]
\centering
\includegraphics[width=11cm]{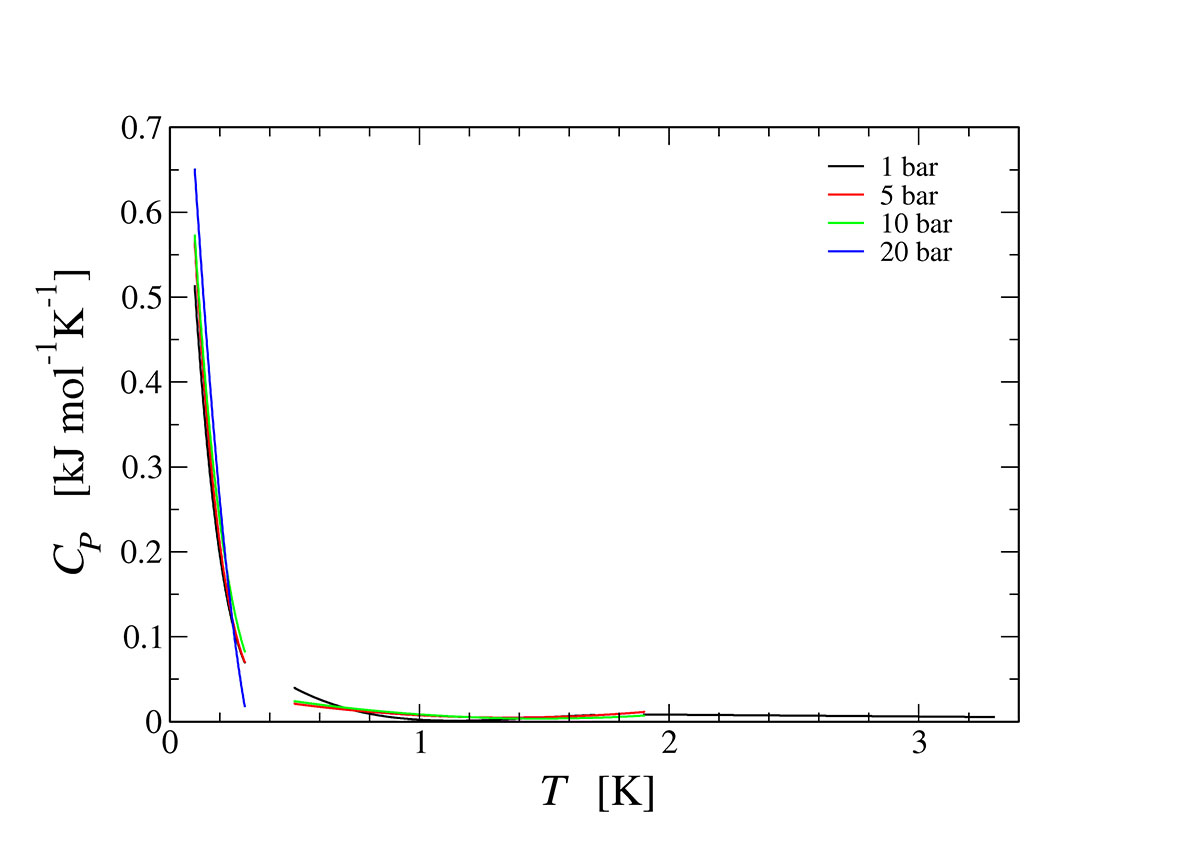}
\caption{\label{fig:SM_Cp}} 
Temperature dependence of isobaric heat capacity $C_P$ obtained from the fitted curves.

\vspace{2mm} \noindent\textbf{Alt Text:} Figure S7 shows the isobaric heat capacity obtained from the fitted enthalpy curves. The heat capacity increases upon isobaric cooling.
\end{figure}

\clearpage
\onecolumngrid
\section{Radial distribution functions}
\label{sec:SMRDF}

\vspace*{\fill}
\begin{figure}[H]
\begin{center}
\includegraphics[width=\columnwidth]{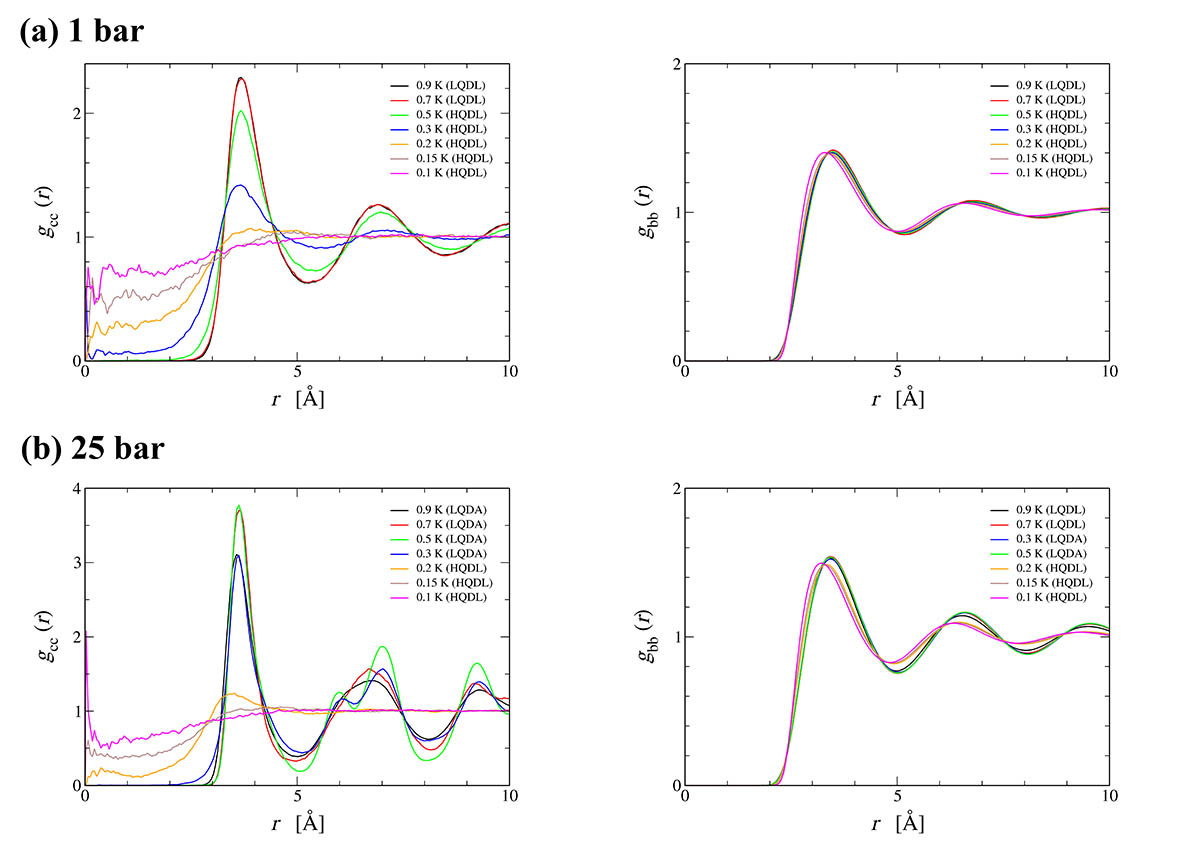}
\caption{\label{fig:SM_RDF}}Radial distribution functions (RDFs) at temperatures from 0.1 to 0.9 K. Left panel: centroid-centroid RDF, $g_{\rm cc}$. Right panel: bead-bead RDF, $g_{\rm bb}$.
\end{center}

\vspace{2mm} \noindent\textbf{Alt Text:} Figure S8 compares centroid-centroid and bead-bead radial distribution functions. High quantum dispersion states exhibit signatures of enhanced quantum delocalization and overlap relative to low quantum dispersion states.
\end{figure}
\vspace*{\fill}

\clearpage
\onecolumngrid
\section{Centroid velocity autocorrelation functions}
\label{sec:SMVAF}

\vspace*{\fill}
\begin{figure}[H]
\begin{center}
    \includegraphics[width=\columnwidth]{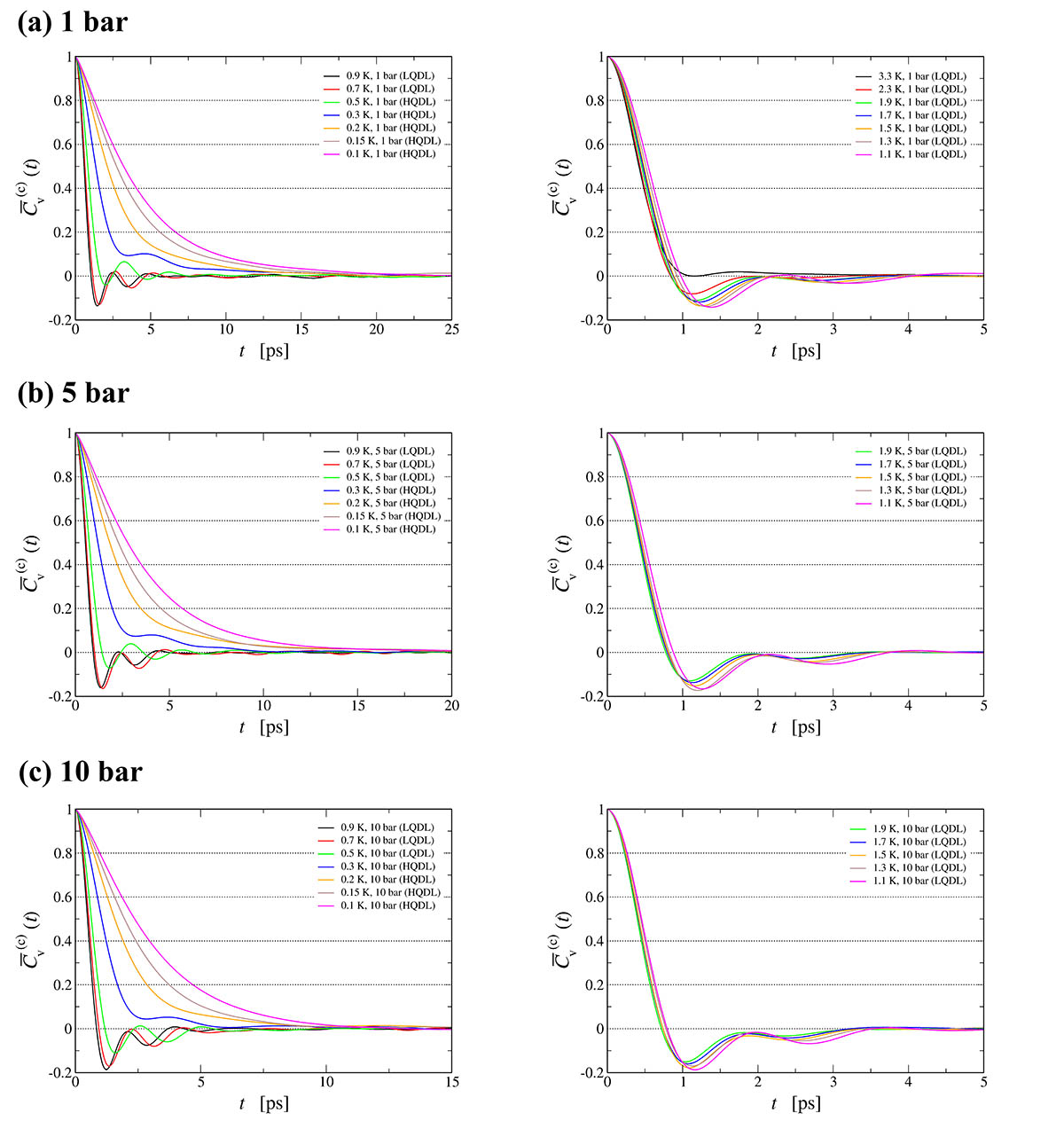}
    \caption{Normalized centroid velocity autocorrelation functions at densities corresponding to 1-10 bar.}
    \label{fig:SM_VAF1}
\end{center}

\vspace{2mm} \noindent\textbf{Alt Text:} Figure S9 shows normalized centroid velocity autocorrelation functions at pressures from 1 to 10 bar. The relaxation behavior evolves from oscillatory liquid-like decay toward slower monotonic decay at lower temperatures.
\end{figure}
\vspace*{\fill}

\clearpage

\vspace*{\fill}
\begin{figure}[H]
\begin{center}
    \includegraphics[width=\columnwidth]{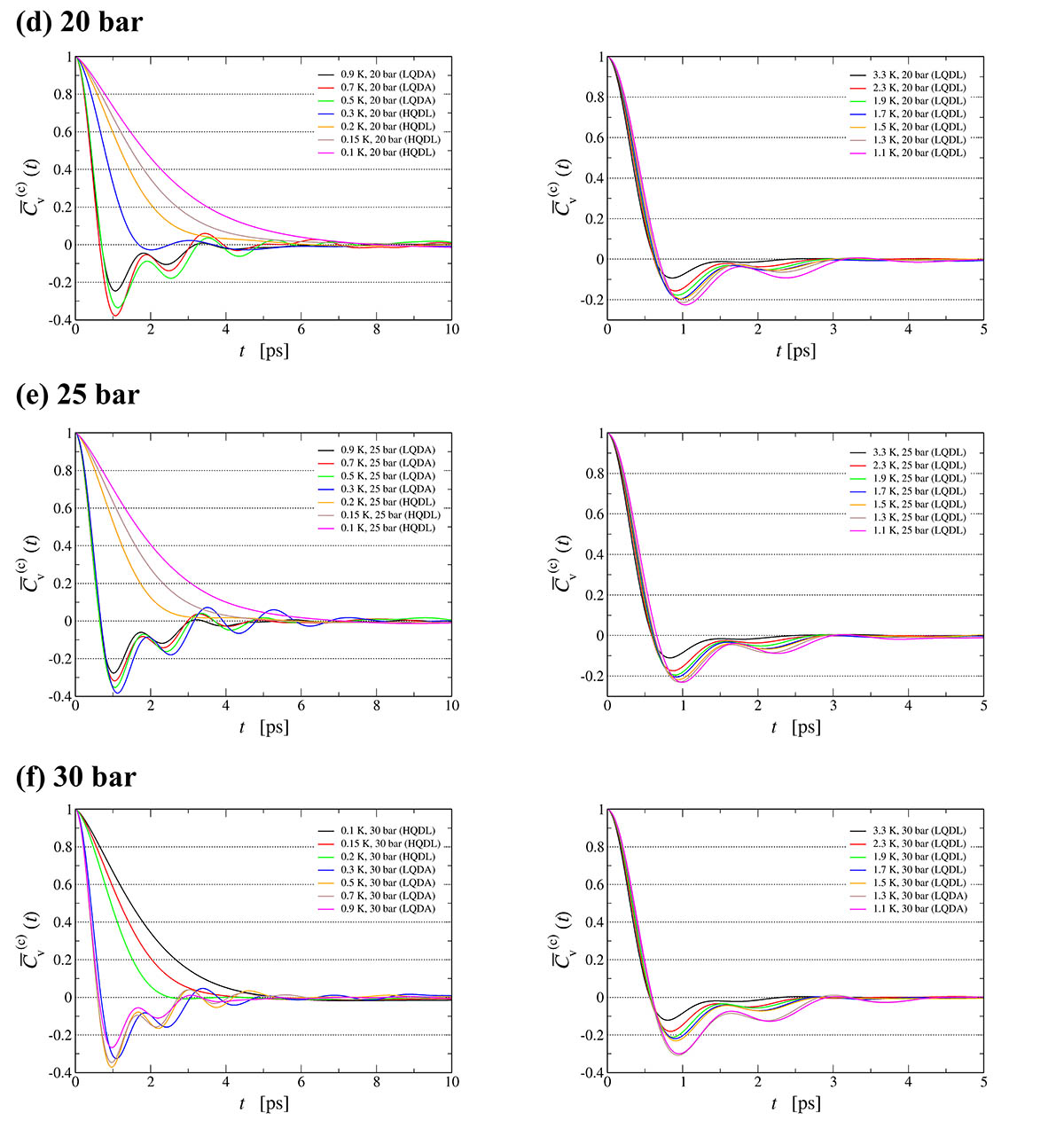}
    \caption{Normalized centroid velocity autocorrelation functions at densities corresponding to 20-30 bar.}
    \label{fig:SM_VAF2}
\end{center}

\vspace{2mm} \noindent\textbf{Alt Text:}
Figure S10 shows normalized centroid velocity autocorrelation functions at pressures from 20 to 30 bar. The results illustrate pressure-dependent changes in the dynamical crossover between liquid-like and gas-like relaxation.
\end{figure}
\vspace*{\fill}

\clearpage

\vspace*{\fill}
\begin{figure}[H]
\begin{center}
   \includegraphics[width=\columnwidth]{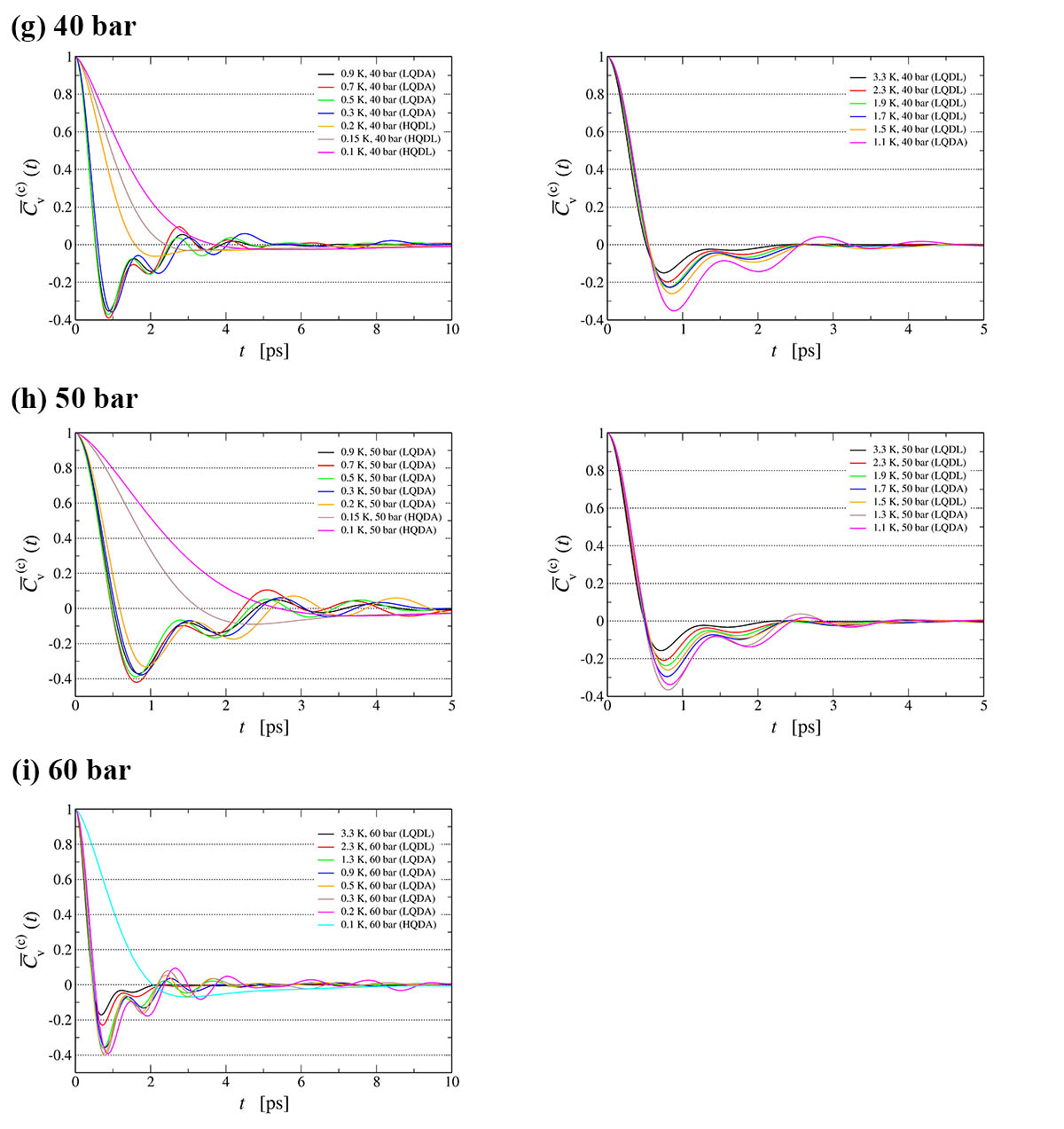}
    \caption{Normalized centroid velocity autocorrelation functions at densities corresponding to 40-60 bar.}
    \label{fig:SM_VAF3}

\vspace{2mm} \noindent\textbf{Alt Text:} Figure S11 shows normalized centroid velocity autocorrelation functions at pressures from 40 to 60 bar. Distinct relaxation behaviors are observed among the low and high quantum dispersion liquids and amorphous states.
\end{center}

\end{figure}
\vspace*{\fill}

\clearpage

\vspace*{\fill}
\begin{figure}[htbp]
\begin{center}
    \includegraphics[width=12cm]{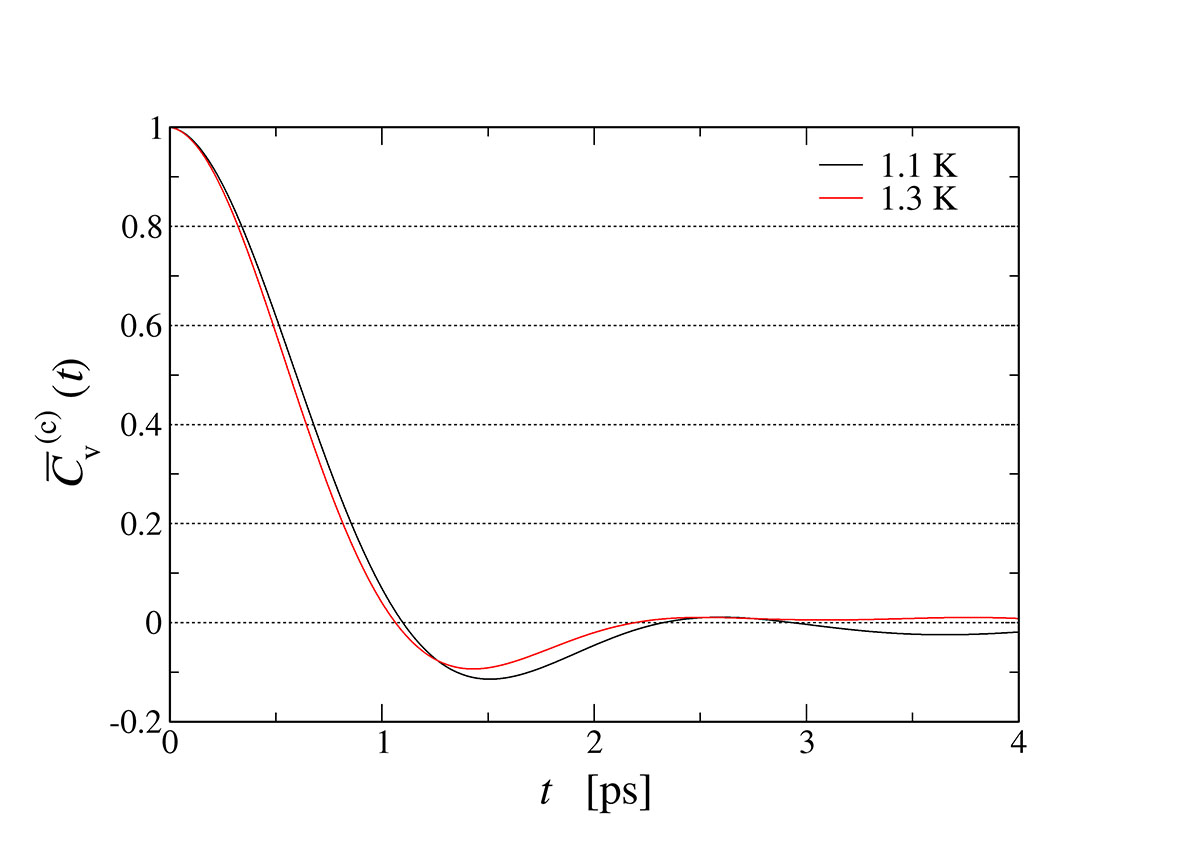}
    \caption{Normalized centroid velocity autocorrelation functions at 
0.145 $\rm{gcm^{-3}}$ obtained from the NVT runs.  The density corresponds to the experimental SVP.}
    \label{fig:nakayamavaf}
\end{center}

\vspace{2mm} \noindent\textbf{Alt Text:} Figure S12 presents normalized centroid velocity autocorrelation functions at the saturated-vapor-pressure density. The relaxation remains oscillatory and is consistent with the low quantum dispersion liquid state.
\end{figure}
\vspace*{\fill}

\vspace*{\fill}
\begin{figure}[htbp]
\begin{center}
    \includegraphics[width=12cm]{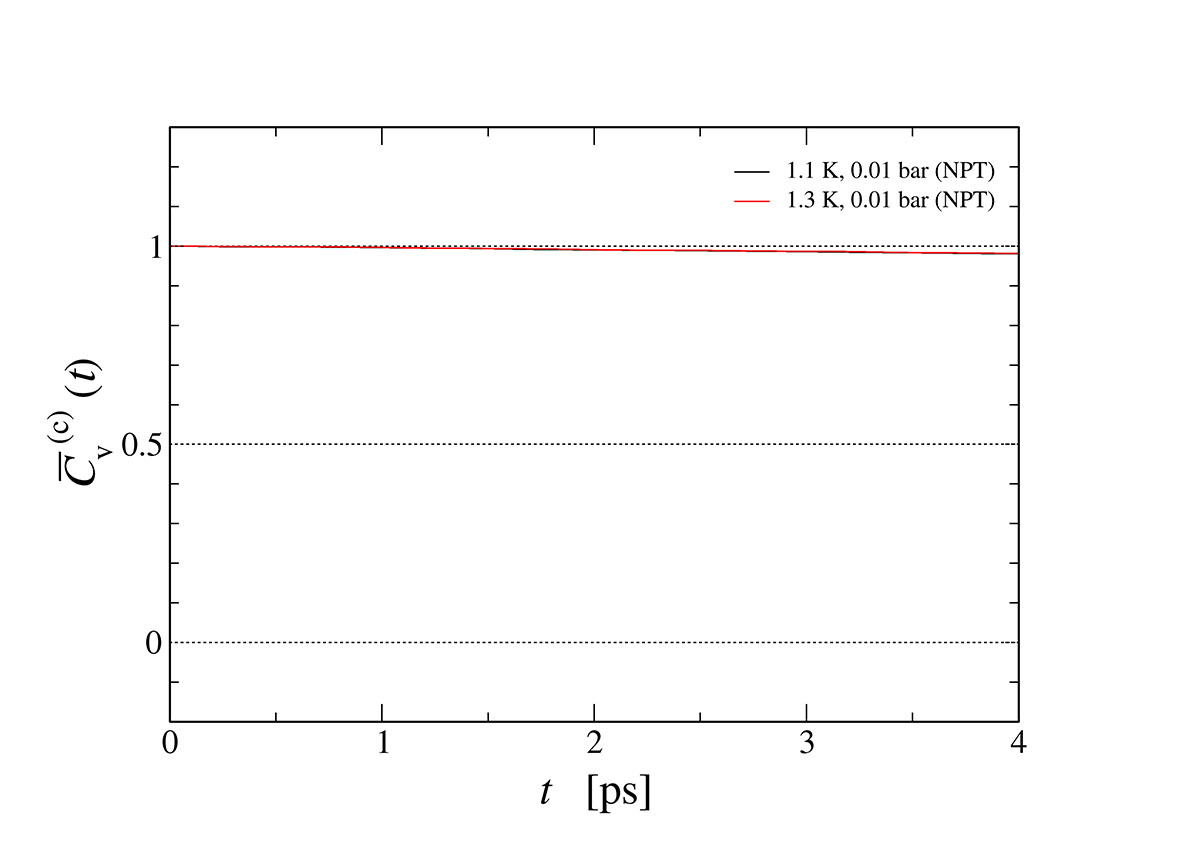}
    \caption{Normalized centroid velocity autocorrelation functions at 
0.01 bar obtained from the NPT runs.}
    \label{fig:tsujimotovaf}
\end{center}

\vspace{2mm} \noindent\textbf{Alt Text:} Figure S13 presents normalized centroid velocity autocorrelation functions from NPT simulations at very low pressure. The  decay-less profile reflects gas-like behavior under these thermodynamic conditions.
\end{figure}
\vspace*{\fill}

\clearpage

\section{Phonon density of states}
\label{sec:SMPDS}

\vspace*{\fill}
\begin{figure}[H]
\begin{center}
    \includegraphics[width=\columnwidth]{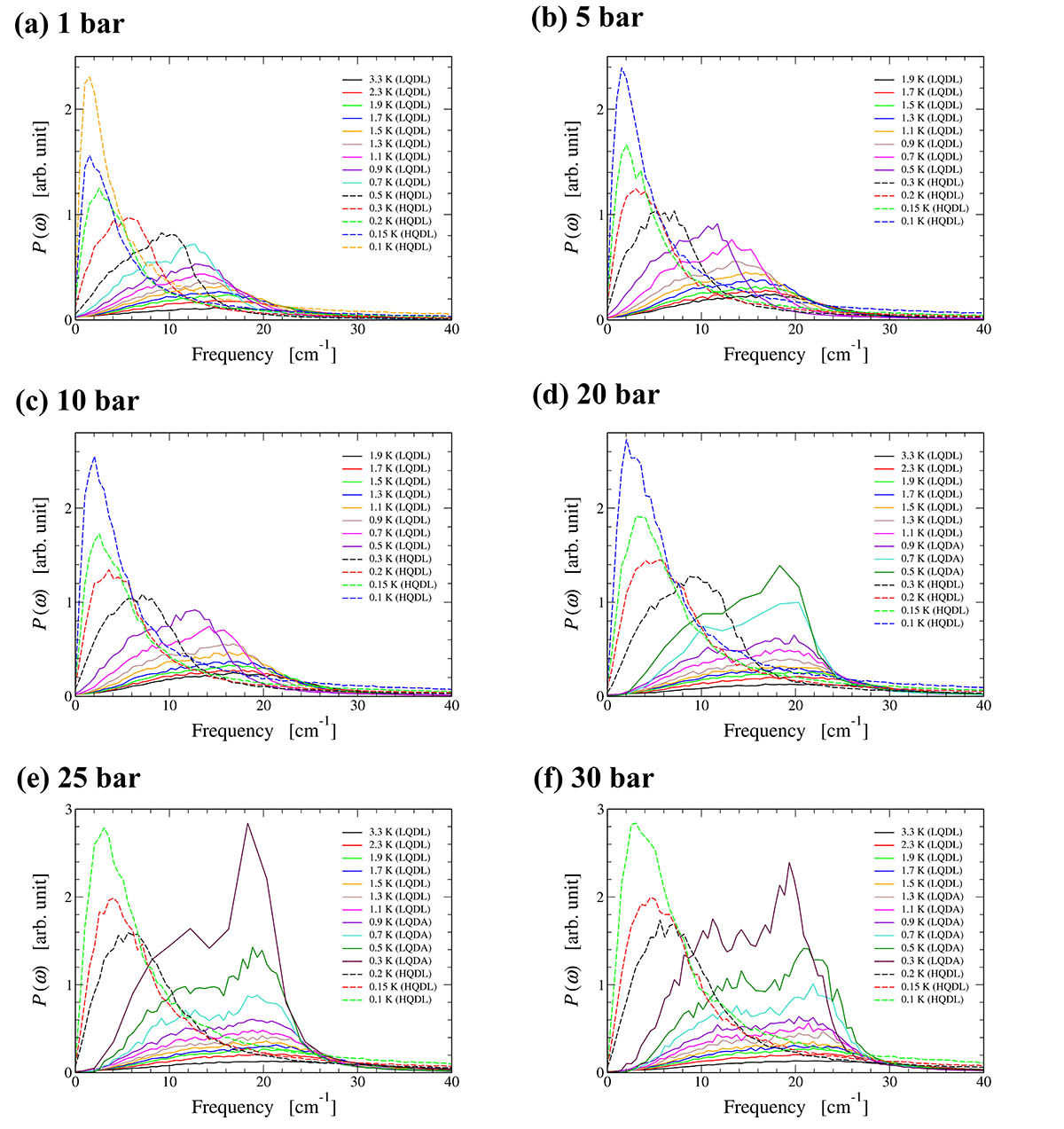}    
    \caption{Quantum vibrational power spectra at densities corresponding to 1-30 bar.}
    \label{fig:SM_FTVAF1}
\end{center}

\vspace{2mm} \noindent\textbf{Alt Text:} Figure S14 shows quantum vibrational power spectra at pressures from 1 to 30 bar. The spectra reveal systematic shifts in vibrational frequencies and clear differences between low and high quantum dispersion liquids.
\end{figure}
\vspace*{\fill}

\clearpage

\vspace*{\fill}
\begin{figure}[H]
\begin{center}
    \includegraphics[width=\columnwidth]{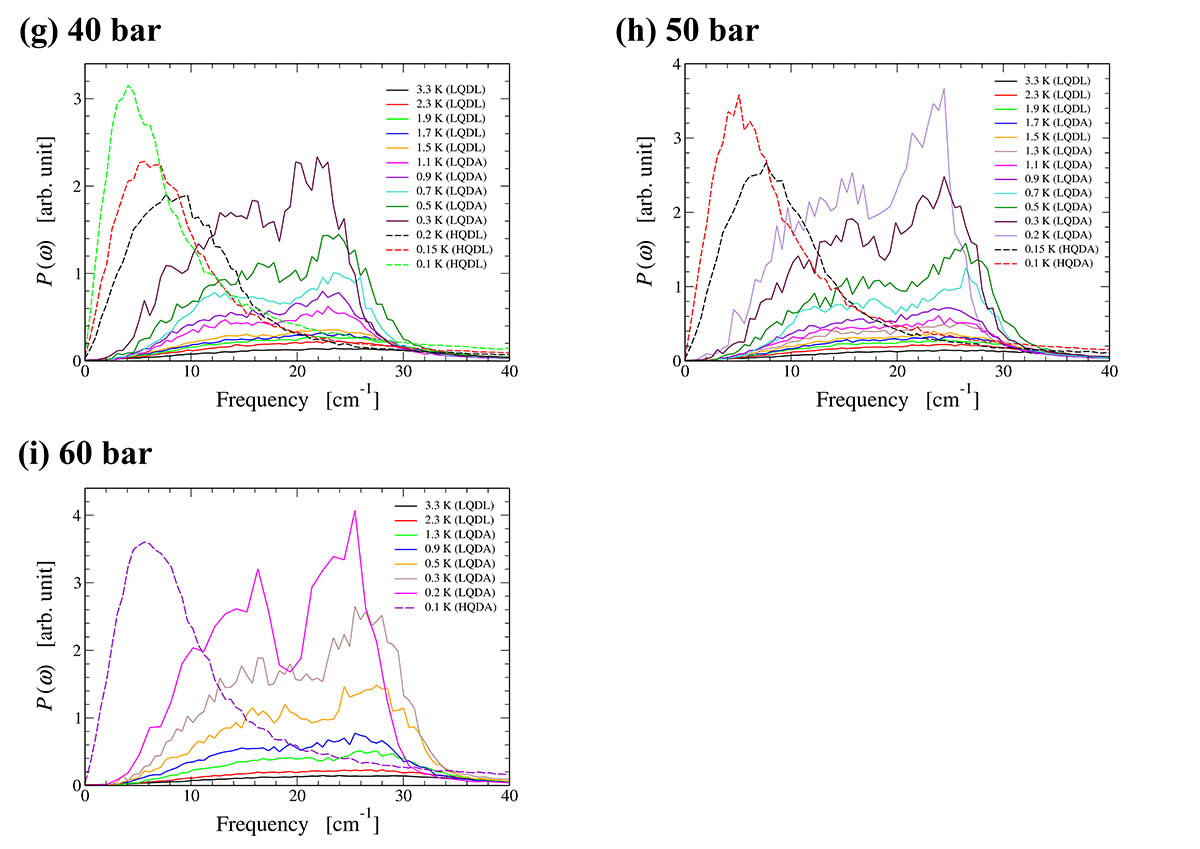}    
    \caption{Quantum vibrational power spectra at densities corresponding to 40-60 bar.}
    \label{fig:SM_FTVAF2}
\end{center}

\vspace{2mm} \noindent\textbf{Alt Text:} Figure S15 shows quantum vibrational power spectra at pressures from 40 to 60 bar. The spectral profiles distinguish the liquid and amorphous states and demonstrate the dominance of low-frequency modes in high quantum dispersion states.
\end{figure}
\vspace*{\fill}

\clearpage

\clearpage
\onecolumngrid
\section{Mean Square displacements}
\label{sec:SMMSD}

\vspace*{\fill}
\begin{figure}[H]
\begin{center}
    \includegraphics[width=\columnwidth]{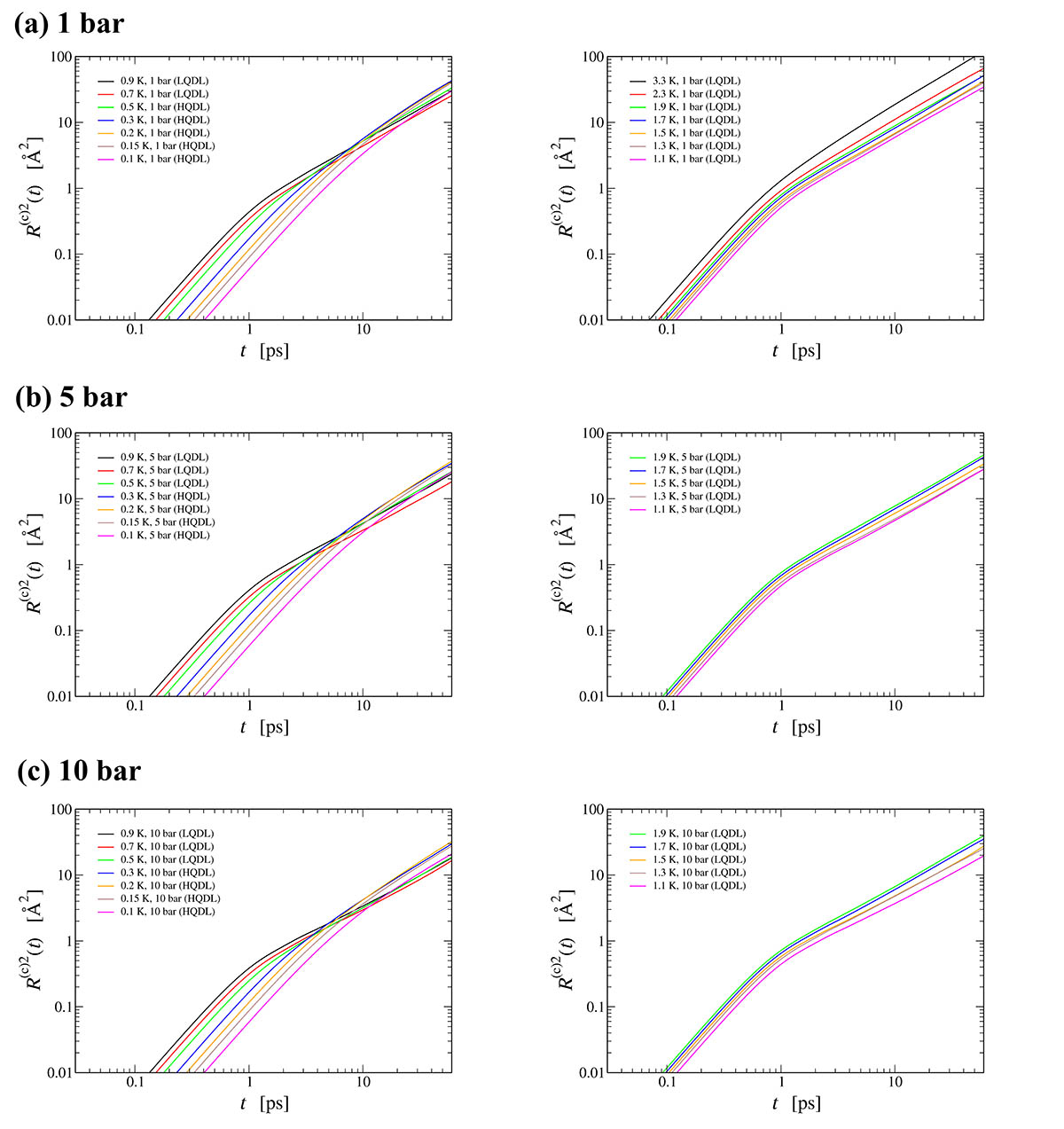}    
    \caption{Mean square displacements at densities corresponding to 1-10 bar.}
    \label{fig:SM_MSD1}
\end{center}

\vspace{2mm} \noindent\textbf{Alt Text:} Figure S16 shows mean square displacements at pressures from 1 to 10 bar. The long-time behavior characterizes normal and superdiffusive atomic motion in the different states.
\end{figure}
\vspace*{\fill}

\clearpage

\vspace*{\fill}
\begin{figure}[H]
\begin{center}
    \includegraphics[width=\columnwidth]{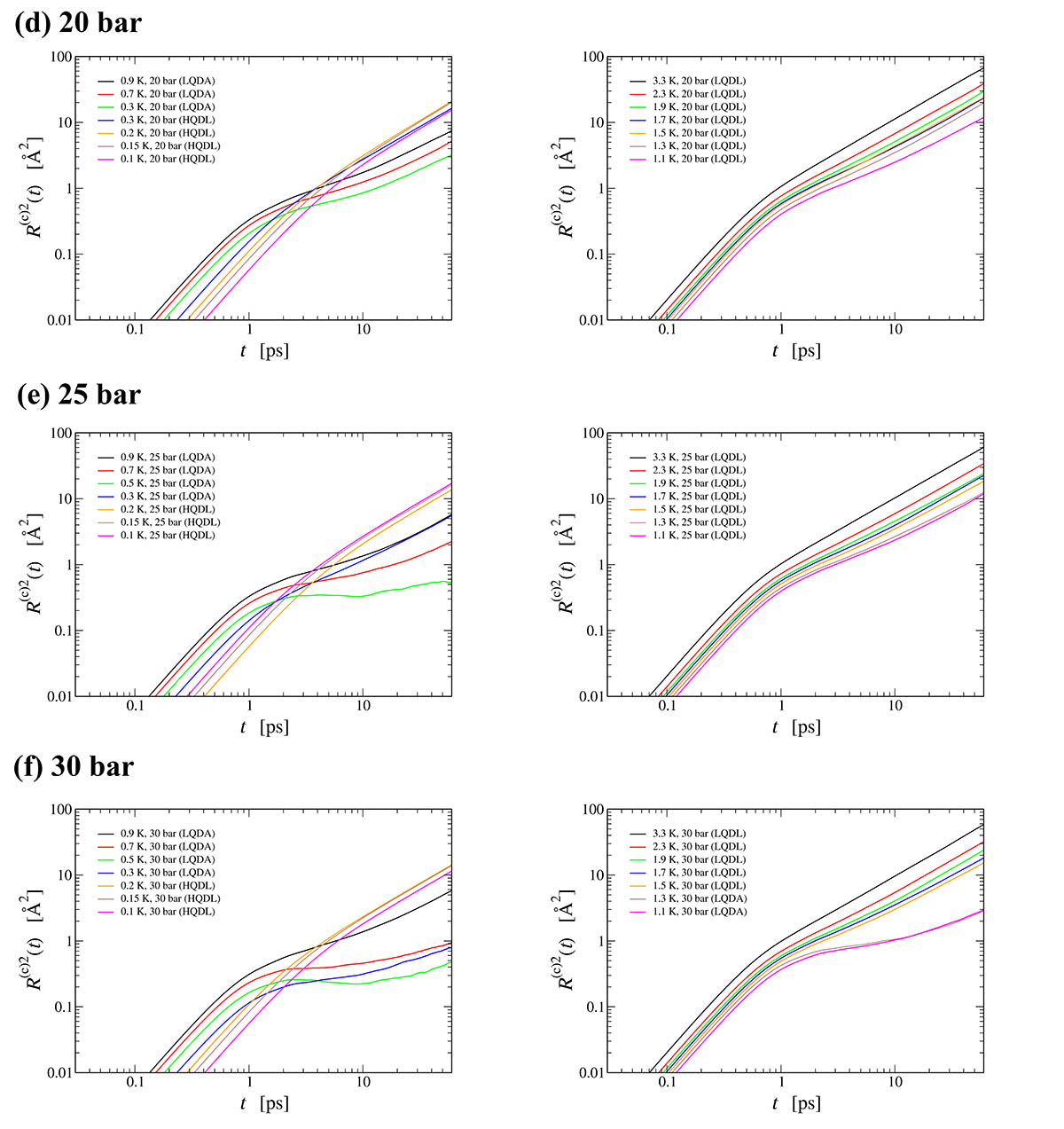}
    \caption{Mean square displacements at densities corresponding to 20-30 bar.}
    \label{fig:SM_MSD2}
\end{center}

\vspace{2mm} \noindent\textbf{Alt Text:} Figure S17 shows mean square displacements at pressures from 20 to 30 bar. The results demonstrate the pressure dependence of diffusive behavior.
\end{figure}
\vspace*{\fill}

\clearpage
\vspace*{\fill}
\begin{figure}[H]
\begin{center}
    \includegraphics[width=\columnwidth]{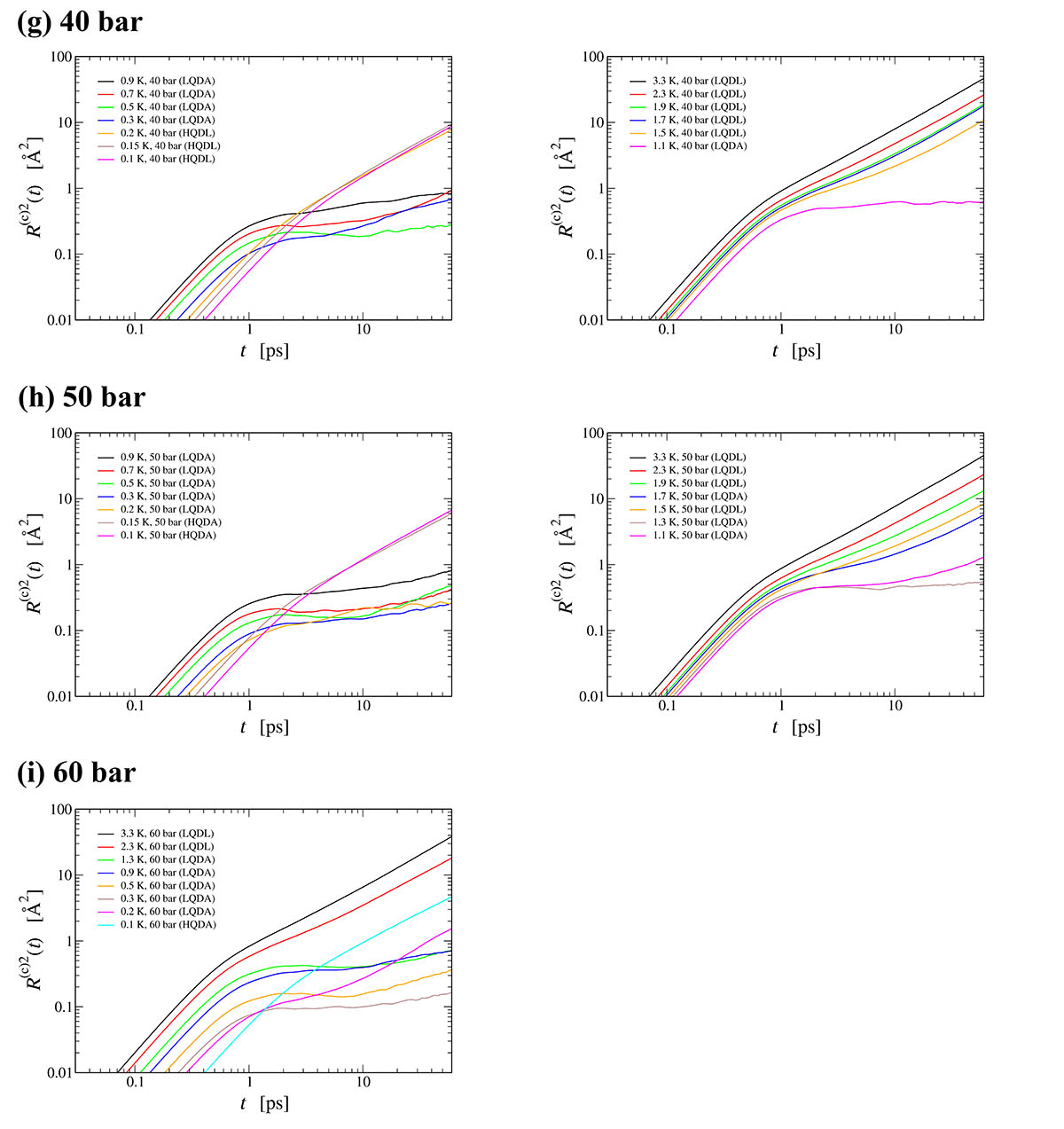}
    \caption{Mean square displacements at densities corresponding to 40-60 bar.}
    \label{fig:SM_MSD3}
\end{center}

\vspace{2mm} \noindent\textbf{Alt Text:} Figure S18 shows mean square displacements at pressures from 40 to 60 bar. Atomic motion becomes increasingly restricted in the amorphous states while remaining relatively mobile in high quantum dispersion regimes.
\end{figure}
\vspace*{\fill}

\clearpage

\clearpage
\onecolumngrid
\section{Non-Gaussian parameters}
\label{sec:SMNGP}

\vspace*{\fill} 
\begin{figure}[H]
\begin{center}
    \includegraphics[width=\columnwidth]{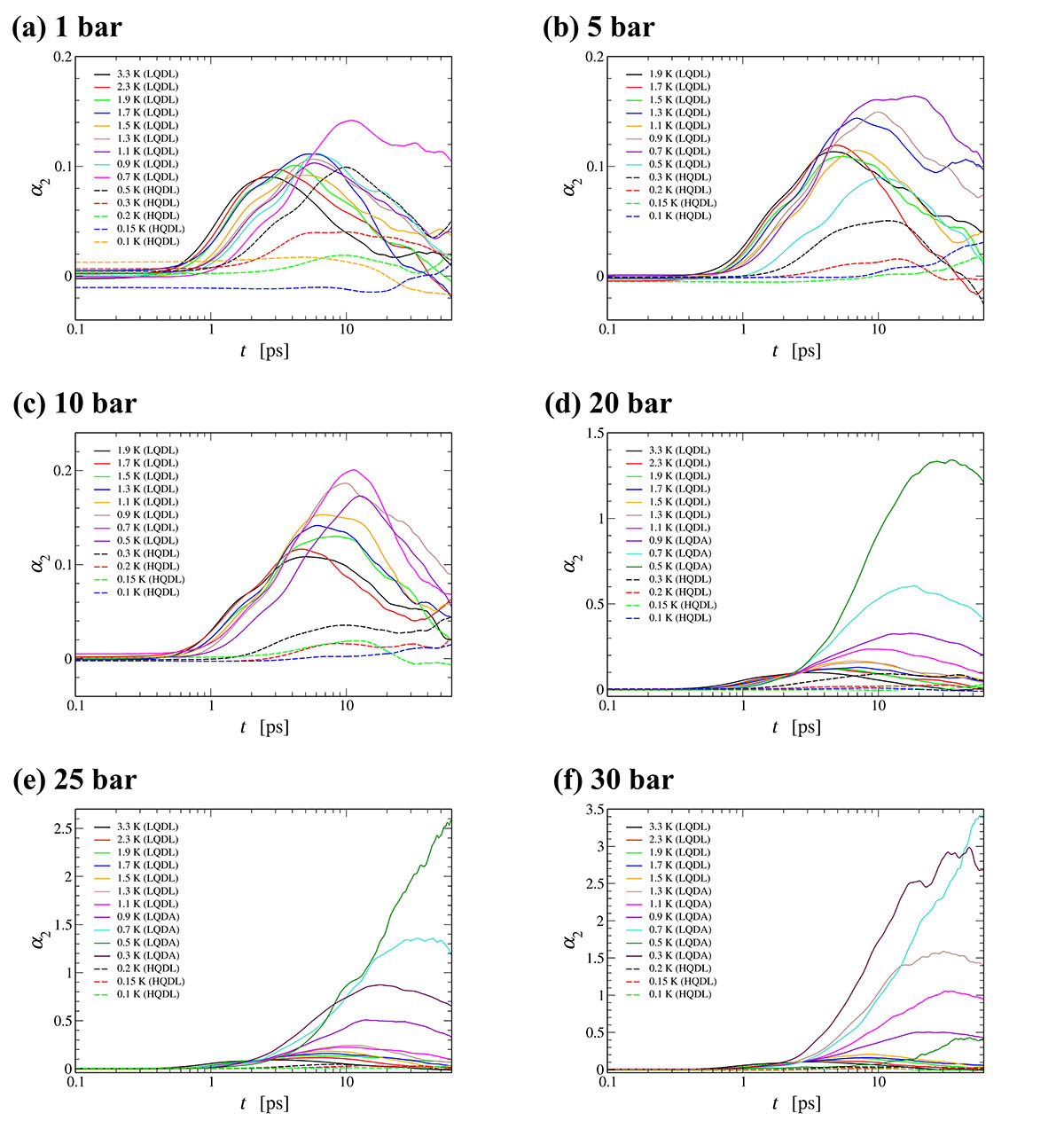}
    \caption{Non-Gaussian parameters at densities corresponding to 1-30 bar.}
    \label{fig:SM_NGP1}
\end{center}

\vspace{2mm} \noindent\textbf{Alt Text:} Figure S19 plots the non-Gaussian parameter at pressures from 1 to 30 bar. Deviations from Gaussian displacement distribution remain small in the liquid states, indicating weak dynamical heterogeneity.
\end{figure}
\vspace*{\fill}

\clearpage

\vspace*{\fill}
\begin{figure}[H]
\begin{center}
    \includegraphics[width=9cm]{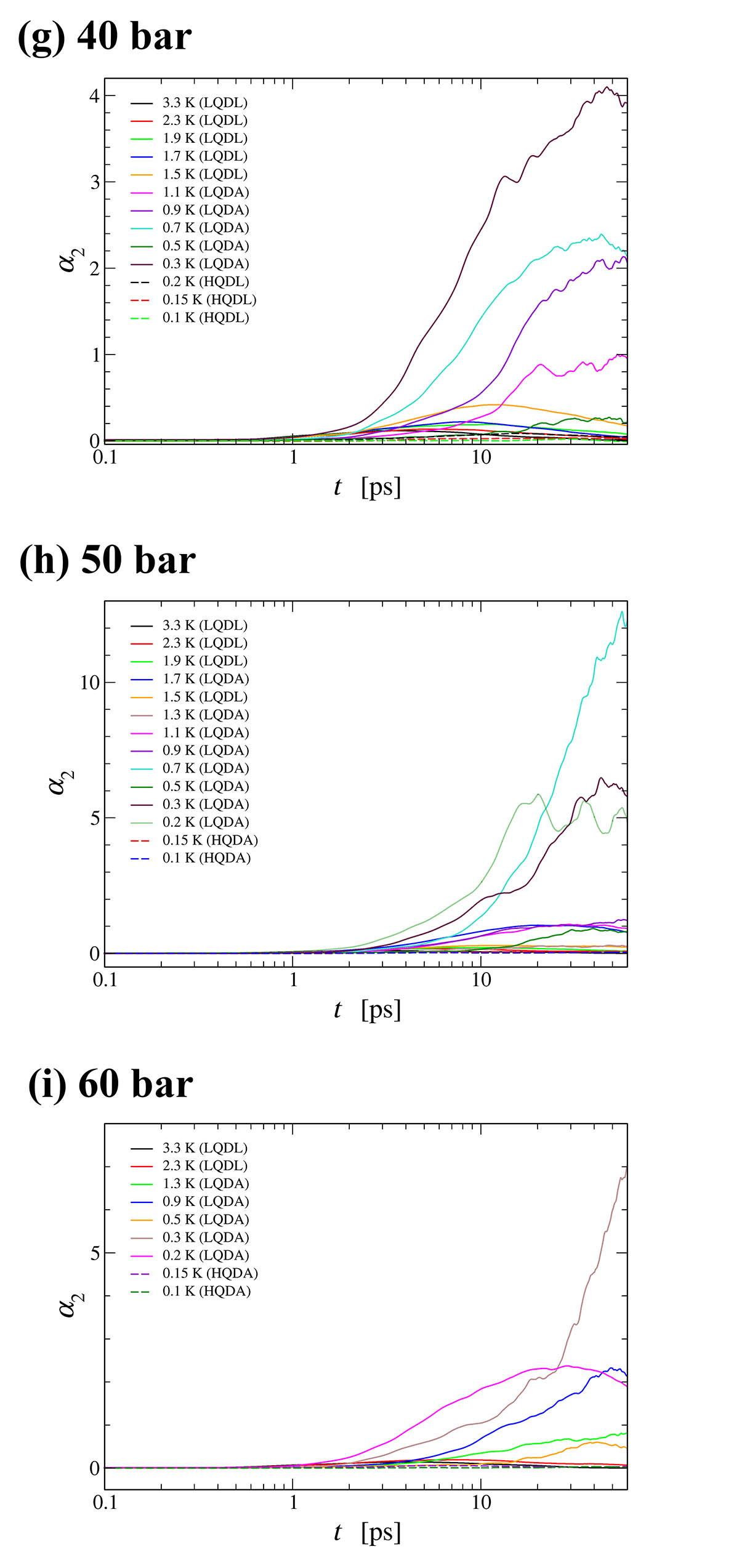}
    \caption{Non-Gaussian parameters at densities corresponding to 40-60 bar.}
    \label{fig:SM_NGP2}
\end{center}

\vspace{2mm} \noindent\textbf{Alt Text:} Figure S20 plots the non-Gaussian parameter at pressures from 40 to 60 bar. Dynamical heterogeneity becomes stronger in the low quantum dispersion amorphous state than in the high quantum dispersion amorphous state.
\end{figure}
\vspace*{\fill}

\clearpage

\section{Off-diagonal stress autocorrelation functions}
\label{sec:SMSAF}

\vspace*{\fill}
\begin{figure}[H]
\begin{center}
    \includegraphics[width=\columnwidth]{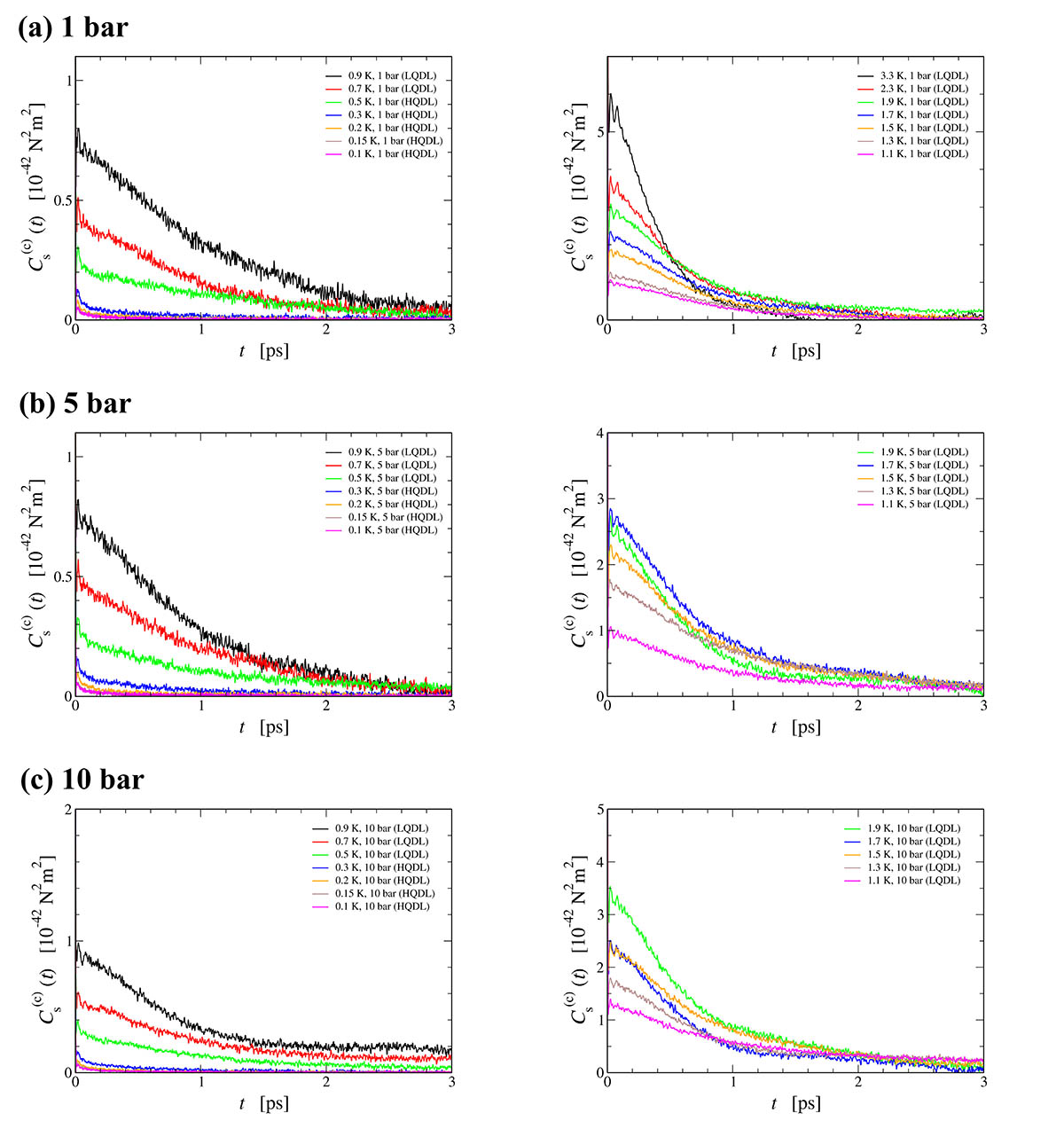}
    \caption{Off-diagonal stress autocorrelation functions at densities corresponding to 1-10 bar.}
    \label{fig:SM_Stress1}
\end{center}

\vspace{2mm} \noindent\textbf{Alt Text:} Figure S21 shows off-diagonal stress autocorrelation functions at pressures from 1 to 10 bar. Stress relaxation becomes increasingly rapid at lower temperatures in the high quantum dispersion liquid.
\end{figure}
\vspace*{\fill}

\clearpage

\vspace*{\fill}
\begin{figure}[H]
\begin{center}
    \includegraphics[width=\columnwidth]{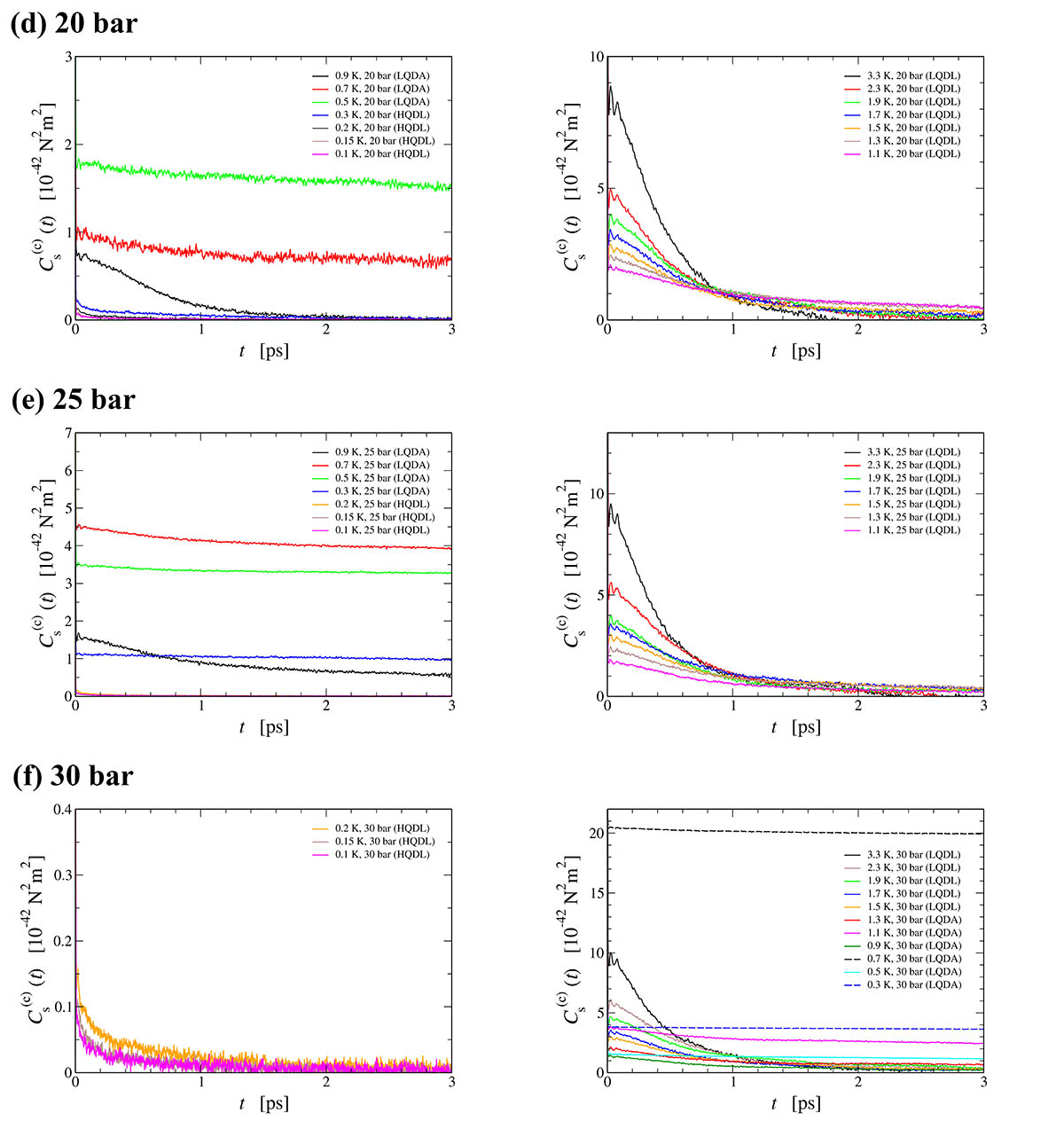}
    \caption{Off-diagonal stress autocorrelation functions at densities corresponding to 20-30 bar.}
    \label{fig:SM_Stress2}
\end{center}

\vspace{2mm} \noindent\textbf{Alt Text:} Figure S22 shows off-diagonal stress autocorrelation functions at pressures from 20 to 30 bar. The relaxation is substantially slow in the low quantum dispersion amorphous state.
\end{figure}
\vspace*{\fill}

\clearpage
\vspace*{\fill}
\begin{figure}[H]
\begin{center}
    \includegraphics[width=\columnwidth]{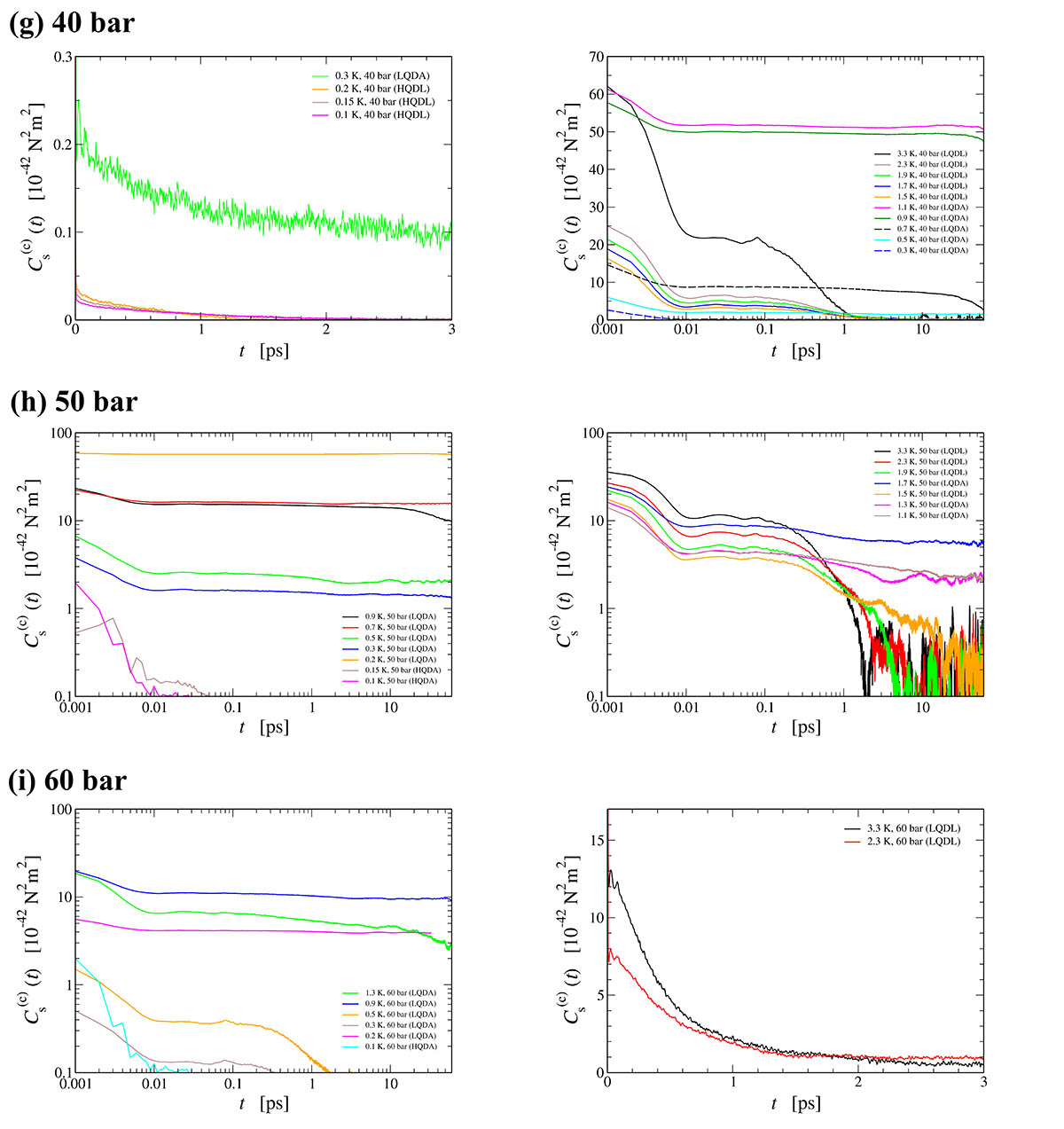}
    \caption{Off-diagonal stress autocorrelation functions at densities corresponding to 40-60 bar.}
    \label{fig:SM_Stress3}
\end{center}

\vspace{2mm} \noindent\textbf{Alt Text:} Figure S23 shows off-diagonal stress autocorrelation functions at pressures from 40 to 60 bar. Slow relaxation in the low quantum dispersion amorphous states contrasts with the faster decay observed in the high quantum dispersion states.
\end{figure}
\vspace*{\fill}

\clearpage

\clearpage
\section{Analysis of shear viscosity}
\label{sec:SMASV}
\onecolumngrid

\vspace*{\fill}

\begin{figure}[htbp]
\includegraphics[width=11cm]{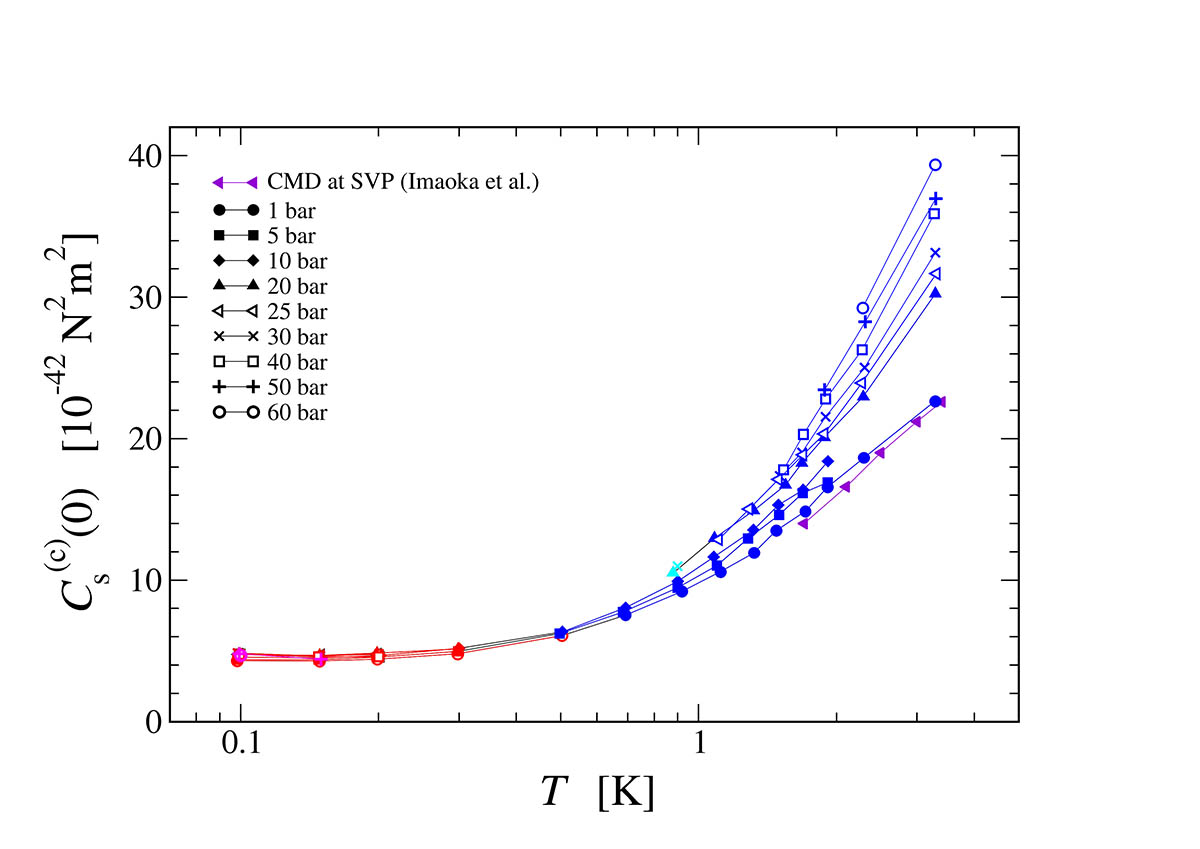}
\caption{\label{fig:seppen_shear} 
Temperature dependence of initial value $C_{\rm s}^{\rm (c)}(0)$ of off-diagonal stress autocorrelation functions: LQDL (blue), LQDA (cyan), HQDL (red), and HQDA (magenta).
}

\vspace{2mm} \noindent\textbf{Alt Text:} Figure S24 plots the initial value of the stress autocorrelation function as a function of temperature. The magnitude generally decreases upon cooling.
\end{figure}

\vspace*{\fill}

\begin{figure}[htbp]
\includegraphics[width=11cm]{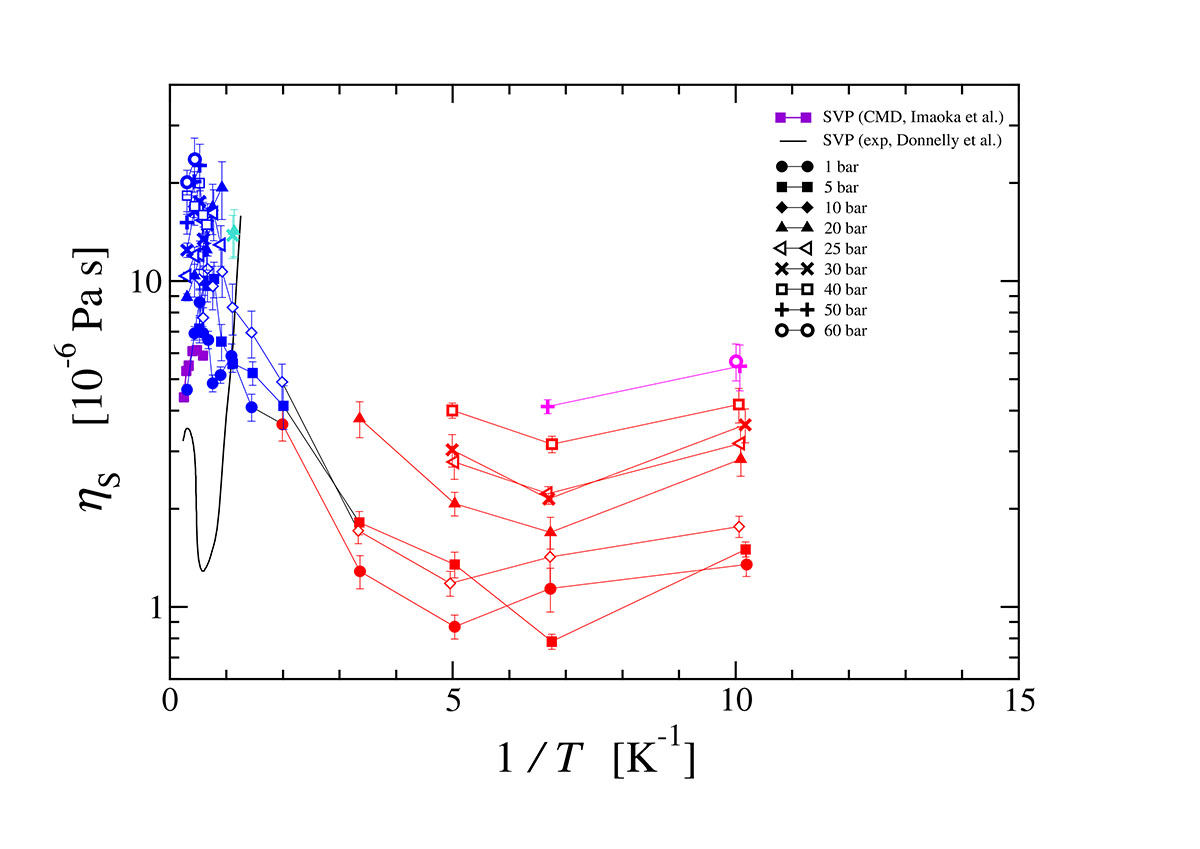}
\caption{\label{fig:Arrhenius} 
Arrhenius plot of shear viscosity of distinguishable liquid $^4$He: LQDL (blue), LQDA (cyan), HQDL (red), and HQDA (magenta).
}

\vspace{2mm} \noindent\textbf{Alt Text:} Figure S25 presents an Arrhenius plot of shear viscosity. The results deviate from simple Arrhenius behavior, indicating a strongly temperature-dependent transport mechanism.
\end{figure}

\vspace*{\fill}
\clearpage
\vspace*{\fill}

\begin{figure}[htbp]
\includegraphics[width=11cm]{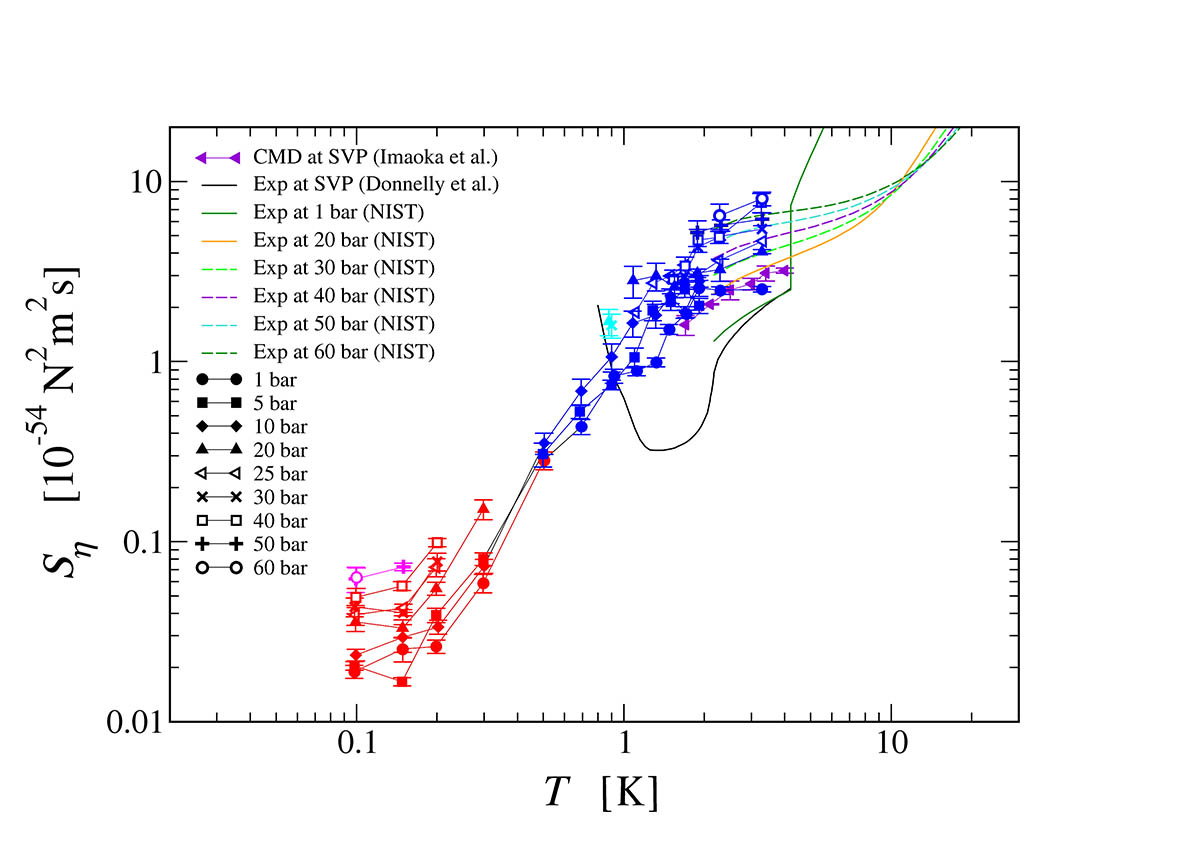}
\caption{\label{fig:sekibun_shear} 
Temperature dependence of time integral $S_{\eta}$ of off-diagonal stress autocorrelation functions: LQDL (blue), LQDA (cyan), HQDL (red), and HQDA (magenta).}

\vspace{2mm} \noindent\textbf{Alt Text:} Figure S26 shows the time integral contribution to the shear viscosity. The integral decreases upon cooling and contributes to the viscosity minimum discussed in the main text.
\end{figure}

\vspace*{\fill}

\begin{figure}[htbp]
\includegraphics[width=11cm]{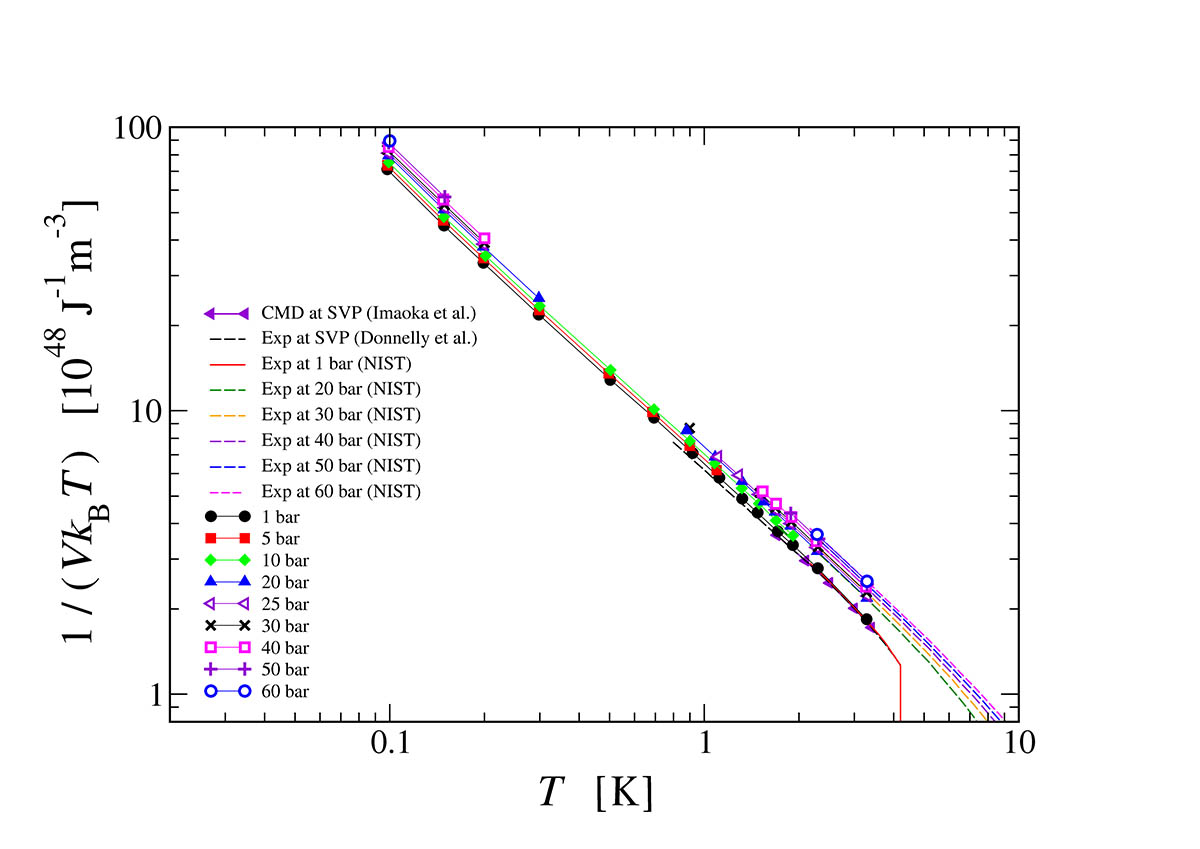}
\caption{\label{fig:prefactor_shear} 
Temperature dependence of prefactor $1/(Vk_{\rm B}T)$ in Eq.  (\ref{eq:shearviscosity}).  No color coding is used to distinguish different states.
}

\vspace{2mm} \noindent\textbf{Alt Text:} Figure S27 plots the prefactor appearing in the shear viscosity expression. The prefactor increases monotonically with decreasing temperature.
\end{figure}

\vspace*{\fill}
\clearpage
\vspace*{\fill}

\begin{figure}[htbp]
\includegraphics[width=11cm]{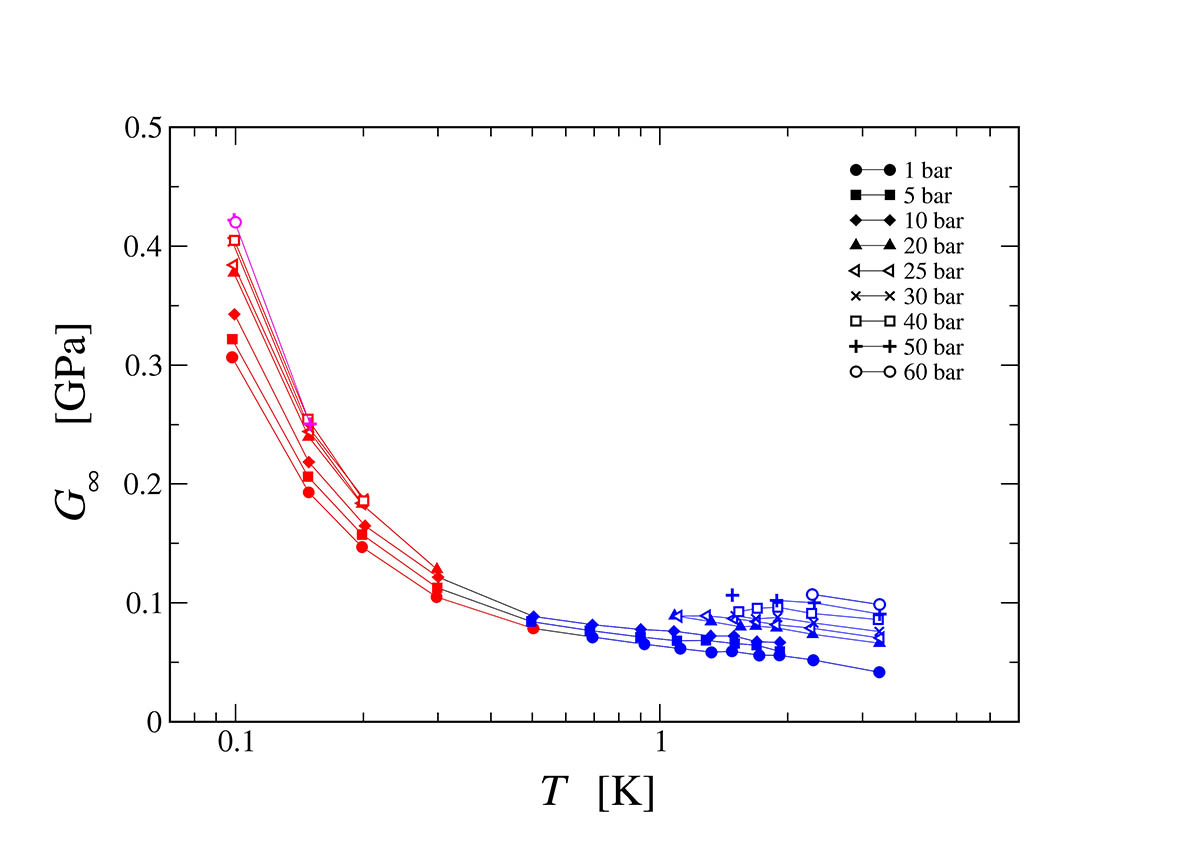}
\caption{\label{fig:highfreq_shear} 
Temperature dependence of high-frequency shear modulus $G_{\infty}$: LQDL (blue), LQDA (cyan), HQDL (red), and HQDA (magenta).
}

\vspace{2mm} \noindent\textbf{Alt Text:} Figure S28 shows the high-frequency shear modulus as a function of temperature. The modulus increases upon cooling, indicating a progressively stronger instantaneous shear response.
\end{figure}

\vspace*{\fill}

\begin{figure}[htbp]
\includegraphics[width=11cm]{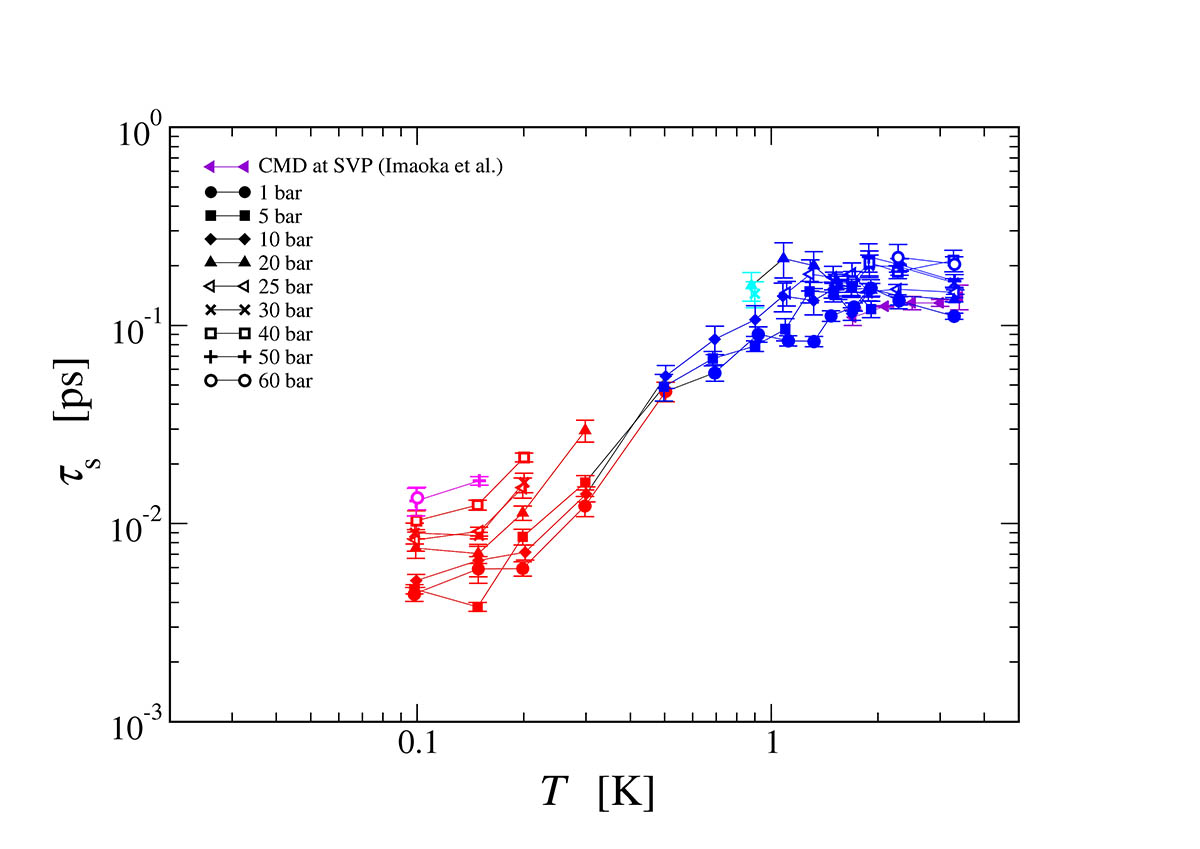}
\caption{\label{fig:kanwa_shear} 
Temperature dependence of relaxation time $\tau_{\rm s}$ of off-diagonal stress autocorrelation function: LQDL (blue), LQDA (cyan), HQDL (red), and HQDA (magenta).
}

\vspace{2mm} \noindent\textbf{Alt Text:} Figure S29 plots the relaxation time of the stress autocorrelation function. The relaxation becomes substantially faster at lower temperatures in the high quantum dispersion regimes.
\end{figure}

\vspace*{\fill}

\clearpage
\onecolumngrid
\section{Energy current autocorrelation functions}
\label{sec:SMECA}

\vspace*{\fill}
\begin{figure}[H]
\begin{center}
    \includegraphics[width=\columnwidth]{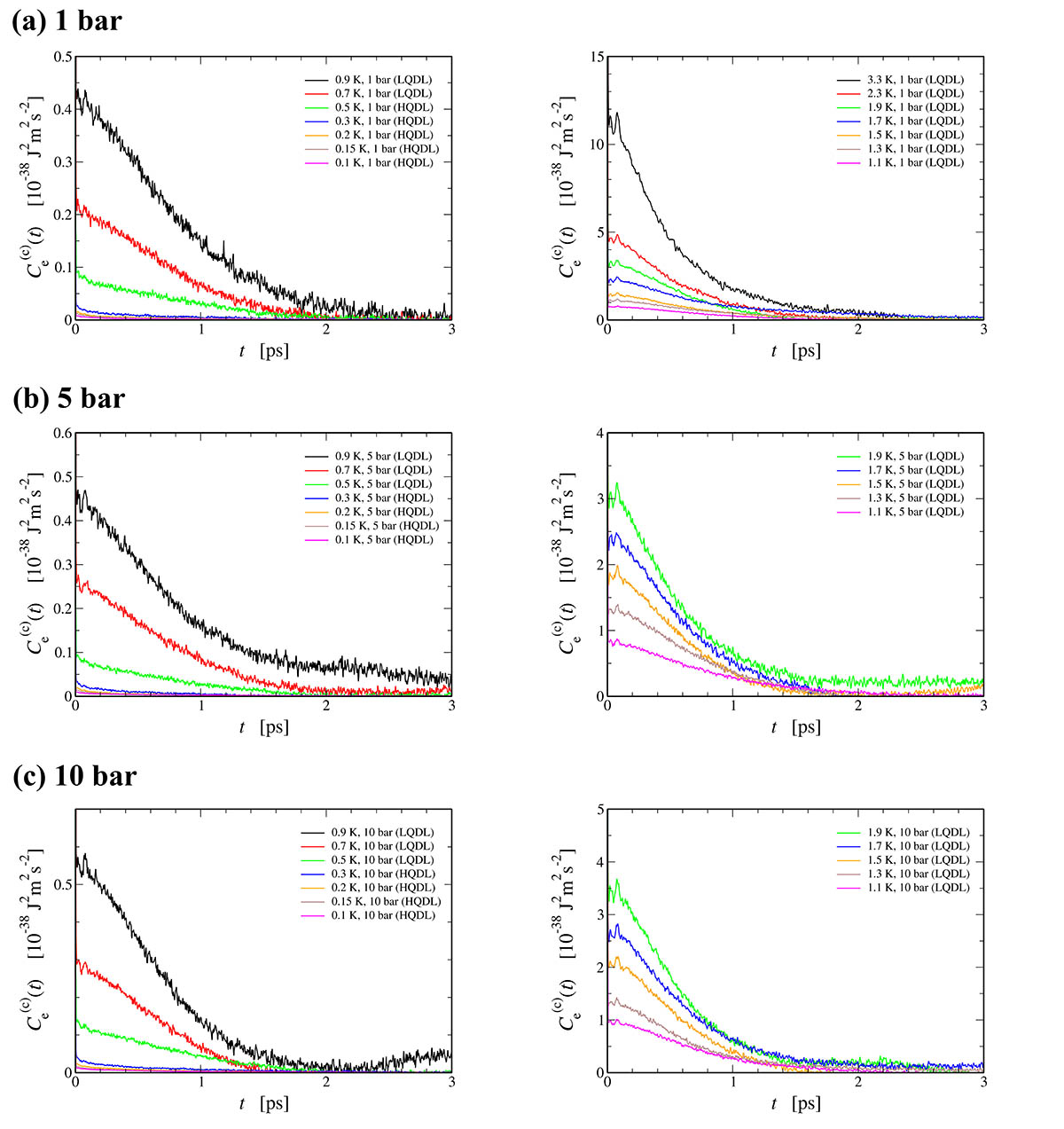}
    \caption{Energy current autocorrelation functions at densities corresponding to 1-10 bar.}
    \label{fig:SM_Ener1}
\end{center}

\vspace{2mm} \noindent\textbf{Alt Text:} Figure S30 shows energy current autocorrelation functions at pressures from 1 to 10 bar. The relaxation behavior governs the thermal conductivity evaluated from the centroid Green-Kubo relation.
\end{figure}
\vspace*{\fill}

\clearpage

\vspace*{\fill}
\begin{figure}[H]
\begin{center}
    \includegraphics[width=\columnwidth]{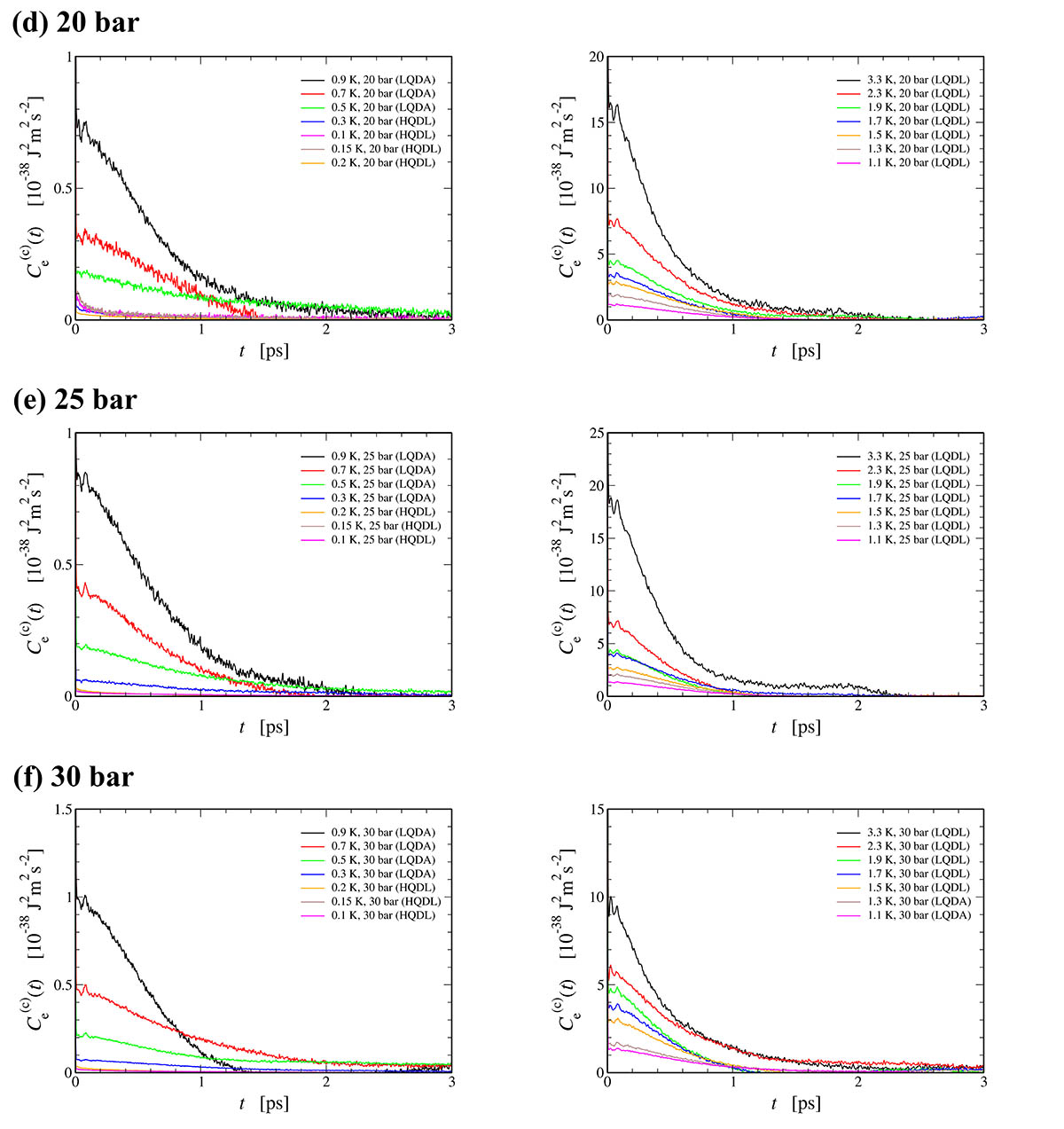}
    \caption{Energy current autocorrelation functions at densities corresponding to 20-30 bar.}
    \label{fig:SM_Ener2}
\end{center}

\vspace{2mm} \noindent\textbf{Alt Text:} Figure S31 shows energy current autocorrelation functions at pressures from 20 to 30 bar. The results illustrate the dependence of the relaxation profiles on both pressure and temperature.
\end{figure}
\vspace*{\fill}

\clearpage

\vspace*{\fill}
\begin{figure}[H]
\begin{center}
    \includegraphics[width=\columnwidth]{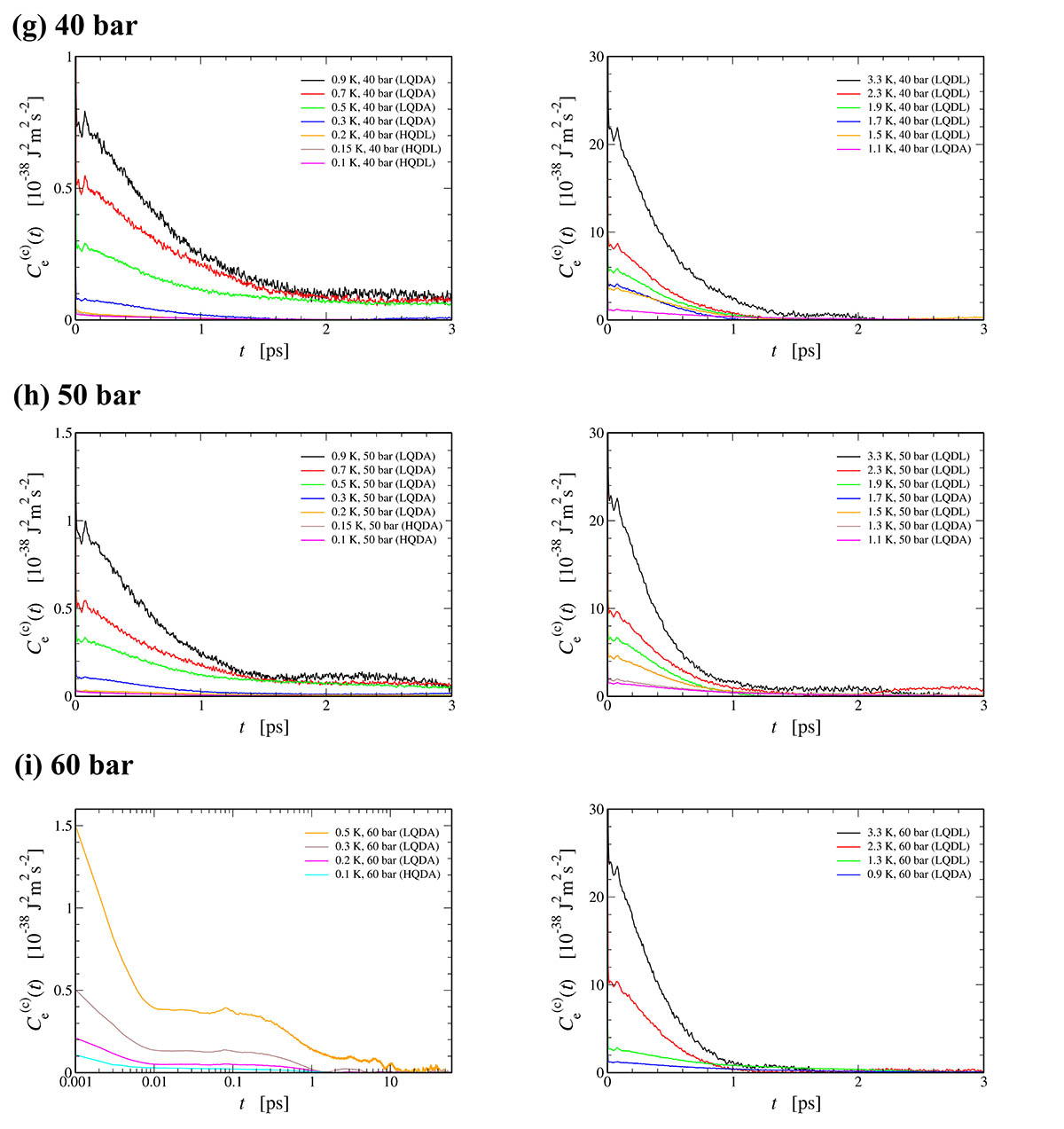}
    \caption{Energy current autocorrelation functions at densities corresponding to 40-60 bar.}
    \label{fig:SM_Ener3}
\end{center}

\vspace{2mm} \noindent\textbf{Alt Text:} Figure S32 shows energy current autocorrelation functions at pressures from 40 to 60 bar. The results illustrate pressure-dependent differences in collective energy transport.
\end{figure}
\vspace*{\fill}

\clearpage

\section{Analysis of thermal conductivity}
\label{sec:SMATC}
\onecolumngrid
\vspace*{\fill}

\begin{figure}[htbp]
\includegraphics[width=11cm]{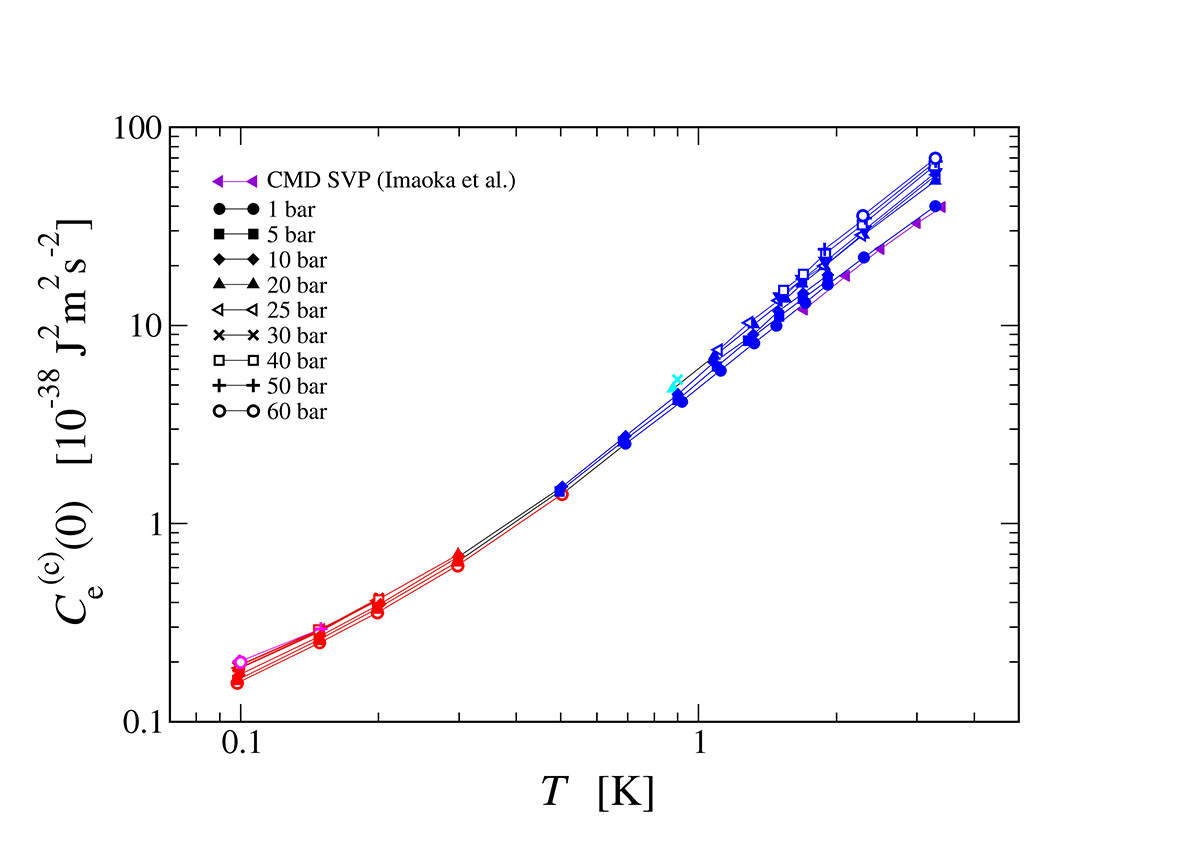}
\caption{\label{fig:seppen_ener} 
Temperature dependence of initial value $C_{\rm e}^{\rm (c)}(0)$ of energy current autocorrelation functions: LQDL (blue), LQDA (cyan), HQDL (red), and HQDA (magenta).
}

\vspace{2mm} \noindent\textbf{Alt Text:} Figure S33 plots the initial value of the energy current autocorrelation function versus temperature. The magnitude decreases markedly as temperature is lowered.
\end{figure}
\vspace*{\fill}

\begin{figure}[htbp]
\includegraphics[width=11cm]{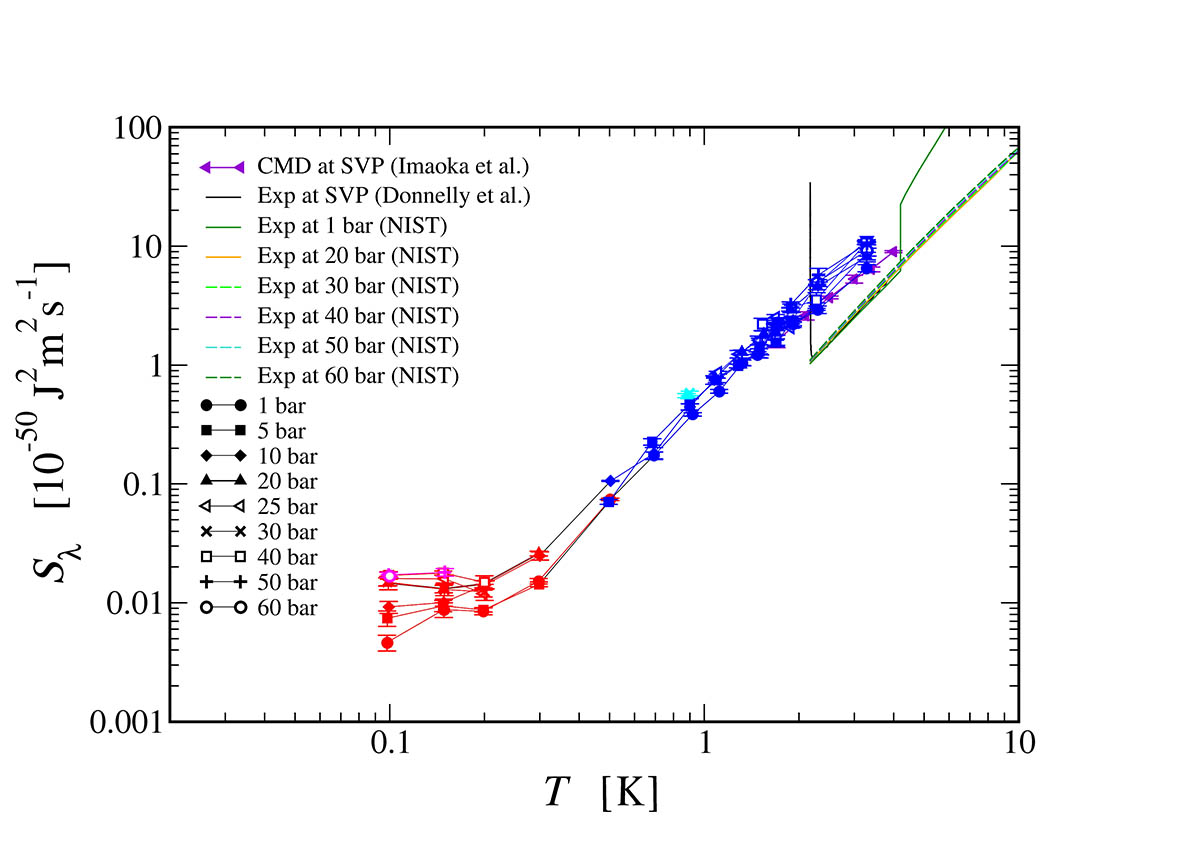}
\caption{\label{fig:sekibun_ener} 
Temperature dependence of time integral $S_{\lambda}$ of energy current autocorrelation functions: LQDL (blue), LQDA (cyan), HQDL (red), and HQDA (magenta).
}

\vspace{2mm} \noindent\textbf{Alt Text:} Figure S34 shows the time integral contribution to the thermal conductivity. The integral generally decreases with cooling and influences the emergence of the conductivity minimum.
\end{figure}
\vspace*{\fill}

\clearpage
\vspace*{\fill}
\begin{figure}[htbp]
\includegraphics[width=11cm]{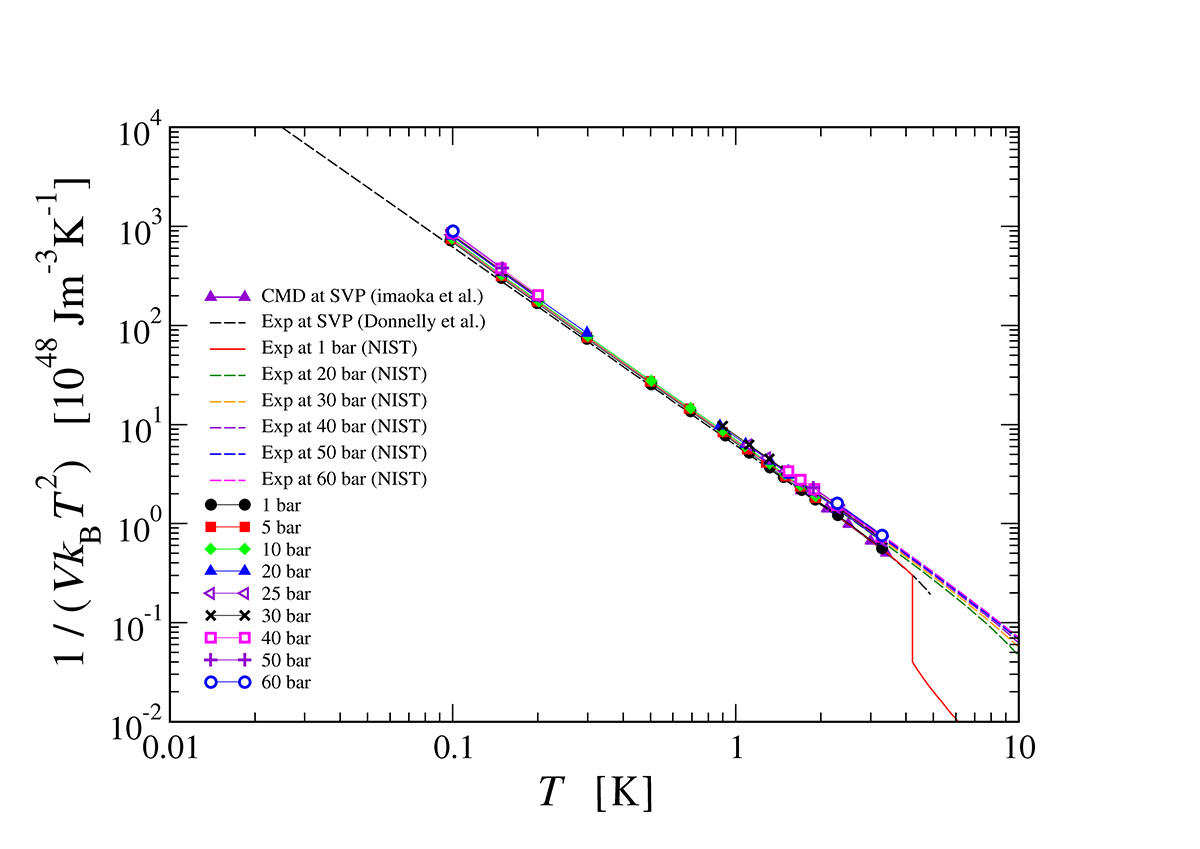}
\caption{\label{fig:prefactor_ener} 
Temperature dependence of prefactor $1/(Vk_{\rm B}T^2)$ in Eq.  (\ref{eq:thermalconductivity}).  No color coding is used to distinguish different states.
}

\vspace{2mm} \noindent\textbf{Alt Text:} Figure S35 plots the prefactor appearing in the thermal conductivity expression. The prefactor increases steadily as temperature decreases.
\end{figure}

\vspace*{\fill}

\begin{figure}[htbp]
\includegraphics[width=11cm]{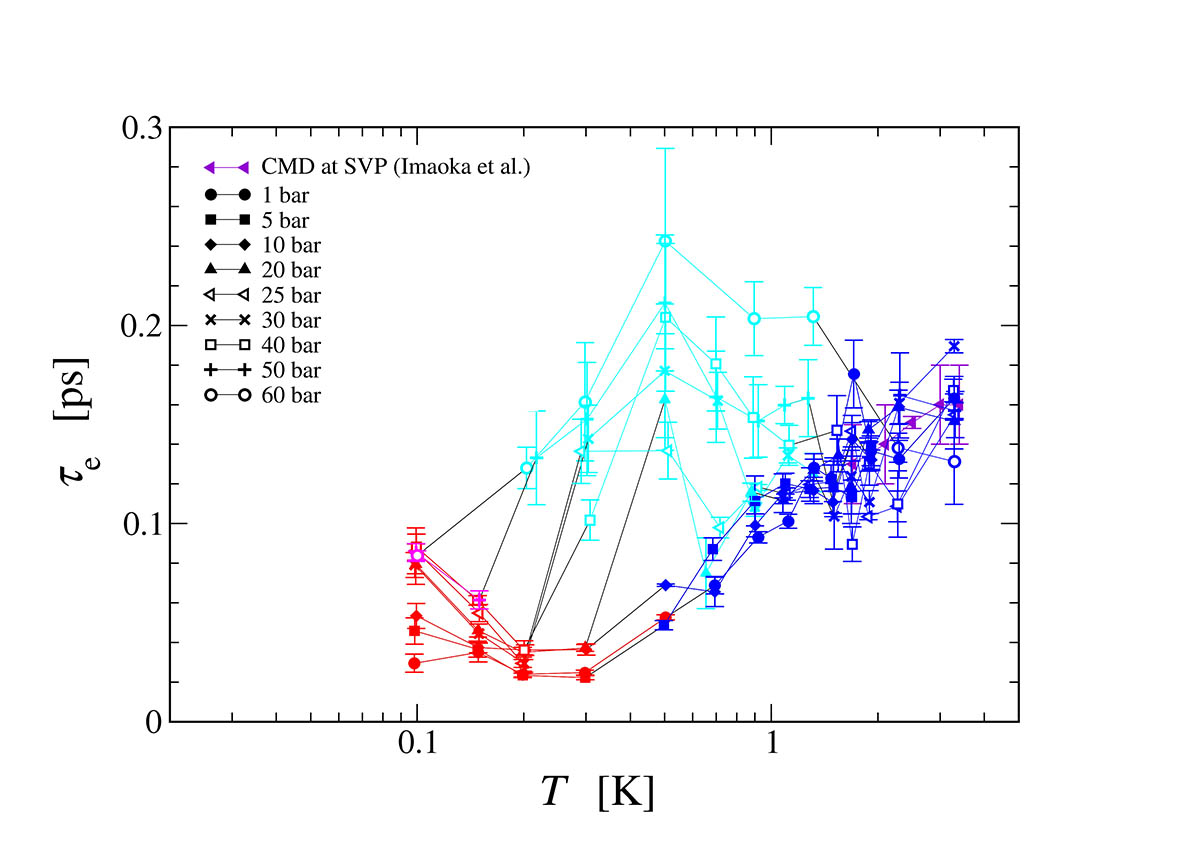}
\caption{\label{fig:kanwa_ener} 
Temperature dependence of relaxation time $\tau_{\rm e}$ of energy current autocorrelation functions: LQDL (blue), LQDA (cyan), HQDL (red), and HQDA (magenta).
}

\vspace{2mm} \noindent\textbf{Alt Text:} Figure S36 shows the relaxation time of the energy current autocorrelation function. The relaxation time varies substantially among the liquid and amorphous states and contributes to differences in thermal conductivity.
\end{figure}

\vspace*{\fill}

\clearpage
\onecolumngrid
\section{Table}
\label{sec:SMTable}

\begin{table}[H]
\caption{\label{tab:xi} The fitted exponent $\xi$ in the fractional Stokes-Einstein relation.}

\vspace{2mm} \noindent\textbf{Alt Text:} Table S1 summarizes the fitted exponent $\xi$ in the fractional Stokes-Einstein relation for the low and high quantum dispersion liquids. The high quantum dispersion liquid exhibits substantially smaller exponents than the low quantum dispersion liquid, indicating a stronger breakdown of the conventional Stokes-Einstein relation.
\\
\begin{ruledtabular}
\begin{tabular}{ccc}
Pressure of the preceding NPT run [bar] &  $\xi_{\rm{HQDL}}$  &   $\xi_{\rm{LQDL}}$   \\
\hline
1 &  0.29 & 0.99  \\
5 &  0.27 & 0.83 \\
10 &  0.30  & 0.98 \\
20 &  0.23  & 1.0\\
\end{tabular}
\end{ruledtabular}
\end{table}

\end{supplemental}

\end{document}